\documentclass{emulateapj}
\usepackage{graphicx}
\usepackage{amssymb,amsfonts,amsmath,amstext,amsgen,amsopn,amsxtra,indentfirst,times}

\begin{document}

\title{Analytical Models of Exoplanetary Atmospheres. \\I. Atmospheric Dynamics via the Shallow Water System}
\author{Kevin Heng\altaffilmark{1}}
\author{Jared Workman\altaffilmark{2}}
\altaffiltext{1}{University of Bern, Center for Space and Habitability, Sidlerstrasse 5, CH-3012, Bern, Switzerland.  Email: kevin.heng@csh.unibe.ch}
\altaffiltext{2}{Colorado Mesa University, 1260 Kennedy Avenue, Grand Junction, CO 81501, U.S.A.  Email: jworkman@coloradomesa.edu}

\begin{abstract}
Within the context of exoplanetary atmospheres, we present a comprehensive linear analysis of forced, damped, magnetized shallow water systems, exploring the effects of dimensionality, geometry (Cartesian, pseudo-spherical and spherical), rotation, magnetic tension and hydrodynamic and magnetic sources of friction.  Across a broad range of conditions, we find that the key governing equation for atmospheres and quantum harmonic oscillators are identical, even when forcing (stellar irradiation), sources of friction (molecular viscosity, Rayleigh drag and magnetic drag) and magnetic tension are included.  The global atmospheric structure is largely controlled by a single, key parameter that involves the Rossby and Prandtl numbers.  This near-universality breaks down when either molecular viscosity or magnetic drag acts non-uniformly across latitude or a poloidal magnetic field is present, suggesting that these effects will introduce qualitative changes to the familiar chevron-shaped feature witnessed in simulations of atmospheric circulation.  We also find that hydrodynamic and magnetic sources of friction have dissimilar phase signatures and affect the flow in fundamentally different ways, implying that using Rayleigh drag to mimic magnetic drag is inaccurate.  We exhaustively lay down the theoretical formalism (dispersion relations, governing equations and time-dependent wave solutions) for a broad suite of models.  In all situations, we derive the steady state of an atmosphere, which is relevant to interpreting infrared phase and eclipse maps of exoplanetary atmospheres.  We elucidate a pinching effect that confines the atmospheric structure to be near the equator.  Our suite of analytical models may be used to decisively develop physical intuition and as a reference point for three-dimensional, magnetohydrodynamic (MHD) simulations of atmospheric circulation. 
\end{abstract}


\section{Introduction}

\subsection{Motivation}

With the atmospheres of exoplanets now accessible to astronomical scrutiny, there is motivation to understand the basic physics governing their structure.  Since highly-irradiated exoplanets are most amenable to atmospheric characterization, a growing body of work has focused on hot Earths/Neptunes/Jupiters, ranging from analytical models to simulations of atmospheric circulation (e.g., \citealt{dd08,showman09,hmp11,tc11,sp11,rm13,showman13,batygin13,rogers14}), in one, two and three dimensions (1D, 2D and 3D).  The path towards full understanding requires the construction of a hierarchy of theoretical models of varying sophistication \citep{held05}.  In this context, analytical models have a vital role to play, since they provide crisp physical insight and are immune to numerical issues (e.g., numerical viscosity, sub-grid physics, spin-up).

Atmospheres behave like heat engines.  Sources of forcing (e.g., stellar irradiation) induce atmospheric motion, which are eventually damped out by sources of friction (e.g., viscosity, magnetic drag).  It is essential to understand atmospheric dynamics, as it sets the background state of velocity, temperature, density and pressure that determines the spectral and temporal appearance of an atmosphere.  It also determines whether an atmosphere attains or is driven away from chemical, radiative and thermodynamic equilibrium.  Even if an atmosphere is not in equilibrium, it must be in a global state of equipoise---sources of forcing and friction negate one another (e.g., \citealt{goodman09}).  In the present study, this is our over-arching physical goal: to analytically derive the global, steady state of an exoplanetary atmosphere (the ``exoclime") in the presence of forcing, friction, rotation and magnetic fields.

\begin{figure}
\begin{center}
\includegraphics[width=\columnwidth]{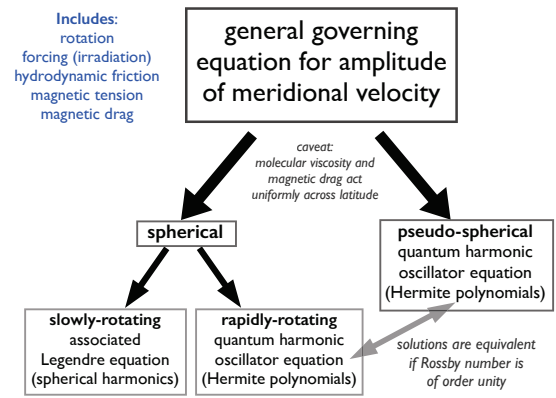}
\end{center}
\vspace{-0.2in}
\caption{Schematic describing the key governing equation in shallow water systems, both on the equatorial $\beta$-plane (pseudo-spherical geometry) and in full spherical geometry.  The key quantity to solve for is the meridional (north-south) velocity, from which the zonal (east-west) velocity, shallow water height perturbation and magnetic field perturbations straightforwardly follow.}
\vspace{0.1in}
\label{fig:schematic}
\end{figure}

Shallow water models are a decisive way of studying exoclimes.  The term ``shallow water" comes from their traditional use in meteorology and oceanography \citep{matsuno66,lindzen67,lh67,lh68,gill80} and refers to the approximation that the horizontal extent modeled far exceeds the vertical/radial one.  They have been used to understand the solar tacocline \citep{gilman00,zara07}, the atmospheres of neutron stars \citep{spit02,hs09} and exoplanetary atmospheres \citep{sp11,hk12}.  Our over-arching technical goal is to perform a comprehensive theoretical survey across dimensionality (1D and 2D), geometry (Cartesian, pseudo-spherical and spherical) and sources of friction (molecular viscosity, Rayleigh drag and magnetic drag) for forced, damped, rotating, magnetized systems.  To retain algebraic tractability, we study the limits of purely vertical/radial or horizontal/toroidal background magnetic fields.

Our main finding is that the global structure of exoplanetary atmospheres is largely controlled by a single, key parameter---at least in the shallow water approximation---which we denote by $\alpha$.  In the hydrodynamic limit, $\alpha$ is directly related to the Rossby and Prandtl numbers.  In forced, damped atmospheres with non-ideal MHD, $\alpha$ encapsulates the effects of molecular viscosity, Rayleigh drag, forcing, magnetic tension and magnetic drag.

\subsection{Terminology}

Due to the technical nature of the present study, we find it useful to concisely summarize a set of terminology that we will use throughout the paper.  

From a set of perturbed equations, we will obtain wave solutions for the velocity, height and magnetic field perturbations.  We assume that the waves have a temporal component of the form $\exp{(-i \omega t)}$, where $\omega$ is the wave frequency.  Generally, the wave frequency has real and imaginary components ($\omega = \omega_{\rm R} + i \omega_{\rm I}$), which describe the oscillatory and growing or decaying parts of the wave, respectively.  For each model, we will obtain a pair of equations describing $\omega_{\rm R}$ and $\omega_{\rm I}$, which we call the ``oscillatory dispersion relation" and the ``growth/decay dispersion relation", respectively.  We will refer to them collectively as ``dispersion relations".  A ``balanced flow" corresponds to the situation when $\omega_{\rm I}=0$.  The ``steady state" of an atmosphere has $\omega_{\rm R}=\omega_{\rm I}=0$.  

We will refer to molecular viscosity, Rayleigh drag and magnetic drag collectively as ``friction".  If we are only considering molecular viscosity and Rayleigh drag, we will use the term ``hydrodynamic friction".  We will use the terms ``friction" and ``damping" interchangeably, but not qualify the latter as being hydrodynamic or MHD in nature.  We will refer to systems as being ``free" if forcing and friction are absent.  Similarly, we will use the terms ``fast" and ``rapid" interchangeably when referring to rotation.  The zeroth order effect of including a magnetic field introduces a term each into the momentum and magnetic induction equations---we will refer to the effects of these ideal-MHD terms as ``magnetic tension".  When non-ideal MHD is considered, we include a resistive term in the induction equation that mathematically resembles diffusion---we will refer to its influence as ``magnetic drag".

We will examine an approximation for including the effects of non-constant rotation, across latitude, on a Cartesian grid, known traditionally as the ``$\beta$-plane approximation" (e.g., \citealt{vallis06} and references therein).  Note that this is \textit{not} the same as a departure from solid body rotation.  Rather, it is an approximation to include the dynamical effects of sphericity.  There are two flavors of this approximation: the simpler version solves for waves that are oscillatory in both spatial dimensions (simply called ``$\beta$-plane"), while a more sophisticated version allows for an arbitrary functional dependence in latitude (``equatorial $\beta$-plane").  Since the equatorial $\beta$-plane treatment more closely approximates the situation on a sphere, we will refer to it as being ``pseudo-spherical".

In constructing the mathematical machinery in this paper, we often have to evaluate long, complex\footnote{Here, it means having both real and imaginary components, rather than ``complicated".} expressions.  To this end, we find it convenient to separate out the real and imaginary components using a series of ``separation functions", which we denote by $\zeta_+$, $\zeta_-$, $\zeta_{\rm R}$, $\zeta_{\rm I}$, $\zeta$ and $\zeta_0$.  The definitions of these dimensionless quantities vary from model to model.  We note that $\zeta_0$ also functions like a generalized friction that includes magnetic tension.

\subsection{Layout of Paper}

In \S\ref{sect:equations}, we state the governing equations and derive their linearized, perturbed forms.  In \S\ref{sect:1d}, we review and extend the 1D models.  We extend our models to 2D Cartesian geometry in \S\ref{sect:2d} and begin to consider the effects of sphericity in \S\ref{sect:2d_beta}.  In \S\ref{sect:spherical}, we present results for 2D models in spherical geometry.  Applications to exoplanetary atmospheres are described in \S\ref{sect:apps}.  We summarize our findings in \S\ref{sect:discussion}.  Table 1 lists the most commonly used quantities and the symbols used to denote them.  Table 2 compares our study with previous analytical work.  Table 3 summarizes the salient lessons learnt from studying each shallow water model.  Figure \ref{fig:schematic} provides a graphical summary of our technical achievements.

\begin{table}
\label{tab:symbols}
\begin{center}
\caption{Commonly Used Symbols}
\begin{tabular}{lcc}
\hline
\hline
Name & Units & Meaning \\
\hline
$g$ & cm s$^{-2}$ & surface gravity \\
$H$ & cm & shallow water height at rest \\
$R$ & cm & radius of exoplanet \\
$\Omega$ & s$^{-1}$ & rotational frequency \\
$t_{\rm dyn}$ & s & dynamical timescale \\
$\beta \equiv 2\Omega\sin\theta/R$ & cm$^{-1}$ s$^{-1}$ & gradient of rotation term ($\beta$-plane) \\
$\bar{B}_j$ & G & background magnetic field strength \\
$c_0 \equiv \left( gH \right)^{1/2}$ & cm s$^{-1}$ & gravity wave speed \\
$v_{\rm A} \equiv \bar{B}_j/2(\pi \rho)^{1/2}$ & cm s$^{-1}$ & Alfv\'{e}n speed \\
$\nu$ & cm$^2$ s$^{-1}$ & molecular viscosity \\
$\eta$ & cm$^2$ s$^{-1}$ & magnetic resistivity/diffusivity \\ 
$k_{\rm x}$ & cm$^{-1}$, $\dagger$ & zonal wavenumber ($\beta$-plane) \\
$n$ & --- & meridional wavenumber ($\beta$-plane) \\
$m$ & --- & zonal wavenumber (spherical) \\
$l$ & --- & meridional wavenumber (spherical) \\
\hline
${\cal R}$ & --- & Reynolds number \\
${\cal R}_{\rm B}$ & --- & magnetic Reynolds number \\
${\cal R}_0$ & --- & Rossby number \\
$\xi \equiv 1/{\cal R}_0^2$ & --- & Lamb's parameter \\
$\zeta_+$, $\zeta_-$, $\zeta_{\rm R}$, $\zeta_{\rm I}$, $\zeta$ & --- & separation functions \\
$\zeta_0$ & --- & generalized friction \\
$\alpha$ & --- & key controlling parameter \\
\hline
$\omega_{\rm drag} \equiv t_{\rm drag}^{-1}$ & s$^{-1}$, $\dagger$ & Rayleigh drag \\
$\omega_\nu$ & s$^{-1}$, $\dagger$ & hydrodynamic friction \\
$F_0 \equiv S_0 - \omega_{\rm rad}$ & s$^{-1}$, $\dagger$ & forcing \\
$\omega_0 \equiv \omega + i \omega_\nu$ & s$^{-1}$, $\dagger$ & --- \\
$\omega_{\rm F} \equiv \omega - i F_0$ & s$^{-1}$, $\dagger$ & --- \\
$\omega_{\rm B_0} \equiv \omega + i \omega_{\rm B}$ & --- & --- \\
\hline
$\tilde{y} \equiv \alpha y$ & --- & transformed latitude ($\beta$-plane) \\
$\mu \equiv \cos\theta$ & --- & cosine of co-latitude (spherical) \\
$\tilde{\mu} \equiv \alpha \mu$ & --- & transformed co-latitude (spherical) \\
${\cal P}^m_l$ & --- & associated Legendre function \\
${\cal H}_n$ or ${\cal H}_l$ & --- & Hermite polynomial \\
$\tilde{{\cal H}}_n \equiv {\cal H}_n(\tilde{y})$ & --- & Hermite polynomial \\
$\tilde{{\cal H}}_l \equiv {\cal H}_l(\tilde{\mu})$ & --- & Hermite polynomial \\
\hline
$v_0$ & --- & arbitrary velocity normalization \\
$v^\prime_{\rm x}$ or $v^\prime_\phi$ & cm s$^{-1}$, $\dagger$ & zonal velocity perturbation \\
$v^\prime_{\rm y}$ or $v^\prime_\theta$ & cm s$^{-1}$, $\dagger$ & meridional velocity perturbation \\
$h^\prime$ & cm, $\dagger$ & shallow water height perturbation \\
$h^\prime_{\rm v} \equiv gh^\prime/2\Omega R$ & cm s$^{-1}$ & --- \\
$b^\prime_{\rm x}$ or $b^\prime_\phi$ & G, $\dagger$ & toroidal magnetic field perturbation \\
$b^\prime_{\rm y}$ or $b^\prime_\theta$ & G, $\dagger$ & poloidal magnetic field perturbation \\
$v_{\rm x_0}$ or $v_{\phi_0}$ & cm s$^{-1}$, $\dagger$ & normalization of $v^\prime_{\rm x}$ or $v^\prime_\phi$ \\
$v_{\rm y_0}$ or $v_{\theta_0}$ & cm s$^{-1}$, $\dagger$ & normalization of $v^\prime_{\rm y}$ or $v^\prime_\theta$ \\
$h_0$ & cm, $\dagger$ & normalization of $h^\prime$ \\
$h_{\rm v_0}$ & cm s$^{-1}$ & normalization of $h^\prime_{\rm v}$ \\
$b_{\rm x_0}$ or $b_{\phi_0}$ & G, $\dagger$ & normalization of $b^\prime_{\rm x}$ or $b^\prime_\phi$ \\
$b_{\rm y_0}$ or $b_{\theta_0}$ & G, $\dagger$ & normalization of $b^\prime_{\rm y}$ or $b^\prime_\theta$ \\
\hline
\end{tabular}
Note: the index $j$ refers to the $x$-, $y$-,$z$-, $r$-, $\theta$- or $\phi$-coordinate.\\
Definitions are given only when they do not vary between models.\\
$\dagger$: quantity may be cast in dimensionless units.\\
\end{center}
\end{table}

\section{Governing Equations of the Shallow Water System}
\label{sect:equations}

\subsection{General Equations}

By including molecular viscosity, Rayleigh drag and magnetic tension in the conservation of linear momentum, we obtain the following equation,
\begin{equation}
\frac{\partial \vec{v}}{\partial t} + \vec{v}.\nabla \vec{v} = -g \nabla h - 2 \vec{\Omega} \times \vec{v} + \nu \nabla^2 \vec{v} - \frac{\vec{v}}{t_{\rm drag}} + \frac{\vec{B}. \nabla\vec{B}}{4\pi \rho},
\label{eq:momentum}
\end{equation}
where $\vec{v}$ is the 2D horizontal velocity vector, $t$ denotes the time, $\vec{\Omega}$ is the planetary rotation vector (pointing north), $\nu$ is the kinematic viscosity, $t_{\rm drag}$ is a constant drag timescale, $\vec{B}$ is the magnetic field strength and $\rho$ is the mass density.  Generally, the momentum equation contains a Lorentz force term, which is $\propto (\nabla \times \vec{B}) \times \vec{B}$ and consists of two terms representing magnetic tension and magnetic pressure.  We have assumed that magnetic pressure (associated with the term $-\nabla P_{\rm B}/\rho$, where $P_{\rm B} = B^2/8\pi$) is negligible compared to thermal pressure.  The remaining magnetic term represents magnetic tension.  Note that equation (\ref{eq:momentum}) is only meaningful in the horizontal directions.  It uses the fact that when $\rho$ is constant, hydrostatic equilibrium yields a linear relationship between $h$ and the pressure $P$,
\begin{equation}
P = P_0 + \rho g \left( h - z \right),
\end{equation}
with $P_0$ being the reference pressure defined at $z=h$ and $z$ being the vertical coordinate.

A convenient hallmark of the shallow water model is that the mass continuity and thermodynamic equations are replaced by an equation for the shallow water height $h$,
\begin{equation}
\frac{\partial h}{\partial t} + \nabla . \left( h \vec{v} \right) = Q,
\label{eq:height}
\end{equation}
where the preceding equation is also only meaningful in the horizontal directions.  The term $Q$ mimics the effect of radiative heating,
\begin{equation}
Q = \frac{h_{\rm eq} - h}{t_{\rm rad}},
\end{equation}
where $h_{\rm eq}$ is the ``equilibrium shallow water height", attained in the event of radiative equilibrium.  Such an approach is often termed ``Newtonian relaxation".  When $Q=0$, equation (\ref{eq:height}) is the 2D analogue of the mass continuity equation in 3D.

When magnetic fields are included, one needs to consider the 3D magnetic induction equation.  We assume the terms associated with ambipolar diffusion and the Hall effect to be negligible, at least for highly-irradiated atmospheres \citep{pmr10a}.  To render the problem analytically tractable, we assume that the magnetic diffusivity/resistivity ($\eta$) has no spatial dependence ($\nabla \eta = 0$).  With these simplications, we have
\begin{equation}
\begin{split}
\frac{\partial \vec{B}}{\partial t} =& \nabla \times \left( \vec{v} \times \vec{B} \right) - \nabla \times \left[ \eta \left( \nabla \times \vec{B} \right) \right]  \\
=& \vec{v} \left( \nabla . \vec{B} \right) - \vec{B} \left( \nabla . \vec{v} \right) + \vec{B} . \nabla\vec{v} - \vec{v} . \nabla\vec{B} \\
&+ \eta \nabla^2 \vec{B} - \eta \nabla\left(\nabla . \vec{B} \right) - \nabla\eta \times \left( \nabla \times \vec{B} \right) \\
=& \vec{v} \left( \nabla . \vec{B} \right) + \vec{B} . \nabla\vec{v} + \eta \nabla^2 \vec{B} . \\
\end{split}
\label{eq:magnetic}
\end{equation}
The $-\vec{B} ( \nabla . \vec{v} )$ term always vanishes for a shallow water system due to the condition of incompressibility, while the $-\vec{v} . \nabla\vec{B}$ term may be neglected for a system perturbed from rest  and if the background magnetic field is constant (as it is then second order in magnitude).  The terms involving $\eta$ describe the effect of magnetic drag.  As a first approximation, we consider only the $\eta \nabla^2 \vec{B}$ term among the ones involving $\eta$.  In atmospheres where a balance between collisional ionization and recombination is attained, $\eta$ has an exponential dependence on the temperature and may vary by orders of magnitude \citep{pmr10a}.  Nevertheless, idealized models with constant $\eta$ provide a starting point for MHD shallow-water investigations and are useful for elucidating the phase behavior of magnetic drag.

To simplify the problem, we will make separate approximations for the first and second derivatives of the components of the magnetic field strengths.  The exact form of these approximations depends on whether we are considering a purely vertical/radial or horizontal background field.

\subsection{Linearized, Perturbed Equations in Cartesian Coordinates}

We perturb the system about a state of rest: $v_{\rm x,y} = v^\prime_{\rm x,y}$ and $h = H + h^\prime$, where $H$ is a constant and $H \gg h^\prime$.  In Cartesian coordinates, the equation for the shallow water height becomes
\begin{equation}
\frac{\partial h^\prime}{\partial t} + H \left( \frac{\partial v^\prime_{\rm x}}{\partial x} + \frac{\partial v^\prime_{\rm y}}{\partial y} \right) = S - \omega_{\rm rad} h^\prime,
\end{equation}
where $\omega_{\rm rad} \equiv 1/t_{\rm rad}$ and the source term is $S \equiv \left( h_{\rm eq} - H \right) \omega_{\rm rad}$.  We assume $S = S_0 h^\prime$ and define $F_0 \equiv S_0 - \omega_{\rm rad}$, as will be made clear in \S\ref{subsect:1d_forcing}.

The forms of the momentum and magnetic induction equations depend on whether one is considering the idealized situation of a purely vertical or horizontal background magnetic field.  We perturb the magnetic field about its constant, background vertical or horizontal value.  Even for an initial field that is purely vertical in nature, disturbances to it introduce finite horizontal perturbations.

\subsubsection{Vertical Background Magnetic Field}

We begin by denoting the components of the \emph{total} magnetic field strength by $B_{\rm x,y,z}$.  We make the following approximations:
\begin{equation}
\begin{split} 
&\frac{\partial v_{\rm x,y}}{\partial z} = 0, \\
&\frac{\partial B_{\rm x,y}}{\partial x} \ll \frac{\partial B_{\rm z}}{\partial z}, ~\frac{\partial B_{\rm x,y}}{\partial y} \ll \frac{\partial B_{\rm z}}{\partial z}, \\
&\frac{\partial B_{\rm x,y}}{\partial z} = -\frac{B_{\rm x,y}}{H}, ~\frac{\partial B_{\rm z}}{\partial z} = \frac{B_{\rm z}}{H}, ~\nabla^2 B_{\rm x,y} \ne 0. \\
\end{split}
\label{eq:vertical_cartesian_approx}
\end{equation}
The first statement in equation (\ref{eq:vertical_cartesian_approx}) is a property of the shallow water system.  These approximations ensure that the dominant magnetic term in the momentum equation produces a restoring force that depends on the background vertical field \citep{hs09}, while the induction equation retains a term that is the magnetic analogue of molecular diffusion.  The negative signs associated with the horizontal field components are intended to provide a restoring, rather than a runaway, force to any perturbation of the magnetic field by fluid motion.  

We represent the constant, \emph{background} magnetic field strengths by $\bar{B}_{\rm x,y,z}$.  We perturb about the horizontal field strengths,
\begin{equation}
B_{\rm x,y} = \bar{B}_{\rm x,y} + b^\prime_{\rm x,y}.
\end{equation}
The background state is simple: at rest, with $\bar{B}_{\rm x,y}=0$ and $B_{\rm z} = \bar{B}_{\rm z}$.  The vertical field is not perturbed.

With these assumptions, the linearized momentum equations read,
\begin{equation}
\begin{split}
\frac{\partial v^\prime_{\rm x}}{\partial t} =& -g \frac{\partial h^\prime}{\partial x} + f v^\prime_{\rm y} + \nu \left(\frac{\partial^2 v^\prime_{\rm x}}{\partial x^2} + \frac{\partial^2 v^\prime_{\rm x}}{\partial y^2} \right) \\
&- \omega_{\rm drag} v^\prime_{\rm x} - \frac{\bar{B}_{\rm z} b^\prime_{\rm x}}{4 \pi \rho H}, \\
\frac{\partial v^\prime_{\rm y}}{\partial t} =& -g \frac{\partial h^\prime}{\partial y} - f v^\prime_{\rm x} + \nu \left(\frac{\partial^2 v^\prime_{\rm y}}{\partial x^2} + \frac{\partial^2 v^\prime_{\rm y}}{\partial y^2} \right) \\
&- \omega_{\rm drag} v^\prime_{\rm y} - \frac{\bar{B}_{\rm z} b^\prime_{\rm y}}{4 \pi \rho H}, \\
\end{split}
\end{equation}
where we have defined a drag frequency $\omega_{\rm drag} \equiv t^{-1}_{\rm drag}$.  The quantity $f$ generalizes upon $2\Omega$ and is the first step towards including the effects of sphericity (e.g., \citealt{vallis06})
\begin{equation}
f = f_0 + \beta y,
\end{equation}
where $f_0 \equiv -2 \Omega \cos\theta$ and $\theta$ denotes the co-latitude (or the polar coordinate in standard spherical coordinates).  The quantity $\beta = 2 \Omega \sin\theta / R$ is the coefficient of the first-order term in the series expansion of $f$.  Physically, it is the gradient of the Coriolis term.  At the poles, we recover $f = f_0$, known as the ``$f$-plane" model.  At the equator, we have $f = \beta y$.  General consideration of $f$ with both terms is known as the ``$\beta$-plane" approximation, which itself comes in two flavors.  The first approximation is to seek sinusoidal solutions in both directions, including for $y$ (see \S\ref{sect:2d}).  A better approximation is to seek sinusoidal solutions only in the $x$-direction and solve for the wave amplitudes as general functions of $y$, which we explore in \S\ref{sect:2d_beta}.

For the linearized magnetic induction equation, the $\vec{v}(\nabla.\vec{B})$ term has a non-zero contribution since a purely vertical field implies a magnetic monopole,
\begin{equation}
\frac{\partial b^\prime_{\rm x,y}}{\partial t} = \frac{\bar{B}_{\rm z} v^\prime_{\rm x,y}}{H} + \eta \left( \frac{\partial^2 b^\prime_{\rm x,y}}{\partial x^2} + \frac{\partial^2 b^\prime_{\rm x,y}}{\partial y^2} \right).
\end{equation}
While magnetic monopoles have never been seen in Nature, this approximation provides a convenient way to study the effect of localized patches of atmospheres where a vertical magnetic field may exist.

Physically, the system starts out from a state of rest with a purely vertical background field that is coupled to the flow.  Any motion of the flow in the horizontal direction induces horizontal magnetic field perturbations that resist this movement.  Since atmospheres are hardly expected to be perfect conductors, this process is expected to be diffusive ($\eta \ne 0$). 

\subsubsection{Horizontal Background Magnetic Field}

For a purely horizontal magnetic field strength, we assume $\nabla.\vec{B}=0$.  We set
\begin{equation}
\frac{\partial v_{\rm x,y}}{\partial z} = 0, 
\end{equation}
which is, as stated previously, a property of the shallow water system.  We set $\bar{B}_{\rm z}=0$.  The background state satisfies:
\begin{equation}
\begin{split}
&\bar{B}_{\rm x} \frac{\partial \bar{B}_{\rm x,y}}{\partial x} + \bar{B}_{\rm y} \frac{\partial \bar{B}_{\rm x,y}}{\partial y} = 0, \\
&\frac{\partial \bar{B}_{\rm x,y}}{\partial t} = \eta \nabla^2 \bar{B}_{\rm x,y}.
\end{split}
\label{eq:background_cartesian}
\end{equation}
Analogous expressions for the velocity would exist if one was perturbing about a moving state.  

The linearized momentum equations read:
\begin{equation}
\begin{split}
\frac{\partial v^\prime_{\rm x}}{\partial t} =& -g \frac{\partial h^\prime}{\partial x} + f v^\prime_{\rm y} + \nu \left(\frac{\partial^2 v^\prime_{\rm x}}{\partial x^2} + \frac{\partial^2 v^\prime_{\rm x}}{\partial y^2} \right) \\
&- \omega_{\rm drag} v^\prime_{\rm x} + \frac{\bar{B}_{\rm x}}{4 \pi \rho} \frac{\partial b^\prime_{\rm x}}{\partial x} + \frac{\bar{B}_{\rm y}}{4 \pi \rho} \frac{\partial b^\prime_{\rm x}}{\partial y}, \\
\frac{\partial v^\prime_{\rm y}}{\partial t} =& -g \frac{\partial h^\prime}{\partial y} - f v^\prime_{\rm x} + \nu \left(\frac{\partial^2 v^\prime_{\rm y}}{\partial x^2} + \frac{\partial^2 v^\prime_{\rm y}}{\partial y^2} \right) \\
&- \omega_{\rm drag} v^\prime_{\rm y} + \frac{\bar{B}_{\rm x}}{4 \pi \rho} \frac{\partial b^\prime_{\rm y}}{\partial x} + \frac{\bar{B}_{\rm y}}{4 \pi \rho} \frac{\partial b^\prime_{\rm y}}{\partial y}. \\
\end{split}
\end{equation}
The linearized magnetic induction equation derives from the $\vec{B}.\nabla\vec{v}$ term,
\begin{equation}
\frac{\partial b^\prime_{\rm x,y}}{\partial t} = \bar{B}_{\rm x} \frac{\partial v^\prime_{\rm x,y}}{\partial x} + \bar{B}_{\rm y} \frac{\partial v^\prime_{\rm x,y}}{\partial y} + \eta \left( \frac{\partial^2 b^\prime_{\rm x,y}}{\partial x^2} + \frac{\partial^2 b^\prime_{\rm x,y}}{\partial y^2} \right).
\end{equation}

\subsection{Equations in Spherical Coordinates}

We employ the $(r,\theta,\phi)$ coordinate system, where $r$ is the radial coordinate, $\theta$ is the co-latitude and $\phi$ is the longitude.  It is instructive to begin by stating the full equations in spherical coordinates,
\begin{equation}
\begin{split}
&\frac{D v_\theta}{D t} - \frac{v_\phi^2}{r \tan\theta} = - \frac{g}{r} \frac{\partial h}{\partial \theta} + 2 \Omega v_\phi \cos\theta \\
&+ \nu \left[ \nabla^2 v_\theta - \frac{1}{r^2\sin^2\theta} \left( v_\theta + 2 \cos\theta \frac{\partial v_\phi}{\partial \phi} \right) \right] \\
&- \omega_{\rm drag} v_\theta + \frac{1}{4\pi \rho} \left[ \vec{B}.\nabla B_\theta + \frac{1}{r} \left( B_r B_\theta - \frac{B^2_\phi}{\tan\theta} \right) \right], \\
&\frac{D v_\phi}{D t} + \frac{v_\theta v_\phi}{r \tan\theta} = - \frac{g}{r \sin\theta} \frac{\partial h}{\partial \phi} - 2 \Omega v_\theta \cos\theta \\
&+ \nu \left[ \nabla^2 v_\phi + \frac{1}{r^2\sin^2\theta} \left(2 \cos\theta \frac{\partial v_\theta}{\partial \phi} - v_\phi \right) \right] \\
&- \omega_{\rm drag} v_\phi + \frac{1}{4\pi \rho} \left[ \vec{B}.\nabla B_\phi + \frac{B_\phi}{r} \left( B_r + \frac{B_\theta}{\tan\theta} \right) \right], \\
&\frac{\partial h}{\partial t} + \frac{1}{r \sin\theta} \left[ \frac{\partial}{\partial \theta} \left( h v_\theta \sin\theta \right) + \frac{\partial}{\partial \phi} \left( h v_\phi \right) \right] = Q, \\
&\frac{\partial B_\theta}{\partial t} = \frac{v_\theta}{r^2} \frac{\partial}{\partial r} \left( r^2 B_r \right) + \frac{v_\theta}{r \sin\theta} \left[ \frac{\partial}{\partial \theta} \left( B_\theta \sin\theta \right) + \frac{\partial B_\phi}{\partial \phi} \right] \\
&+ \vec{B}.\nabla v_\theta - \frac{B_\phi v_\phi}{r \tan\theta} - \vec{v}.\nabla\vec{B}\vert_\theta + \eta \nabla^2 B_\theta \\
&+ \frac{\eta}{r^2} \left[ 2 \frac{\partial B_r}{\partial \theta} - \frac{1}{\sin^2\theta} \left( B_\theta + 2 \cos\theta \frac{\partial B_\phi}{\partial \phi} \right) \right], \\
&\frac{\partial B_\phi}{\partial t} = \frac{v_\phi}{r^2} \frac{\partial}{\partial r} \left( r^2 B_r \right) + \frac{v_\phi}{r \sin\theta} \left[ \frac{\partial}{\partial \theta} \left( B_\theta \sin\theta \right) + \frac{\partial B_\phi}{\partial \phi} \right] \\
&+ \vec{B}.\nabla v_\phi + \frac{B_\phi v_\theta}{r \tan\theta} - \vec{v}.\nabla\vec{B}\vert_\phi + \eta \nabla^2 B_\phi \\
&+ \frac{\eta}{r^2 \sin^2\theta} \left[ 2 \left( \frac{\partial B_r}{\partial \phi} + \cos\theta \frac{\partial B_\theta}{\partial \phi} \right) - B_\phi \right].
\end{split}
\end{equation}
We have made the approximation that the terms involving the radial velocity ($v_r$) and its gradients are sub-dominant and may be neglected.  A myriad of geometric terms appear from taking the gradient and Laplacian of $\vec{v}$ and $\vec{B}$, although we have kept the $-\vec{v}.\nabla \vec{B}$ term in its vectorial form because it vanishes for a system perturbed from rest and for a constant background magnetic field.  Similarly, the geometric terms appearing in the momentum equations vanish for a system perturbed from rest, because they generally depend on the square of velocity components.

The equation for the shallow water height perturbation may be stated without knowledge of the magnetic field geometry,
\begin{equation} 
\frac{\partial h^\prime}{\partial t} + \frac{H}{R \sin \theta} \left[ \frac{\partial}{\partial \theta} \left( v_\theta^\prime \sin\theta \right) + \frac{\partial v_\phi^\prime}{\partial \phi} \right] = F_0 h^\prime,
\end{equation}
where we have evaluated the radial coordinate at the (constant) radius of the exoplanet ($r=R$).

\subsubsection{Radial Background Magnetic Field}

We employ the same procedure as in the case of Cartesian coordinates.  We first represent the components of the \emph{total} magnetic field strength by $B_{r,\theta,\phi}$ and make the following approximations:
\begin{equation}
\begin{split}
&\frac{\partial v_{\rm \theta,\phi}}{\partial r} = 0, \\
&\frac{\partial B_{r,\theta,\phi}}{\partial \theta} \ll \frac{\partial B_r}{\partial r}, ~\frac{\partial B_{r,\theta,\phi}}{\partial \phi} \ll \frac{\partial B_r}{\partial r}, \\
&\frac{\partial B_{\theta,\phi}}{\partial r} = -\frac{B_{\theta,\phi}}{H}, ~\frac{\partial B_r}{\partial r} = \frac{B_r}{H}, ~\nabla^2 B_{\theta,\phi} \ne 0.
\end{split}
\end{equation}
We now represent the components of the \emph{background} magnetic field by $\bar{B}_{r,\theta,\phi}$.  The background state is again at rest and with $B_r=\bar{B}_r$ and $\bar{B}_{\theta,\phi}=0$.  The radial field is not perturbed.

The perturbed equations are
\begin{equation}
\begin{split}
\frac{\partial b^\prime_{\theta,\phi}}{\partial t} =& \frac{\bar{B}_r v^\prime_{\theta,\phi}}{H} -  \frac{\eta b^\prime_{\theta,\phi}}{R^2\sin^2\theta} \\
&+ \frac{\eta}{R^2} \left( \frac{1}{\tan\theta} \frac{\partial b_{\theta,\phi}}{\partial \theta} +  \frac{\partial^2 b_{\theta,\phi}}{\partial \theta^2} + \frac{1}{\sin^2\theta} \frac{\partial^2 b_{\theta,\phi}}{\partial \phi^2} \right), \\
\frac{\partial v_\theta^\prime}{\partial t} =& - \frac{g}{R} \frac{\partial h^\prime}{\partial \theta} + 2 \Omega v^\prime_\phi \cos\theta - \omega_{\rm drag} v^\prime_\theta \\
&+ \frac{\nu}{R^2} \left( \frac{1}{\tan\theta} \frac{\partial v^\prime_\theta}{\partial \theta} + \frac{\partial^2 v^\prime_\theta}{\partial \theta^2} + \frac{1}{\sin^2\theta} \frac{\partial^2 v^\prime_\theta}{\partial \phi^2} \right) \\
&- \frac{\nu}{R^2 \sin^2\theta} \left( v^\prime_\theta + 2 \cos\theta \frac{\partial v^\prime_\phi}{\partial \phi} \right) - \frac{\bar{B}_r b^\prime_\theta}{4 \pi \rho H}, \\
\frac{\partial v_\phi^\prime}{\partial t} =& - \frac{g}{R \sin\theta} \frac{\partial h^\prime}{\partial \phi} - 2 \Omega v^\prime_\theta \cos\theta - \omega_{\rm drag} v^\prime_\phi \\
&+ \frac{\nu}{R^2} \left( \frac{1}{\tan\theta} \frac{\partial v^\prime_\phi}{\partial \theta} + \frac{\partial^2 v^\prime_\phi}{\partial \theta^2} + \frac{1}{\sin^2\theta} \frac{\partial^2 v^\prime_\phi}{\partial \phi^2} \right) \\
&+ \frac{\nu}{R^2 \sin^2\theta} \left( 2 \cos\theta \frac{\partial v^\prime_\theta}{\partial \phi} - v^\prime_\phi \right) - \frac{\bar{B}_r b^\prime_\phi}{4 \pi \rho H}. \\
\end{split}
\end{equation}
Several terms vanish because $1/R \ll 1/H$.

\subsubsection{Horizontal Background Magnetic Field}

We set
\begin{equation}
\frac{\partial v_{\theta,\phi}}{\partial r}= 0.
\end{equation}
The background state satisfies:
\begin{equation}
\begin{split}
&\bar{B}_\theta \sin\theta \frac{\partial \bar{B}_\theta}{\partial \theta} + \bar{B}_\phi \frac{\partial \bar{B}_\theta}{\partial \phi} = \bar{B}^2_\phi \cos\theta, \\
&\bar{B}_\theta \sin\theta \frac{\partial \bar{B}_\phi}{\partial \theta} + \bar{B}_\phi \frac{\partial \bar{B}_\phi}{\partial \phi} = - \bar{B}_\theta \bar{B}_\phi \cos\theta, \\
&\frac{\partial \bar{B}_{\theta,\phi}}{\partial t} = \eta \nabla^2 \bar{B}_{\theta,\phi} - \frac{\eta \bar{B}_{\theta,\phi}}{r^2 \sin^2\theta}.
\end{split}
\end{equation}
The first two expressions reduce to their Cartesian counterparts at the equator ($\theta=90^\circ$).

The perturbed equations are
\begin{equation}
\begin{split}
\frac{\partial b^\prime_\theta}{\partial t} =& \frac{1}{R} \left( \bar{B}_\theta \frac{\partial v^\prime_\theta}{\partial \theta} + \frac{\bar{B}_\phi}{\sin\theta} \frac{\partial v^\prime_\theta}{\partial \phi} - \frac{\bar{B}_\phi v^\prime_\phi}{\tan\theta} \right) \\
&+ \frac{\eta}{R^2} \left(  \frac{1}{\tan\theta} \frac{\partial b^\prime_\theta}{\partial \theta} + \frac{\partial^2 b^\prime_\theta}{\partial \theta^2} + \frac{1}{\sin^2\theta} \frac{\partial^2 b^\prime_\theta}{\partial \phi^2} \right) \\
&- \frac{\eta}{R^2 \sin^2\theta} \left( b^\prime_\theta + 2 \cos\theta \frac{\partial b^\prime_\phi}{\partial \phi} \right), \\
\frac{\partial b^\prime_\phi}{\partial t} =& \frac{1}{R} \left( \bar{B}_\theta \frac{\partial v^\prime_\phi}{\partial \theta} + \frac{\bar{B}_\phi}{\sin\theta} \frac{\partial v^\prime_\phi}{\partial \phi} + \frac{\bar{B}_\phi v^\prime_\theta}{\tan\theta} \right) \\
&+ \frac{\eta}{R^2} \left(  \frac{1}{\tan\theta} \frac{\partial b^\prime_\phi}{\partial \theta} + \frac{\partial^2 b^\prime_\phi}{\partial \theta^2} + \frac{1}{\sin^2\theta} \frac{\partial^2 b^\prime_\phi}{\partial \phi^2} \right) \\
&+ \frac{\eta}{R^2 \sin^2\theta} \left( 2 \cos\theta \frac{\partial b^\prime_\theta}{\partial \phi} - b^\prime_\phi \right), \\
\frac{\partial v_\theta^\prime}{\partial t} =& - \frac{g}{R} \frac{\partial h^\prime}{\partial \theta} + 2 \Omega v^\prime_\phi \cos\theta - \omega_{\rm drag} v^\prime_\theta \\
&+ \frac{\nu}{R^2} \left( \frac{1}{\tan\theta} \frac{\partial v^\prime_\theta}{\partial \theta} + \frac{\partial^2 v^\prime_\theta}{\partial \theta^2} + \frac{1}{\sin^2\theta} \frac{\partial^2 v^\prime_\theta}{\partial \phi^2} \right) \\
&- \frac{\nu}{R^2 \sin^2\theta} \left( v^\prime_\theta + 2 \cos\theta \frac{\partial v^\prime_\phi}{\partial \phi} \right) \\
&+ \frac{1}{4 \pi \rho R} \left( \bar{B}_\theta \frac{\partial b^\prime_\theta}{\partial \theta} + \frac{\bar{B}_\phi}{\sin\theta} \frac{\partial b^\prime_\theta}{\partial \phi} - \frac{2\bar{B}_\phi b^\prime_\phi}{\tan\theta} \right), \\
\frac{\partial v_\phi^\prime}{\partial t} =& - \frac{g}{R \sin\theta} \frac{\partial h^\prime}{\partial \phi} - 2 \Omega v^\prime_\theta \cos\theta - \omega_{\rm drag} v^\prime_\phi \\
&+ \frac{\nu}{R^2} \left( \frac{1}{\tan\theta} \frac{\partial v^\prime_\phi}{\partial \theta} + \frac{\partial^2 v^\prime_\phi}{\partial \theta^2} + \frac{1}{\sin^2\theta} \frac{\partial^2 v^\prime_\phi}{\partial \phi^2} \right) \\
&+ \frac{\nu}{R^2 \sin^2\theta} \left( 2 \cos\theta \frac{\partial v^\prime_\theta}{\partial \phi} - v^\prime_\phi \right) \\
&+ \frac{1}{4 \pi \rho R} \left( \bar{B}_\theta \frac{\partial b^\prime_\phi}{\partial \theta} + \frac{\bar{B}_\phi}{\sin\theta} \frac{\partial b^\prime_\phi}{\partial \phi} \right) \\
& + \frac{1}{4 \pi \rho R \tan\theta} \left( \bar{B}_\theta b^\prime_\phi + \bar{B}_\phi b^\prime_\theta \right) . \\
\end{split}
\end{equation}

\subsection{Seeking Wave Solutions}
\label{subsect:seek_wave}

\begin{figure*}
\begin{center}
\vspace{-0.2in}
\includegraphics[width=\columnwidth]{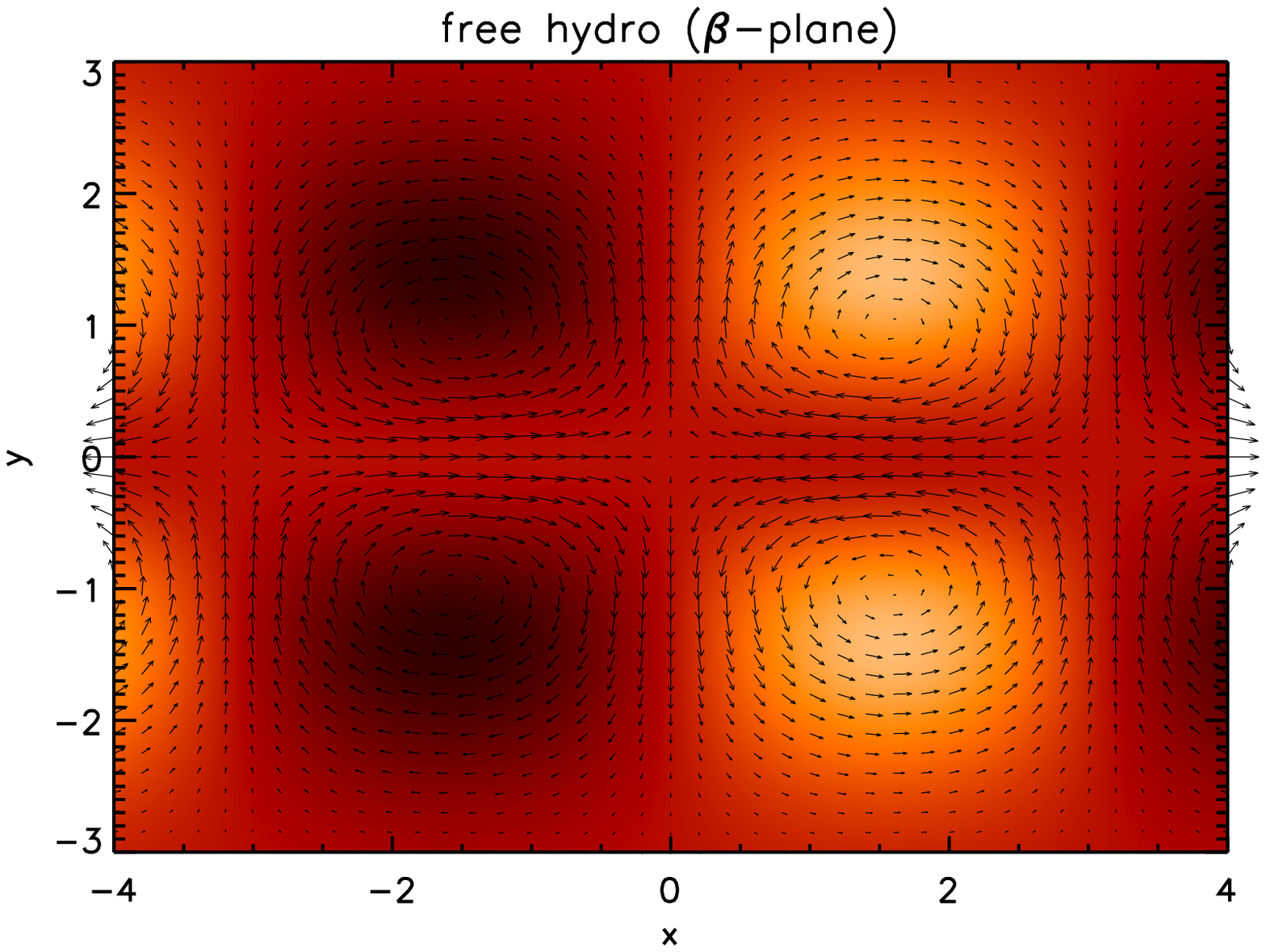}
\includegraphics[width=\columnwidth]{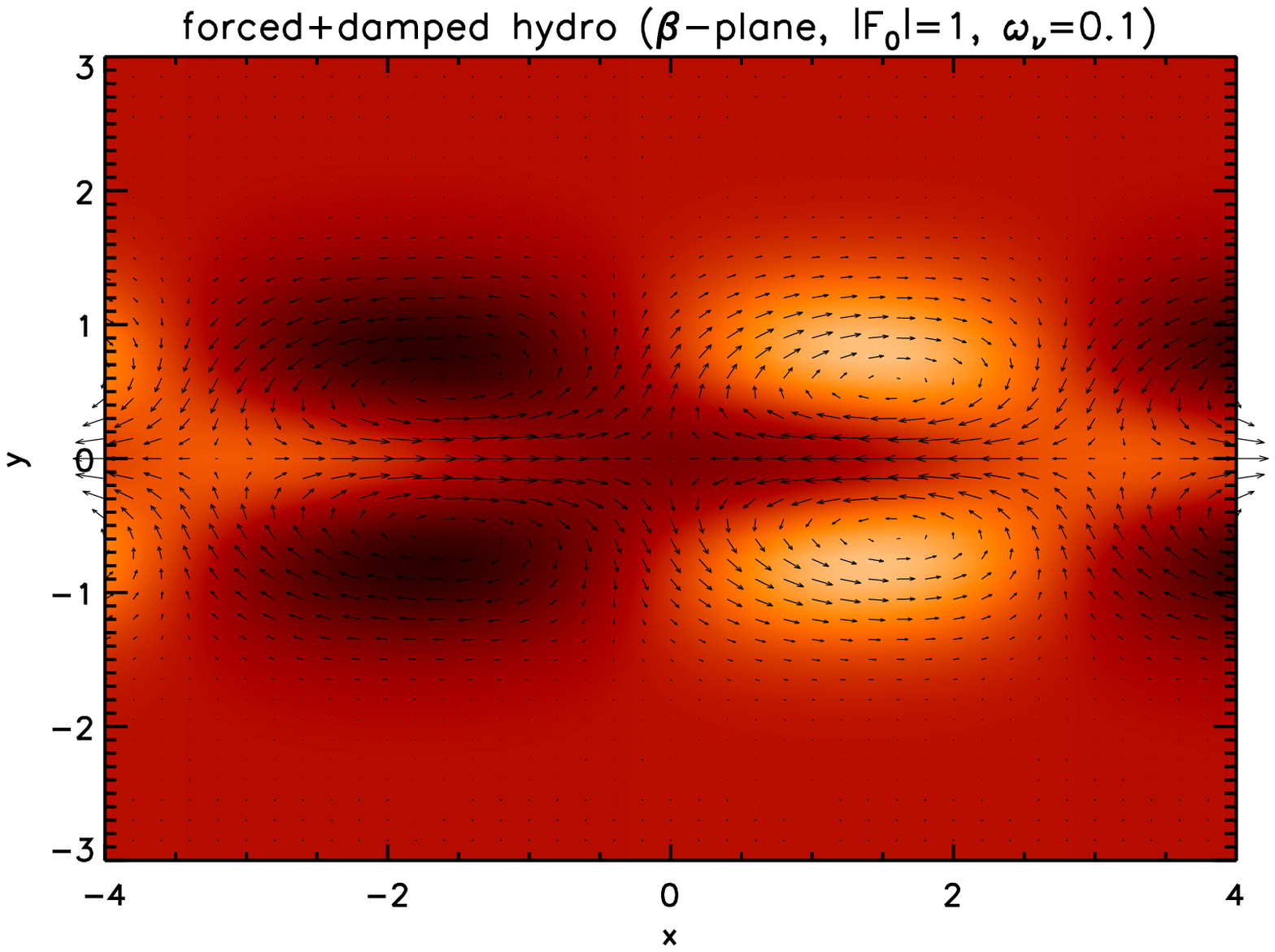}
\includegraphics[width=\columnwidth]{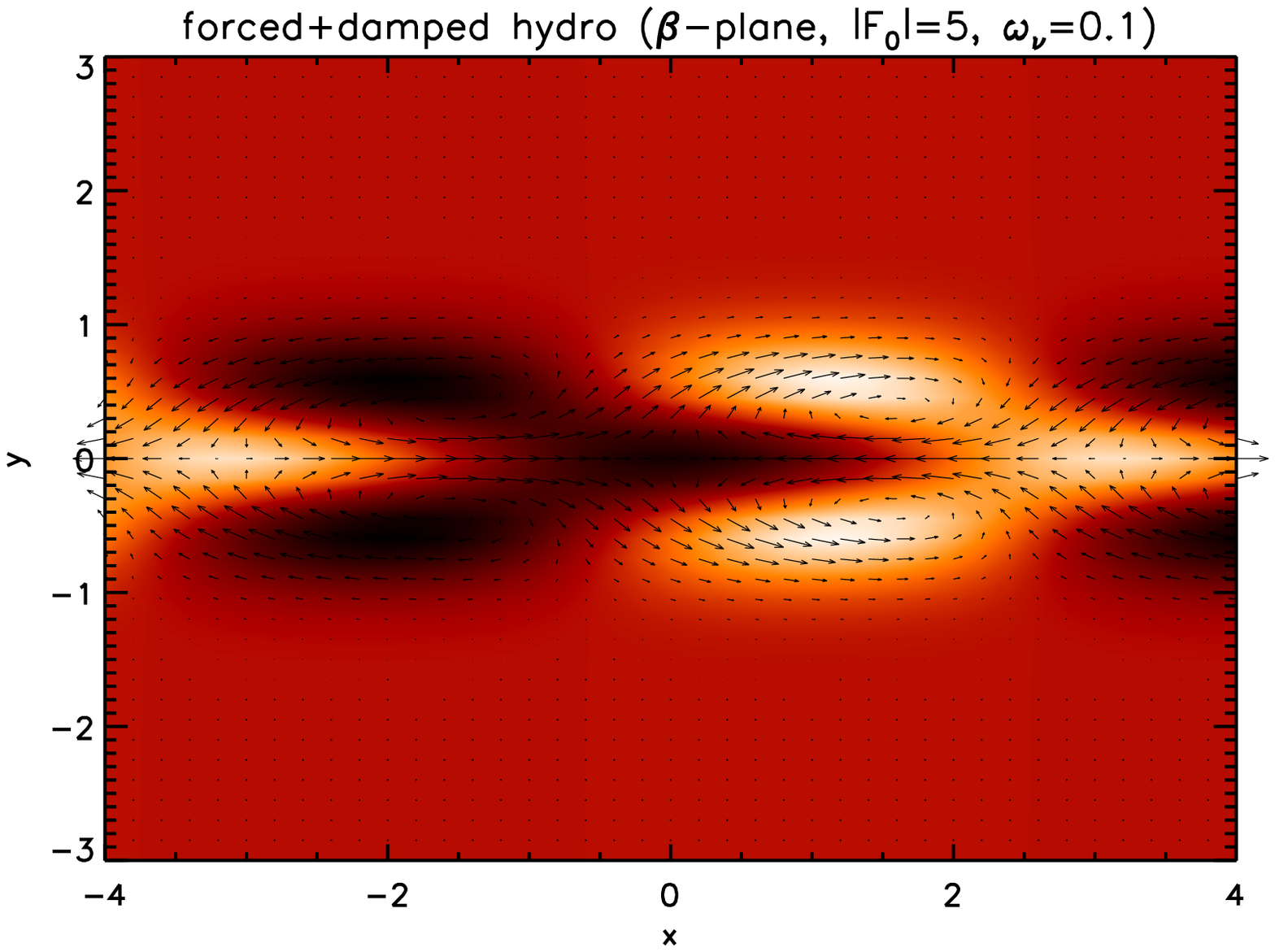}
\includegraphics[width=\columnwidth]{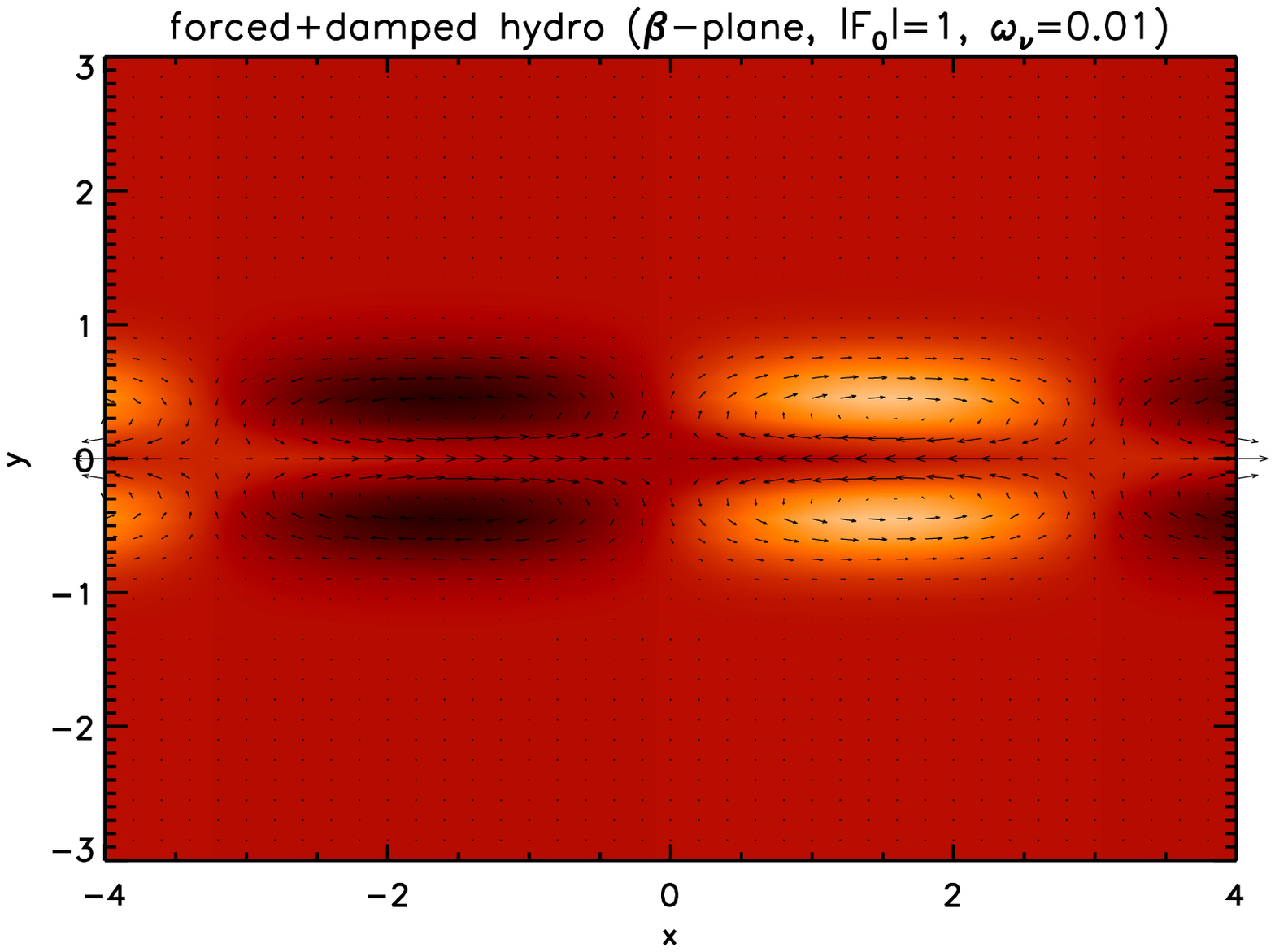}
\end{center}
\vspace{-0.2in}
\caption{Montage of plots of velocity (arrows) and water height (contours) perturbations for steady-state, hydrodynamic systems in the equatorial $\beta$-plane approximation for $n=k_{\rm x}=1$.  The different panels are for different strengths of forcing ($\vert F_0 \vert$) and hydrodynamic friction ($\omega_\nu$).  All quantities are computed in terms of an arbitrary velocity normalization ($v_0$).  Bright and dark colors correspond to positive and negative height perturbations, respectively.}
\vspace{0.1in}
\label{fig:beta_hydro_forced}
\end{figure*}

\subsubsection{Cartesian and Pseudo-Cartesian Geometries}

For a perturbed quantity $X$, we will generally seek wave solutions of the following form,
\begin{equation} 
X = X_0 \Psi_0 \Psi,
\end{equation}
where $X_0$ is an arbitrary normalization factor and we have defined
\begin{equation}
\Psi \equiv \exp{\left[ i \left( k_{\rm x} x - \omega t \right) \right]} = \exp{\left( \omega_{\rm I} t \right)} \exp{\left[ i \left( k_{\rm x} x - \omega_{\rm R} t \right) \right]},
\end{equation}
where $k_{\rm x}$ is the wavenumber in the $x$- or zonal direction.  For 1D models, we set $\Psi_0=1$.  For 2D Cartesian models, we set $\Psi_0 = \exp{(i k_{\rm y} y)}$, where $k_{\rm y}$ is the wavenumber in the $y$- or meridional direction.  For 2D pseudo-spherical models, we will allow $X_0 \Psi_0$ to retain a general functional dependence on $y$ and solve for this dependence.

\subsubsection{Spherical Geometry}

On a sphere, we define
\begin{equation} 
\Psi \equiv \exp{\left[ i \left( m \phi - \omega t_0 \right) \right]},
\end{equation}
where $m$ is the azimuthal or zonal wavenumber, while $\omega$ and $t_0 \equiv 2 \Omega t$ are the wave frequency and time, respectively, in dimensionless units.  Analogous to the pseudo-spherical case, we solve for the $\theta$-dependence of $X_0 \Psi_0$.

\section{1D Models}
\label{sect:1d}

Due to the relative simplicity of the algebra involved, 1D models have the virtue that they provide clean insights into the effects of each piece of physics.  Our main finding in this section is that some of these effects may couple in non-intuitive ways.

In order to compare the 1D and 2D models, we have cast the velocity and height perturbations in terms of the velocity amplitude ($v_{\rm x_0}$).  As we will see in \S\ref{sect:2d_beta} and \S\ref{sect:spherical}, it is more natural to use $v_{\rm x_0}$, rather than the height perturbation amplitude ($h_0$), because the key governing equation is for the meridional velocity.    

\subsection{Hydrodynamic}

\subsubsection{Basic}
\label{subsect:1d_basic}

To develop intuition, we review the most basic model: 1D with no rotation, viscosity, drag or magnetic fields.  Since this is a free system (with no sources of forcing or friction), the wave frequency is real, i.e., $\omega = \omega_{\rm R}$.  The velocity amplitude is
\begin{equation}
v_{\rm x_0} = \frac{g k_{\rm x} h_0}{\omega_{\rm R}} = \frac{\omega_{\rm R} h_0}{k_{\rm x} H}.
\end{equation}
The velocity perturbation has the solution, 
\begin{equation}
v_{\rm x}^\prime = v_{\rm x_0} ~\cos{\left( k_{\rm x} x - \omega_{\rm R}t \right)},
\end{equation}
while the height perturbation takes on the form, 
\begin{equation}
h^\prime = \frac{k_{\rm x} H v_{\rm x_0}}{\omega_{\rm R}} ~\cos{\left( k_{\rm x} x - \omega_{\rm R}t \right)}.
\end{equation}
Since no sources of dissipation exist, the velocity and height perturbations are perfectly in phase.  The dispersion relation is
\begin{equation}
\omega_{\rm R} = \pm \left( gH \right)^{1/2} k_{\rm x}.
\end{equation}
Only gravity waves exist in this most basic of systems and they have the special property that they are non-dispersive (i.e., they possess the same group velocity regardless of wavelength).

\subsubsection{Kinematic (Molecular) Viscosity}
\label{subsect:1d_viscous}

For the second simplest model, we add a source of friction: kinematic viscosity ($\nu$).  It is usually associated with molecular viscosity, although it has sometimes been used to mimic the presence of large-scale, turbulent viscosity (e.g., \citealt{dd08}).  Strictly speaking, the $\nu \nabla^2\vec{v}$ term is non-negligible only when the Reynolds number is of order unity (or less).

The dispersion relation now has real and imaginary parts,
\begin{equation}
i \omega^2 - \nu \omega k_{\rm x}^2 = i g H k^2_{\rm x}.
\end{equation}
To properly evaluate it, we need to substitute $\omega = \omega_{\rm R} + i \omega_{\rm I}$ into the preceding expression, which yields separate equations for the oscillatory ($\omega_{\rm R}$) and decaying ($\omega_{\rm I}$) parts of the wave solutions.  The former has the solution,
\begin{equation}
\omega_{\rm R} = \pm \left[ gH k^2_{\rm x} - \left(\frac{\nu k_{\rm x}^2}{2}\right)^2 \right]^{1/2},
\end{equation}
implying that viscosity damps the wave frequency.  Viscosity also tends to act on smaller scales, since
\begin{equation}
\omega_{\rm I} = -\frac{\nu k_{\rm x}^2}{2},
\end{equation}
on a characteristic viscous timescale of $t_\nu \sim \vert \omega_{\rm I} \vert^{-1}$.  In fact, from examining the expression for $\omega_{\rm R}$, one may derive a viscous length scale ($L_\nu$) and Reynolds number (${\cal R}$) at which gravity waves are completely damped out, 
\begin{equation}
L_\nu = \frac{\pi \nu}{c_0}, ~{\cal R} \equiv \frac{c_0 L_\nu}{\nu} = \pi,
\end{equation}
where $c_0 \equiv ( gH )^{1/2}$.

The wave solutions are 
\begin{equation}
\begin{split}
v^\prime_{\rm x} &= v_{\rm x_0} ~\exp{\left( \omega_{\rm I} t \right)} ~\cos{\left( k_{\rm x} x - \omega_{\rm R}t \right)},\\
h^\prime &= \frac{v_{\rm x_0}}{g k_{\rm x}} ~\exp{\left( \omega_{\rm I} t \right)} \\
&\times \left[ \omega_{\rm R} \cos{\left( k_{\rm x} x - \omega_{\rm R}t \right)} - \frac{\nu k_{\rm x}^2}{2} \sin{\left( k_{\rm x} x - \omega_{\rm R}t \right)} \right].
\end{split}
\end{equation}
We will use this form for $v_{\rm x}^\prime$ throughout \S\ref{sect:1d}.  In the presence of viscosity, the velocity and height perturbations are now out of phase.  Specifically, there is a viscous term that is $90^\circ$ out of phase with the non-viscous terms.

A ``hyperviscosity" has sometimes been used in simulations of atmospheric circulation of exoplanets (e.g., \citealt{hmp11}).  It asserts that the viscous term in the momentum equation can be expressed as $\nu \nabla^{n_\nu} \vec{v}$ with $\nu$ having to take on the appropriate physical units for it to retain the physical units of acceleration.  It follows that
\begin{equation}
\omega_{\rm I} = \frac{\left( -1 \right)^{n_\nu/2} \nu k^{n_\nu}}{2}, ~n_\nu \in 2\mathbb{Z}.
\end{equation}
Oddly enough, it means that certain values of $n_\nu$ result in growing wave solutions that are unstable, e.g., $n_\nu=8$ (a value often used in 3D simulations).  Odd values of $n_\nu$ are unphysical, at least for the shallow water system, because they produce spurious oscillatory modes.

\subsubsection{Rayleigh Drag}
\label{subsect:1d_drag}

The term $-\vec{v}/t_{\rm drag}$ is often inserted into the momentum equation to mimic large-scale friction in atmospheres.  It is called ``Rayleigh drag".  On Earth, it mimics the effect of the planetary boundary layer, where the atmosphere transitions from a ``free slip" to a ``no slip" boundary condition.  In atmospheric circulation simulations of exoplanets, it has been used to mimic the effects of magnetic drag \citep{pmr10a,rm13}.  A key difference between Rayleigh drag and molecular viscosity is that the former acts \emph{equally on all scales}, since
\begin{equation}
\omega_{\rm I} = -\frac{1}{2t_{\rm drag}} \equiv -\frac{\omega_{\rm drag}}{2}.
\end{equation}
Like viscosity, Rayleigh drag damps the frequency of gravity waves,
\begin{equation}
\omega_{\rm R} = \pm \left[ gH k^2_{\rm x} - \left(\frac{\omega_{\rm drag}}{2}\right)^2 \right]^{1/2}.
\end{equation}
Analogous to the viscous length scale, there is a drag length scale on which gravity waves are completely damped out,
\begin{equation}
L_{\rm drag} = 4 \pi c_0 t_{\rm drag}.
\end{equation}
The Taylor number is then ${\cal T} = ( t_{\rm drag} c_0 / L_{\rm drag} )^2 = 1/16\pi^2 \ll 1$, as expected for flows under strong drag.

Again, analogous to the case of viscosity, Rayleigh drag damps the wave amplitudes and introduces an out-of-phase component to the wave solutions, since
\begin{equation}
\begin{split}
h^\prime &= \frac{v_{\rm x_0}}{gk_{\rm x}} ~\exp{\left( \omega_{\rm I} t \right)} \\
&\times \left[ \omega_{\rm R} \cos{\left( k_{\rm x} x - \omega_{\rm R}t \right)} - \frac{\omega_{\rm drag}}{2} \sin{\left( k_{\rm x} x - \omega_{\rm R}t \right)} \right],
\end{split}
\end{equation}
with the difference being that $\omega_{\rm drag}$ has no $k_{\rm x}$-dependence.

\subsubsection{Molecular Viscosity and Rayleigh Drag}
\label{subsect:1d_friction}

When both viscosity and Rayleigh drag are included, the mathematical form of the dispersion relation and wave solutions are identical, except that one replaces $\nu k^2_{\rm x}$ or $\omega_{\rm drag}$ by
\begin{equation}
\omega_\nu \equiv \nu k^2_{\rm x} + \omega_{\rm drag}.
\end{equation}
Friction acting upon the flow now has two components: a scale-free part and one directed strongly at small scales.

\subsubsection{Forcing}
\label{subsect:1d_forcing}

In the presence of external forcing (e.g., stellar irradiation), the equation for the shallow water height has a non-zero source term ($Q$).  In its perturbed form, it involves two terms: $S - \omega_{\rm rad} h^\prime$.  In seeking wave solutions, we adopt
\begin{equation}
S = S_0 h_0 \Psi,
\label{eq:forcing}
\end{equation}
where $S_0$ is a dimensionless number.  As expected, forcing results in a growing wave solution, since
\begin{equation}
\omega_{\rm I} = \frac{S_0 - \omega_{\rm rad}}{2} \equiv \frac{F_0}{2}.
\end{equation}
We have specifically defined $F_0 \equiv S_0 - \omega_{\rm rad}$ as the ``forcing".  It vanishes when heating is balanced by radiative cooling ($S_0 = \omega_{\rm rad}$).

Forcing introduces an out-of-phase component to the wave solution,
\begin{equation} 
\begin{split}
h^\prime &= \frac{v_{\rm x_0}}{g k_{\rm x}} ~\exp{\left( \omega_{\rm I} t \right)} \\
&\times \left[ \omega_{\rm R} \cos{\left( k_{\rm x} x - \omega_{\rm R}t \right)} - \frac{F_0}{2} \sin{\left( k_{\rm x} x - \omega_{\rm R}t \right)} \right].
\end{split}
\end{equation}
Non-intuitively, forcing \emph{decreases} the wave frequency, since
\begin{equation}
\omega_{\rm R} = \pm \left( g H k^2_{\rm x} - \frac{F^2_0}{4} \right)^{1/2}.
\end{equation}
Gravity waves are most strongly affected by forcing on a length scale
\begin{equation}
L_{\rm F} = \frac{4\pi c_0}{F_0}.
\end{equation}

No stable solution for the flow exists unless $F_0=0$.  

\subsubsection{Forcing with Molecular Viscosity and Rayleigh Drag}
\label{subsect:1d_forcing_friction}

With the framework developed in \S\ref{subsect:1d_basic}--\ref{subsect:1d_forcing}, we can now explore a 1D hydrodynamic, forced system with friction.  This system admits both growing and decaying wave solutions,
\begin{equation}
\omega_{\rm I} = \frac{F_0 - \omega_\nu}{2}.
\end{equation}
Globally, irradiated atmospheres are in a state of equipoise---forcing balances friction, such that there is stability.  In such a \emph{balanced flow}, we have $\omega_{\rm I}=0$ and $F_0 = \omega_\nu$.  Globally, a stable atmosphere needs to be damped exactly at the same frequency at which it is being forced.  In three-dimensional, non-linear systems, equipoise might be absent at local scales.  In our simplified, 1D, linear model, we assume equipoise to be both a local and a global condition.  By enforcing $F_0 = \omega_\nu$, the wave frequency becomes
\begin{equation}
\begin{split}
\omega_{\rm R} &= \pm \left[ g H k^2_{\rm x} - \left( \frac{F_0 + \omega_\nu}{2}\right)^2 \right]^{1/2} \\
&=\pm \left( g H k^2_{\rm x} - \omega_\nu^2 \right)^{1/2}.
\end{split}
\end{equation}
Forcing and friction combine to damp the wave frequency more strongly than if either was acting alone.  In a balanced flow, we have 
\begin{equation}
h^\prime = \frac{v_{\rm x_0}}{g k_{\rm x}} \left[ \omega_{\rm R} \cos{\left( k_{\rm x} x - \omega_{\rm R}t \right)} - \omega_\nu \sin{\left( k_{\rm x} x - \omega_{\rm R}t \right)} \right].
\end{equation}

\subsection{Magnetohydrodynamic}

\subsubsection{Basic (Ideal MHD)}
\label{subsect:1d_basic_mhd}

In 1D, the simplest instance of a MHD system is one in which all forms of friction (molecular viscosity, Rayleigh drag, magnetic drag) are absent.  (By definition, rotation is a 2D phenomenon.)  For a vertical background field, the dispersion relation reads
\begin{equation}
\omega_{\rm R} = \pm \left[ gH k_{\rm x}^2 + \left( \frac{v_{\rm A}}{H} \right)^2 \right]^{1/2},
\end{equation}
where $v_{\rm A} \equiv \bar{B}_{\rm z}/2\sqrt{\pi \rho}$ is the Alfv\'{e}n speed.  Evidently, the presence of a magnetic field enhances the wave frequency relative to the hydrodynamic case.  These are called ``magneto-gravity waves".  Their wave solutions are 
\begin{equation}
\begin{split}
h^\prime &= \frac{k_{\rm x} H v_{\rm x_0}}{\omega_{\rm R}} \cos{\left( k_{\rm x} x - \omega_{\rm R}t \right)}, \\
b^\prime_{\rm x} &= -\frac{\bar{B}_{\rm z} v_{\rm x_0}}{\omega_{\rm R} H} \sin{\left( k_{\rm x} x - \omega_{\rm R}t \right)}.
\end{split}
\end{equation}
The height and velocity perturbations remain in phase, but they are out of phase with the magnetic field by $90^\circ$.

The qualitative effects of a horizontal background field are equivalent, except that they act more strongly at smaller scales,
\begin{equation}
\begin{split}
\omega_{\rm R} &= \pm \left[ gH k^2_{\rm x} + \left( v_{\rm A} k_{\rm x} \right)^2 \right]^{1/2}, \\
h^\prime &= \frac{k_{\rm x} H v_{\rm x_0}}{\omega_{\rm R}} \cos{\left( k_{\rm x} x - \omega_{\rm R}t \right)}, \\
b^\prime_{\rm x} &= -\frac{k_{\rm x} \bar{B}_{\rm x} v_{\rm x_0}}{\omega_{\rm R}} \cos{\left( k_{\rm x} x - \omega_{\rm R}t \right)},
\end{split}
\end{equation}
and the Alfv\'{e}n speed is now $v_{\rm A} \equiv \bar{B}_{\rm x}/2\sqrt{\pi \rho}$.

\subsubsection{Magnetic Drag}
\label{subsect:1d_magdrag}

A 1D system with magnetic drag ($\eta \ne 0$) is the first example where an analytical solution for $\omega_{\rm I}$ cannot be obtained, although the equation describing it may be stated (for a vertical background field),
\begin{equation}
\begin{split}
8 \omega^3_{\rm I} &+ 8 \eta k^2_{\rm x} \omega^2_{\rm I} + 2 \omega_{\rm I} \left[ gH k_{\rm x}^2 + \left( \frac{v_{\rm A}}{H} \right)^2 + \left( \eta k^2_{\rm x} \right)^2 \right] \\
&+ \eta \left( \frac{k_{\rm x} v_{\rm A}}{H} \right)^2 = 0.
\end{split}
\end{equation}
From examining the preceding equation for small and large $\omega_{\rm I}$, one may infer that two physically meaningful solutions of $\omega_{\rm I} < 0$ exist.  The oscillatory part of the wave solution has the frequency,
\begin{equation}
\omega_{\rm R} = \pm \left[ gH k^2_{\rm x} + \left( \frac{v_{\rm A}}{H} \right)^2 + \omega_{\rm I} \left( 3 \omega_{\rm I} + 2 \eta k^2_{\rm x} \right) \right]^{1/2}.
\end{equation}
Even without obtaining an analytical solution for $\omega_{\rm I}$, we see that in a balanced flow ($\omega_{\rm I}=0$) the 1D, free, MHD limit is obtained---the expression for $\omega_{\rm R}$ does not involve $\eta$.  

For a horizontal background field, the governing equations for $\omega_{\rm I}$ and $\omega_{\rm R}$ are similar,
\begin{equation}
\begin{split}
8 \omega^3_{\rm I} &+ 8 \eta k^2_{\rm x} \omega^2_{\rm I} + 2 \omega_{\rm I} \left[ gH k_{\rm x}^2 + \left( v_{\rm A} k_{\rm x} \right)^2 + \left( \eta k^2_{\rm x} \right)^2 \right] \\
&+ \eta \left( v_{\rm A} k_{\rm x}^2 \right)^2 = 0,\\
\omega_{\rm R} &= \pm \left[ gH k^2_{\rm x} + \left( v_{\rm A} k_{\rm x} \right)^2 + \omega_{\rm I} \left( 3 \omega_{\rm I} + 2 \eta k^2_{\rm x} \right) \right]^{1/2}. \\
\end{split}
\end{equation}

In other words, a balanced flow may only exist when $\eta=0$.  Our derivation is somewhat unsatisfactory, because it does not allow us to study the effects of $\eta$ on the phases of the wave solutions.

\subsubsection{Forcing}
\label{subsect:1d_forcing_mhd}

When forcing is added to the 1D MHD system, a similar situation occurs in the case when magnetic drag is present.  A cubic equation for $\omega_{\rm I}$ ensues (for a vertical background field),
\begin{equation}
\begin{split}
8\omega_{\rm I}^3 &- 8 F_0 \omega_{\rm I}^2 + 2 \omega_{\rm I} \left[ F^2_0 + g H k^2_{\rm x} + \left( \frac{v_{\rm A}}{H} \right)^2 \right] \\
&- F_0 g H k^2_{\rm x} = 0.
\end{split}
\end{equation}
When $F_0=0$, the solution for $\omega_{\rm I}$ is unphysical ($\omega_{\rm I}^2 < 0$) and may be disregarded.  The other two solutions, at small and large $\omega_{\rm I}$, are positive, implying growing modes as expected.  The wave frequency is
\begin{equation}
\omega_{\rm R} = \pm \left[ g H k^2_{\rm x} + \left( \frac{v_{\rm A}}{H} \right)^2 + \omega_{\rm I} \left( 3\omega_{\rm I} - 2 F_0 \right) \right]^{1/2}.
\end{equation}
No balanced flow exists unless $F_0=0$.  Again, we caution that we are interpreting this to be both a local and a global condition for our simplified, 1D, linear models.

With a horizontal background field, we reach a similar conclusion, since
\begin{equation}
\begin{split}
8\omega_{\rm I}^3 &- 8 F_0 \omega_{\rm I}^2 + 2 \omega_{\rm I} \left[ F^2_0 + g H k^2_{\rm x} + \left( v_{\rm A}k_{\rm x} \right)^2 \right] \\
&- F_0 g H k^2_{\rm x} = 0,\\
\omega_{\rm R} &= \pm \left[ g H k^2_{\rm x} + \left( v_{\rm A}k_{\rm x} \right)^2 + \omega_{\rm I} \left( 3\omega_{\rm I} - 2 F_0 \right) \right]^{1/2}.\\
\end{split}
\end{equation}

\subsubsection{Molecular Viscosity and Rayleigh Drag}
\label{subsect:1d_friction_mhd}

Before we examine the 1D MHD case with forcing and all three forms of friction, it is instructive to examine the basic magnetic case with just hydrodynamic friction present.  Specifically, we set $F_0 = \eta = 0$.  The wave solutions are decaying, since (for a vertical background field)
\begin{equation}
\omega_{\rm I} = - \frac{\omega_\nu}{2},
\end{equation}
while the wave frequencies are damped,
\begin{equation}
\omega_{\rm R} = \pm \left[ g H k_{\rm x}^2 + \left( \frac{v_{\rm A}}{H} \right)^2 - \left(\frac{\omega_\nu}{2}\right)^2 \right]^{1/2}.
\end{equation}
The wave solutions are
\begin{equation}
\begin{split}
h^\prime =& \frac{k_{\rm x} H v_{\rm x_0}}{gH k^2_{\rm x} + \left( v_{\rm A}/H \right)^2} ~\exp{\left( \omega_{\rm I} t \right)} \\
&\times \left[ \omega_{\rm R} \cos{\left( k_{\rm x} x - \omega_{\rm R}t \right)} - \frac{\omega_\nu}{2} \sin{\left( k_{\rm x} x - \omega_{\rm R}t \right)} \right], \\
b^\prime_{\rm x} =& - \frac{ \bar{B}_{\rm z} v_{\rm x_0}}{H \left[ gH k^2_{\rm x} + \left( v_{\rm A}/H \right)^2 \right]} ~\exp{\left( \omega_{\rm I} t \right)} \\
&\times \left[ \frac{\omega_\nu}{2} \cos{\left( k_{\rm x} x - \omega_{\rm R}t \right)} + \omega_{\rm R} \sin{\left( k_{\rm x} x - \omega_{\rm R}t \right)} \right], \\
\end{split}
\end{equation}
For a horizontal background field, we have
\begin{equation}
\begin{split}
\omega_{\rm I} =& - \frac{\omega_\nu}{2}, \\
\omega_{\rm R} =& \pm \left[ g H k_{\rm x}^2 + \left( v_{\rm A} k_{\rm x} \right)^2 - \left(\frac{\omega_\nu}{2}\right)^2 \right]^{1/2}, \\
h^\prime =& \frac{k_{\rm x} H v_{\rm x_0}}{g H k_{\rm x}^2 + \left( v_{\rm A} k_{\rm x} \right)^2} ~\exp{\left( \omega_{\rm I} t \right)} \\
&\times \left[ \omega_{\rm R} \cos{\left( k_{\rm x} x - \omega_{\rm R}t \right)} - \frac{\omega_\nu}{2} \sin{\left( k_{\rm x} x - \omega_{\rm R}t \right)} \right], \\
b^\prime_{\rm x} =& - \frac{k_{\rm x} \bar{B}_{\rm x} v_{\rm x_0}}{g H k_{\rm x}^2 + \left( v_{\rm A} k_{\rm x} \right)^2} ~\exp{\left( \omega_{\rm I} t \right)} \\
&\times \left[ \omega_{\rm R} \cos{\left( k_{\rm x} x - \omega_{\rm R}t \right)} - \frac{\omega_\nu}{2} \sin{\left( k_{\rm x} x - \omega_{\rm R}t \right)} \right], \\
\end{split}
\end{equation}

In summary, hydrodynamic friction introduces a viscous component to the velocity perturbation that is $90^\circ$ out of phase with the non-viscous component.  By contrast, we will see that magnetic drag introduces a component that has a different phase signature from hydrodynamic friction.

\subsubsection{Friction (Molecular Viscosity, Rayleigh Drag and Magnetic Drag)}
\label{subsect:1d_friction_all}

When both hydrodynamic (viscosity and Rayleigh drag) and magnetic sources of friction are included, something curious happens.  The wave frequency becomes (for a vertical background field), 
\begin{equation}
\omega_{\rm R}^2 = g H k^2_{\rm x} + \left( \frac{v_{\rm A}}{H} \right)^2 + \omega_\nu \eta k^2_{\rm x} + \omega_{\rm I} \left[ 2 \left( \eta k^2_{\rm x} + \omega_\nu \right) + 3 \omega_{\rm I} \right],
\end{equation}
implying that an extra contribution ($\omega_\nu \eta k_{\rm x}^2$) is present even for a balanced flow ($\omega_{\rm I}=0$).  For this term to be non-vanishing, \emph{both} magnetic and hydrodynamic drag have to be present.  The growing or decaying part of the wave solution is described by
\begin{equation}
\begin{split}
&8 \omega_{\rm I}^3 + 8 \left( \eta k_{\rm x}^2 + \omega_\nu \right) \omega_{\rm I}^2 \\
&+ 2 \left[ g H k^2_{\rm x} + \left( \frac{v_{\rm A}}{H} \right)^2 + \omega_\nu \eta k^2_{\rm x} + \left( \eta k_{\rm x}^2 + \omega_\nu \right)^2 \right] \omega_{\rm I}\\
&+ \left( \eta k_{\rm x}^2 + \omega_\nu \right) \left[ g H k^2_{\rm x} + \left( \frac{v_{\rm A}}{H} \right)^2 + \omega_\nu \eta k^2_{\rm x} \right] - g H k^4_{\rm x} \eta = 0.\\
\end{split}
\end{equation}
The preceding equation has the curious property that when $\omega_{\rm I} = 0$, one obtains an expression relating $\nu$ and $\omega_\nu$, suggesting that magnetic drag and hydrodynamic friction are balancing out each other.  In a balanced flow, one obtains
\begin{equation}
\begin{split}
\omega_{\rm R}^2 =& \frac{g H k^4_{\rm x} \eta}{\eta k^2_{\rm x} + \omega_\nu}, \\
h^\prime =& \frac{k_{\rm x} H v_{\rm x_0}}{\omega_{\rm R}} \cos{\left( k_{\rm x} x - \omega_{\rm R}t \right)}, \\
b^\prime_{\rm x} =& \frac{\bar{B}_{\rm z} v_{\rm x_0}}{H \left[ \left( \eta k^2_{\rm x} \right)^2 + \omega^2_{\rm R} \right]} \\
& \times \left[ \eta k^2_{\rm x} \cos{\left( k_{\rm x} x - \omega_{\rm R}t \right)} - \omega_{\rm R} \sin{\left( k_{\rm x} x - \omega_{\rm R}t \right)} \right].
\end{split}
\end{equation}
As anticipated, there is an in-phase component to the magnetic field perturbation (due solely to magnetic drag) and an out-of-phase component that depends on all forms of friction.  In other words, magnetic and hydrodynamic drag possess different phase signatures, producing magnetic field perturbations that negate each other.  Such a finding suggests that one should not use hydrodynamic drag as a proxy for magnetic drag (e.g., \citealt{pmr10a}).

For a horizontal background field, we get similar results,
\begin{equation}
\begin{split}
\omega_{\rm R}^2 =& \frac{g H k^4_{\rm x} \eta}{\eta k^2_{\rm x} + \omega_\nu}, \\
h^\prime =& \frac{k_{\rm x} H v_{\rm x_0}}{\omega_{\rm R}} \cos{\left( k_{\rm x} x - \omega_{\rm R}t \right)}, \\
b^\prime_{\rm x} =& - \frac{k_{\rm x} \bar{B}_{\rm x} v_{\rm x_0}}{\left( \eta k^2_{\rm x} \right)^2 + \omega^2_{\rm R}} \\
& \times \left[ \omega_{\rm R} \cos{\left( k_{\rm x} x - \omega_{\rm R}t \right)} + \eta k^2_{\rm x} \sin{\left( k_{\rm x} x - \omega_{\rm R}t \right)} \right].
\end{split}
\end{equation}
Here, both components in the magnetic field perturbation are out of phase with the height and velocity perturbations.

In summary, magnetic and hydrodynamic drag may negate each other in a balanced flow.

\subsubsection{Forcing with Friction (Molecular Viscosity, Rayleigh Drag and Magnetic Drag)}
\label{subsect:1d_mhd_all}

We are now ready to examine the most complex 1D model, which includes magnetic fields and all forms of friction.  This is the first example where different equations for $\omega_{\rm R}$ and $\omega_{\rm I}$ cannot be cleanly obtained (for a vertical background field),
\begin{equation}
\begin{split}
&\omega^3_{\rm I} - 3 \omega_{\rm R}^2 \omega_{\rm I} - \left( \omega_{\rm R}^2 - \omega_{\rm I}^2 \right) \left( \omega_\nu + \eta k^2_{\rm x} - F_0 \right) \\
&+ \omega_{\rm I} \left[ gH k^2_{\rm x} + \left( \frac{v_{\rm A}}{H} \right)^2 + \omega_\nu \eta k^2_{\rm x} - F_0 \left( \omega_\nu + \eta k^2_{\rm x} \right) \right] \\
&+ g H k^4_{\rm x} \eta - F_0 \left[ \left( \frac{v_{\rm A}}{H} \right)^2 + \omega_\nu \eta k^2_{\rm x} \right] = 0, \\
&\omega^2_{\rm R} - 3 \omega^2_{\rm I} - 2 \omega_{\rm I} \left( \omega_\nu + \eta k^2_{\rm x} - F_0 \right) \\
&- \left[ gH k^2_{\rm x} + \left( \frac{v_{\rm A}}{H} \right)^2 + \omega_\nu \eta k^2_{\rm x} - F_0 \left( \omega_\nu + \eta k^2_{\rm x} \right) \right] = 0.
\end{split}
\label{eq:dispersion_1d_general}
\end{equation}
The ``mixing" of the terms involving $\eta$, $\omega_\nu$ and $F_0$ suggests that a series of couplings are occurring between the out-of-phase components of these different effects.  The first expression describes $\omega_{\rm I}$.  For example, one may verify that in the limit of $v_{\rm A}=\eta=\omega_{\rm I}=0$, one obtains $\omega^2_{\rm R} (F_0 - \omega_\nu)=0$, which implies $F_0=\omega_\nu$ for a balanced flow.  In the limit of $F_0 = \omega_\nu = \omega_{\rm I} = 0$, one may also verify that a balanced flow only exists when $\eta=0$, since $\eta k^2_{\rm x} ( gH k^2_{\rm x} - \omega_{\rm R}^2) = 0$.  The second expression in equation (\ref{eq:dispersion_1d_general}) has two roots for the eastward- and westward-propagating gravity waves, albeit modified by forcing, friction and magnetic tension.  

In the most complex 1D model, the system is described by four parameters related to forcing and friction: $F_0$, $\omega_\nu$, $\eta$ and $v_{\rm A}$.  When three of them are specified, the first expression in equation (\ref{eq:dispersion_1d_general}) yields the value of the fourth parameter, via
\begin{equation}
\begin{split}
&gHk^2_{\rm x} \left( \omega_\nu - F_0 \right) + \left[ \left( \frac{v_{\rm A}}{H} \right)^2 + \omega_\nu \eta k^2_{\rm x} \right]\left( \omega_\nu + \eta k^2_{\rm x} \right) \\
&- F_0 \left( \omega_\nu + \eta k^2_{\rm x} \right) \left(\omega_\nu + \eta k^2_{\rm x} - F_0 \right) = 0.
\end{split}
\end{equation}
Algebraic amenability suggests that it is easiest to specify $F_0$, $\omega_\nu$ and $\eta$ and use the preceding expression to solve for $v_{\rm A}$.

The wave solutions are 
\begin{equation}
\begin{split}
h^\prime =& \frac{k_{\rm x} H v_{\rm x_0}}{\omega^2_{\rm R} + F_0^2} \left[ \omega_{\rm R} \cos{\left( k_{\rm x} x - \omega_{\rm R}t \right)} - F_0 \sin{\left( k_{\rm x} x - \omega_{\rm R}t \right)} \right], \\
b^\prime_{\rm x} =& \frac{\bar{B}_{\rm z} v_{\rm x_0}}{H \left[ \left( \eta k^2_{\rm x} \right)^2 + \omega^2_{\rm R} \right]} \\
& \times [ \eta k^2_{\rm x} \cos{\left( k_{\rm x} x - \omega_{\rm R}t \right)} - \omega_{\rm R} \sin{\left( k_{\rm x} x - \omega_{\rm R}t \right)} ].
\end{split}
\end{equation}

For a horizontal background field, the algebra is somewhat more tractable.  The oscillatory part of the wave solutions is described by
\begin{equation}
\begin{split}
\omega_{\rm R}^2 =& k^2_{\rm x} \left[ gH + v_{\rm A}^2 + \eta \left( \omega_\nu - F_0 \right) \right] - F_0 \omega_\nu \\
&+ \omega_{\rm I} \left[ 3 \omega_{\rm I} + 2 \left( \eta k_{\rm x}^2 + \omega_\nu - F_0 \right) \right],
\end{split}
\end{equation}
from which it is possible to obtain an expression for $\omega_{\rm R}$ for a balanced flow.  Requiring $\omega_{\rm R}^2 \ge 0$ sets a condition on the strength of forcing,
\begin{equation}
F_0 \le \frac{k^2_{\rm x} \left[ gH + v_{\rm A}^2 + \eta \omega_\nu \right]}{\eta k^2_{\rm x} + \omega_\nu}.
\end{equation}
If balanced flow is interpreted as a local condition, then the preceding equation specifies the maximum amount of forcing that can be applied to maintain it.  An equation for $\omega_{\rm I}$ may be obtained,
\begin{equation}
\begin{split}
8\omega_{\rm I}^3 &+ 8 \left( \omega_\nu + \eta k^2_{\rm x} - F_0 \right) \omega^2_{\rm I} \\
&+ 2 k^2_{\rm x} \left[ gH + v^2_{\rm A} + \eta \left( \omega_\nu - F_0 \right) \right] \omega_{\rm I} \\
& + 2\left[ \left( \omega_\nu + \eta k^2_{\rm x} - F_0 \right)^2 - F_0 \omega_\nu \right] \omega_{\rm I} \\
& + \left( \omega_\nu + \eta k^2_{\rm x} - F_0 \right) \left\{ k^2_{\rm x} \left[ gH + v^2_{\rm A} + \eta \left( \omega_\nu - F_0 \right) \right] \right\} \\
&- F_0 \omega_\nu \left( \omega_\nu + \eta k^2_{\rm x} - F_0 \right) - gHk^4_{\rm x} \eta \\
&+ F_0 k^2_{\rm x} \left( v^2_{\rm A} + \omega_\nu \eta \right) = 0.\\
\end{split}
\end{equation}
Even in a balanced flow ($\omega_{\rm I}=0$), the preceding expression demonstrates that non-vanishing terms remain because of the various couplings between forcing and friction.  The wave solutions are 
\begin{equation}
\begin{split}
h^\prime =& \frac{k_{\rm x} H v_{\rm x_0}}{\omega^2_{\rm R} + F_0^2} \left[ \omega_{\rm R} \cos{\left( k_{\rm x} x - \omega_{\rm R}t \right)} - F_0 \sin{\left( k_{\rm x} x - \omega_{\rm R}t \right)} \right], \\
b^\prime_{\rm x} =& -\frac{k_{\rm x} \bar{B}_{\rm x} v_{\rm x_0}}{\left( \eta k^2_{\rm x} \right)^2 + \omega^2_{\rm R}} \\
& \times [ \omega_{\rm R} \cos{\left( k_{\rm x} x - \omega_{\rm R}t \right)} + \eta k^2_{\rm x} \sin{\left( k_{\rm x} x - \omega_{\rm R}t \right)} ].
\end{split}
\end{equation}

\section{2D Models (Cartesian)}
\label{sect:2d}

\begin{figure*}
\begin{center}
\vspace{-0.2in}
\includegraphics[width=\columnwidth]{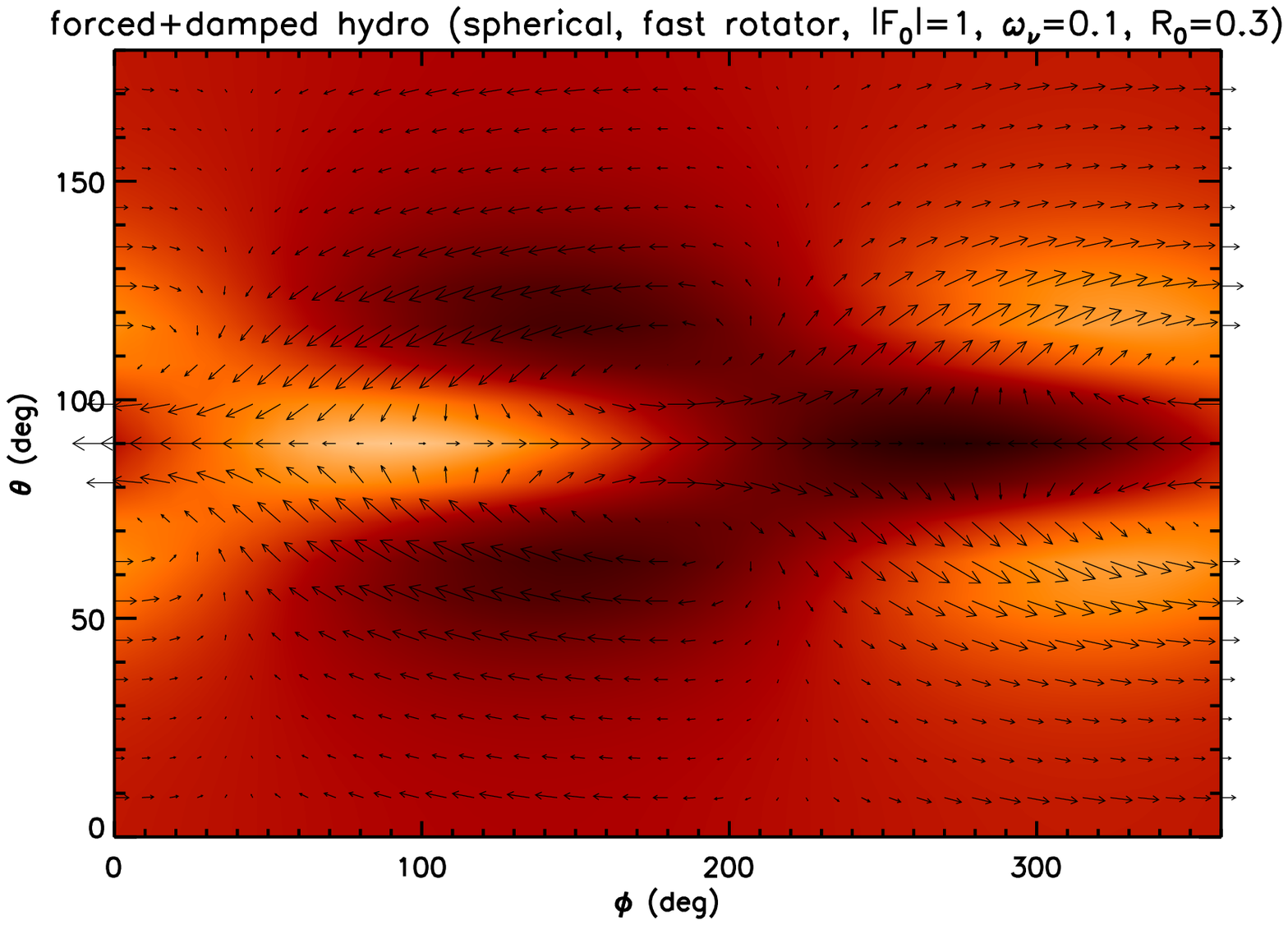}
\includegraphics[width=\columnwidth]{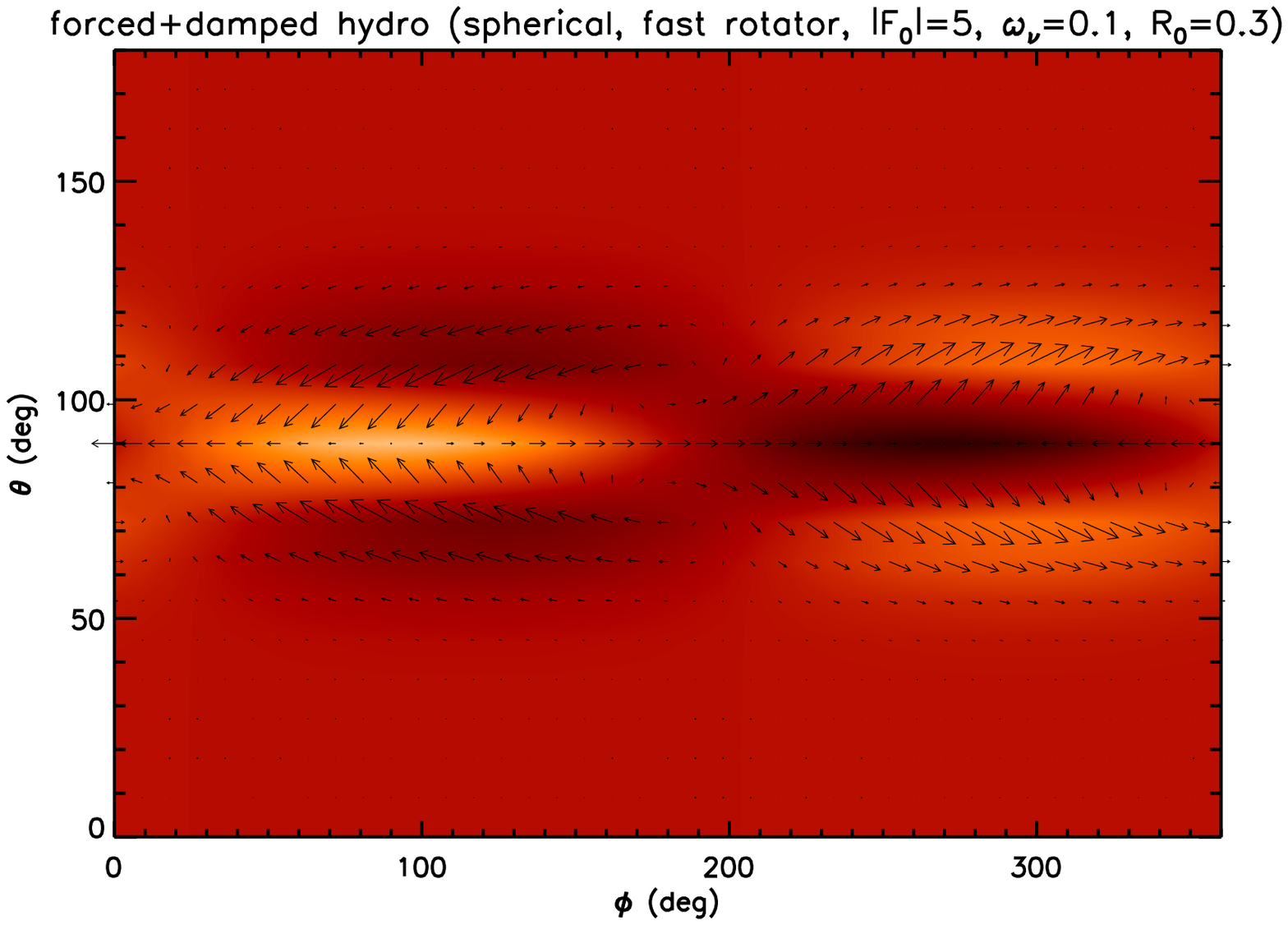}
\includegraphics[width=\columnwidth]{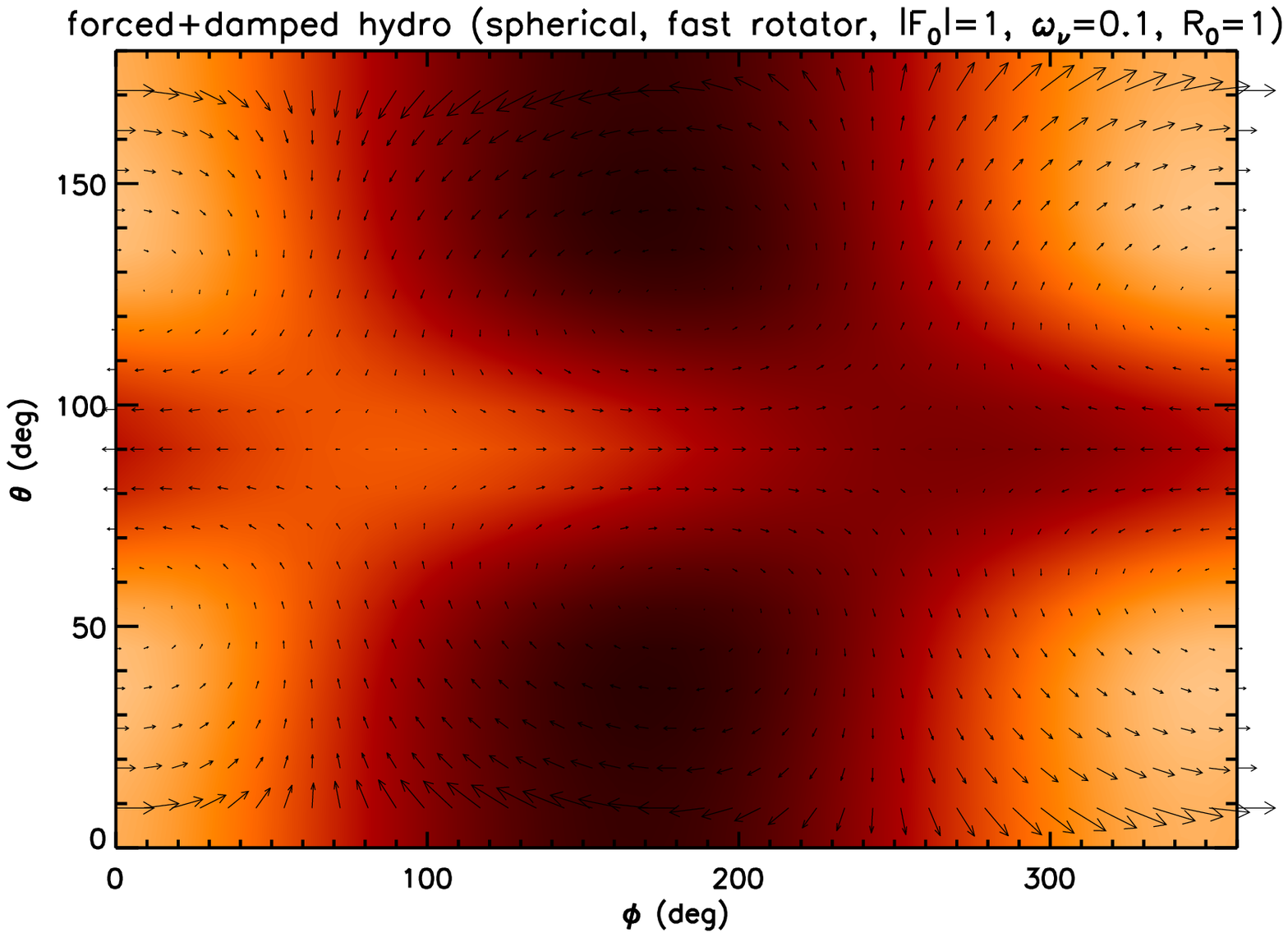}
\includegraphics[width=\columnwidth]{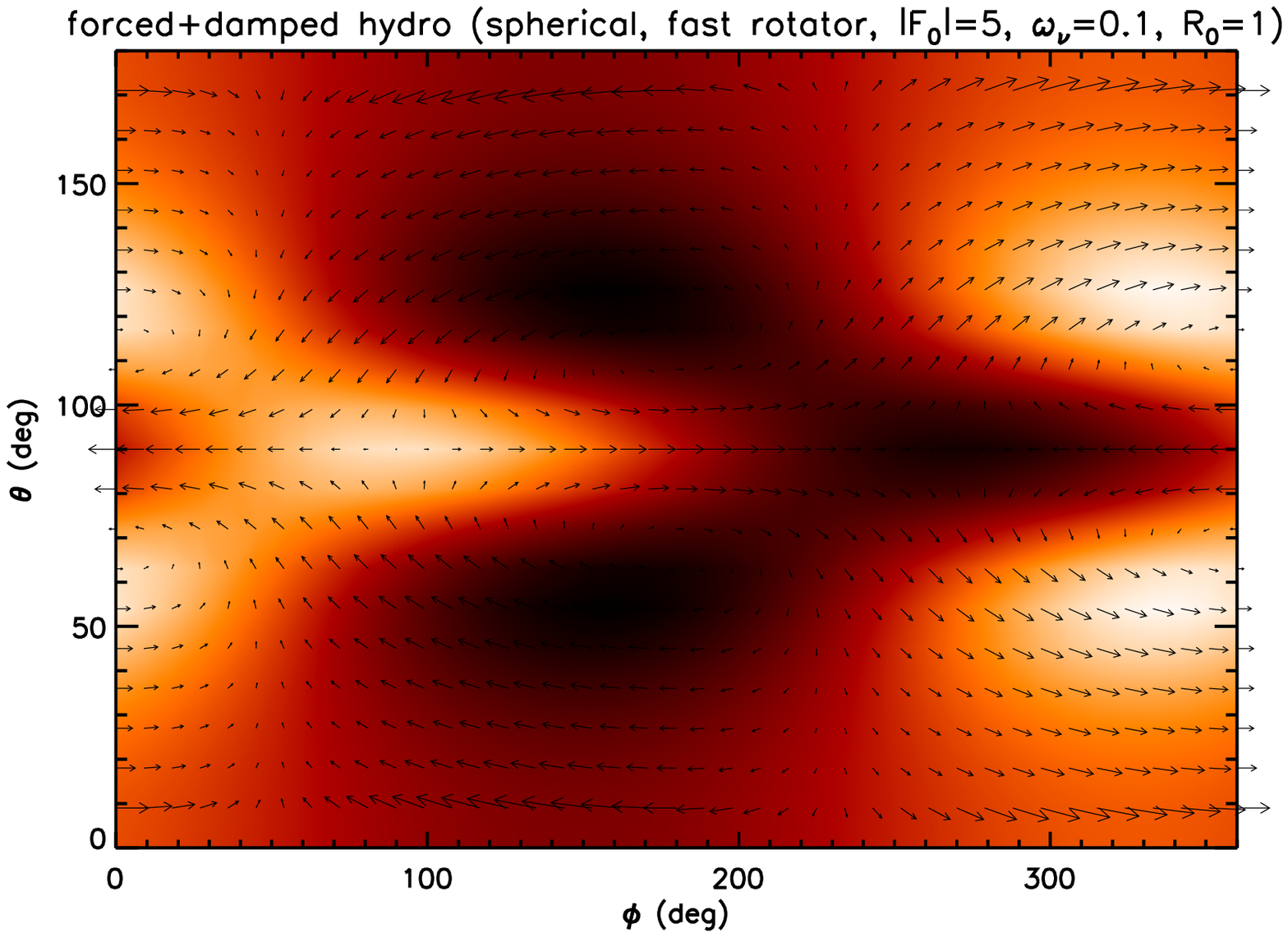}
\includegraphics[width=\columnwidth]{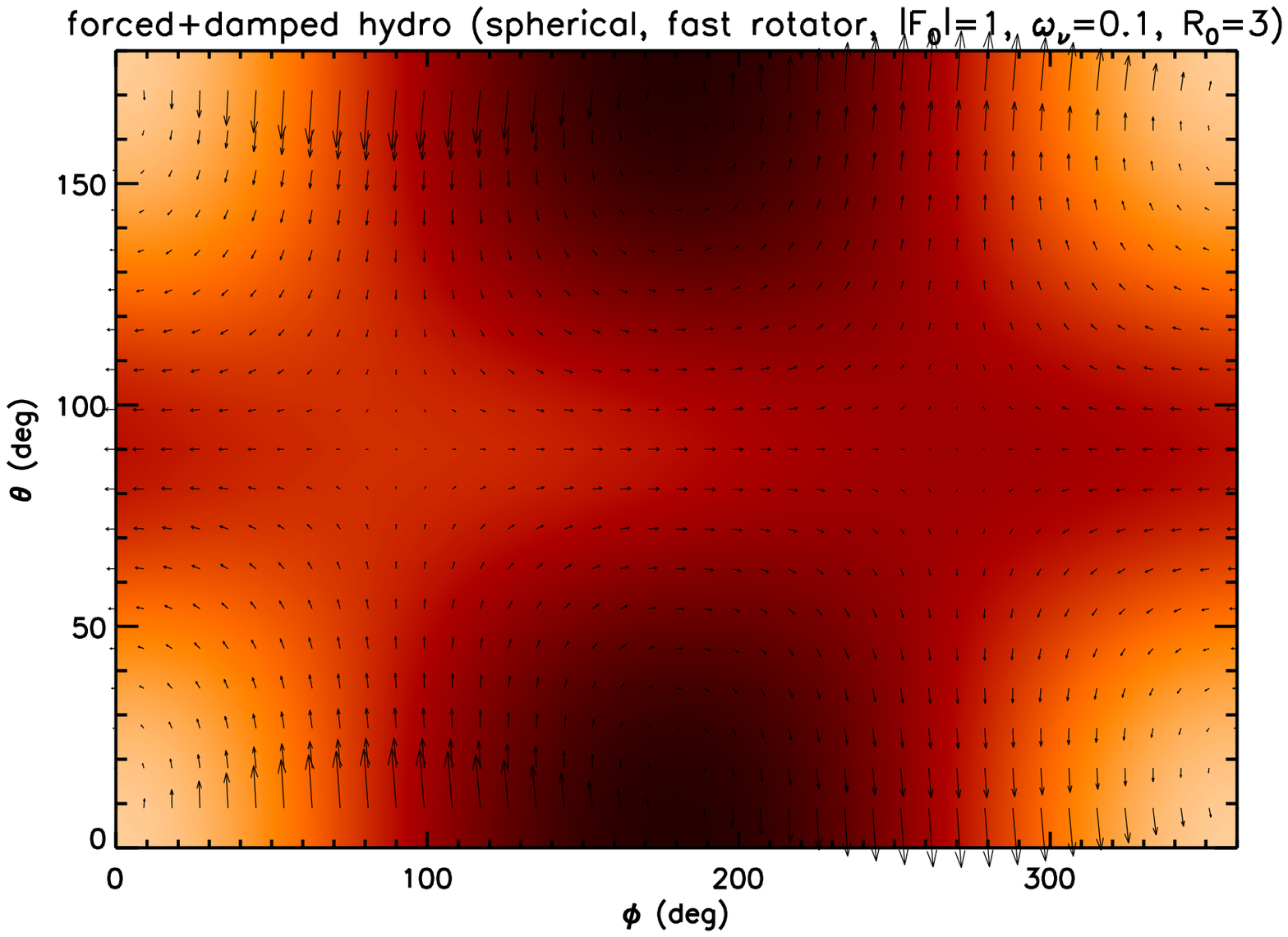}
\includegraphics[width=\columnwidth]{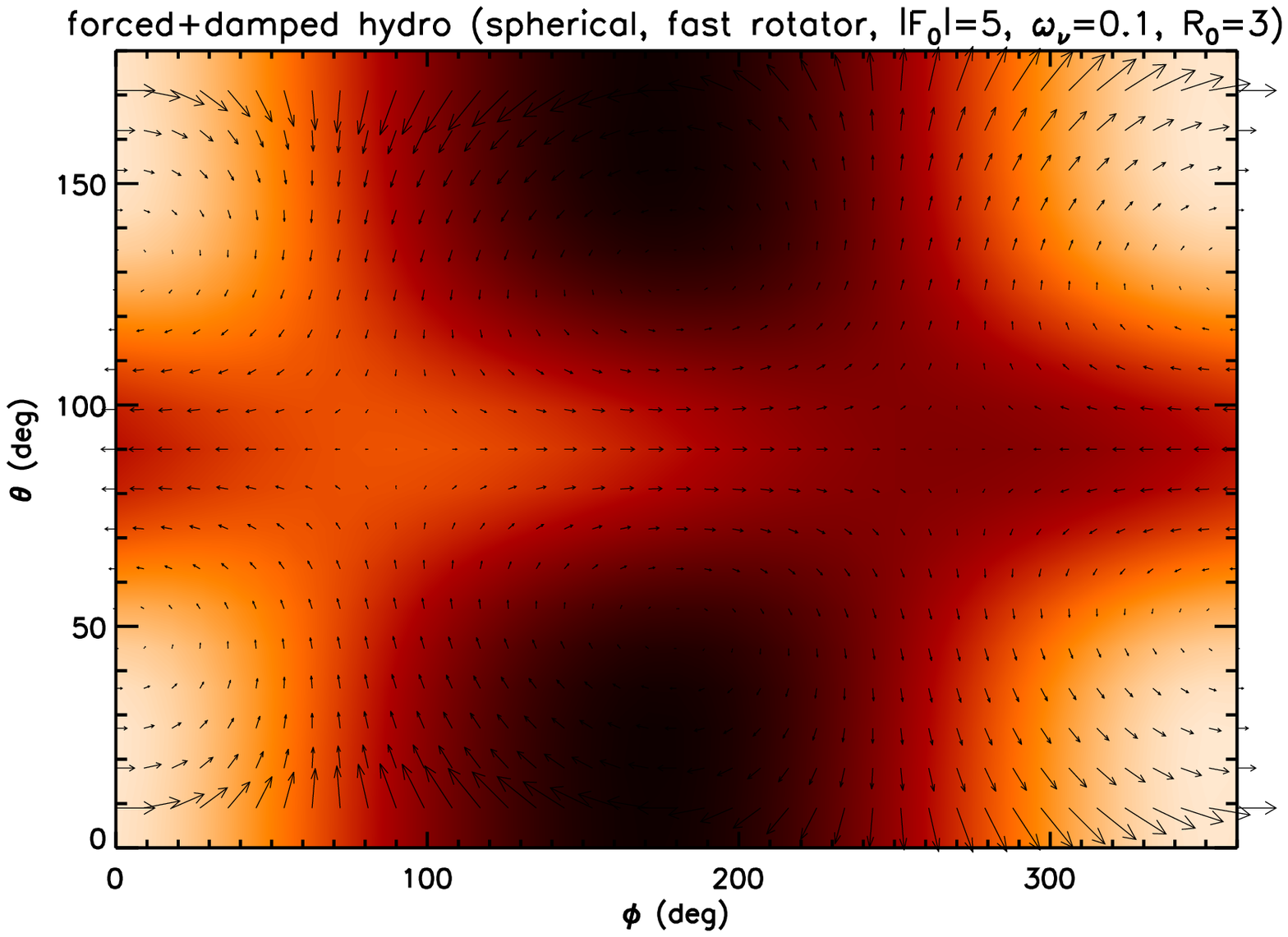}
\end{center}
\vspace{-0.2in}
\caption{Montage of plots of velocity perturbations (arrows) and $h^\prime_{\rm v}$ (contours) for steady-state, hydrodynamic systems in full spherical geometry for $l=m=1$, exploring the effects of rotation (via variation of the Rossby number) and forcing.  All quantities are computed in terms of an arbitrary velocity normalization ($v_0$).  Bright and dark colors correspond to positive and negative height perturbations, respectively.}
\vspace{0.1in}
\label{fig:sphere_montage}
\end{figure*}

Rotation is intrinsically a 2D effect and exerts non-trivial effects on the atmospheric flow.  It generally modifies the condition corresponding to a balanced flow away from its 1D counterpart.  In this section, we seek wave solutions with $\Psi_0 = \exp{(i k_{\rm y} y)}$ (see \S\ref{subsect:seek_wave}).  This approach of seeking sinusoidal solutions in both the $x$- and $y$-directions is less general than solving for the $y$-dependence of each quantity (see \S\ref{sect:2d_beta}).  We shall simply refer to it as the ``$\beta$-plane approximation".  Our main goal in this section is to extract dispersion relations in the presence of rotation, in anticipation of the more general treatments in \S\ref{sect:2d_beta} and \S\ref{sect:spherical}.

Generally, we find that the hydrodynamic and MHD systems are described by dispersion relations with the same mathematical forms, but with the wave frequency generalized to complex frequencies involving forcing, friction and magnetic field strength.

\subsection{Forcing with Hydrodynamic Friction}

A useful mathematical trick is to is to differentiate the equation for $v^\prime_{\rm x}$ with respect to $y$ before seeking wave solutions \citep{hs09},
\begin{equation}
\begin{split}
\frac{\partial^2 v^\prime_{\rm x}}{\partial t \partial y} =& -g \frac{\partial^2 h^\prime}{\partial x \partial y} + \beta v^\prime_{\rm y} + f \frac{\partial v^\prime_{\rm y}}{\partial y} \\
&+ \nu \left( \frac{\partial^3 v^\prime_{\rm x}}{\partial x^2 \partial y} + \frac{\partial^3 v^\prime_{\rm x}}{\partial y^3} \right) - \omega_{\rm drag} \frac{\partial v^\prime_{\rm x}}{\partial y},
\end{split}
\end{equation}
where we note that $\partial f/\partial y = \beta$.  Along with the other two perturbed equations, we seek wave solutions and arrange them into the following form,
\[ \hat{{\cal M}} \left( \begin{array}{c}
v_{\rm x_0} \\
v_{\rm y_0} \\
h_0 \\
\end{array} \right) = 0.\]
The preceding expression is non-trivial only if the determinant of the matrix,
\[ \hat{{\cal M}} = \left( \begin{array}{ccc}
k_{\rm y} \omega_0 & - \left( \beta + i f k_{\rm y} \right) & -g k_{\rm x} k_{\rm y} \\
f & -i \omega_0 & i g k_{\rm y} \\
k_{\rm x} H & k_{\rm y} H & - \omega_{\rm F} \end{array} \right),\]
is zero.  To express the matrix more compactly, we have defined
\begin{equation}
\begin{split}
\omega_\nu &\equiv \nu k^2 + \omega_{\rm drag}, \\
\omega_0 &\equiv \omega + i\omega_\nu, \\
\omega_{\rm F} &\equiv \omega - i F_0, \\
\end{split}
\end{equation}
where $k^2 = k_{\rm x}^2 + k_{\rm y}^2$.  

When one sets det$\hat{{\cal M}}=0$, it follows that
\begin{equation}
\begin{split}
&\omega_{\rm I}^3 + \omega_{\rm I} \left[ gH k^2 + f^2 + \omega_\nu \left( \omega_\nu - 2 F_0 \right) - 3 \omega^2_{\rm R} \right] \\
&+ \left( \omega_{\rm R}^2 - \omega_{\rm I}^2 \right) \left( F_0 - 2 \omega_\nu \right) - \frac{\beta f \omega_{\rm R}}{k_{\rm y}} - F_0 \left( \omega^2_\nu + f^2 \right) \\
&+ gH k^2 \omega_\nu = 0, \\
&\omega^3_{\rm R} - \omega_{\rm R}  \left[ gH k^2 + f^2 + \omega_\nu \left( \omega_\nu - 2 F_0 \right) + 3 \omega^2_{\rm I} \right] \\
&- 2 \left( 2\omega_\nu - F_0 \right) \omega_{\rm R} \omega_{\rm I} - g H k_{\rm x} \beta - \frac{\beta f \left( F_0 - \omega_{\rm I} \right)}{k_{\rm y}} = 0.
\end{split}
\label{eq:2d_cartesian_dispersion}
\end{equation}
As in 1D, the first equation describes how to obtain the condition for a balanced flow ($\omega_{\rm I}=0$).  The second equation is the dispersion relation for $\omega_{\rm R}$.  In general, it is challenging to solve this pair of coupled equations, but we are primarily interested in the $\omega_{\rm I}=0$ limit, for which the first expression in (\ref{eq:2d_cartesian_dispersion}) becomes
\begin{equation}
F_0 \omega^2_{\rm R} + \frac{\beta f \omega_{\rm R}}{k_{\rm y}} + F_0 \left[ \left( F_0^2 + f^2 \right) - gH k^2 \right]=0.
\end{equation}
An important implication of the preceding expression is that $F_0 = \omega_\nu$ is generally \emph{not} a consequence of $\omega_{\rm I}=0$.  However, when one additionally sets $k_{\rm y}=0$, one obtains $\beta f \omega_{\rm R} = 0$, which implies that $F_0 = \omega_\nu$ is a consequence of $\omega_{\rm I}=0$ either at the poles ($\beta=0$, since $\beta = 2\Omega \sin\theta/R$) or when rotation is absent ($f=0$).  These inferences also obtain in free systems: setting $F_0=0$ yields $\beta f \omega_{\rm R} = 0$.

In general, given specified values of $F_0$ and $\omega_\nu$, one may solve the coupled pair of equations in (\ref{eq:2d_cartesian_dispersion}) for the values of $\omega_{\rm R}$ and $\omega_{\rm I}$.  Such an approach yields solutions for flows that are generally unbalanced.  In this four-dimensional space, there exists solutions for balanced flows ($\omega_{\rm I}=0$).  An easier way to proceed is to \emph{assume} a balanced flow by setting $\omega_{\rm I}=0$ and solving the equations in (\ref{eq:2d_cartesian_dispersion}).  One obtains the usual three values of $\omega_{\rm R}$ for the eastward- and westward-propagating Poincar\'{e} waves and the westward-propagating Rossby waves (e.g., \citealt{kundu}).\footnote{Here, ``eastward" and ``westward" refer to waves propagating along and against the direction of rotation, respectively.}  Poincar\'{e} waves are rotationally-modified gravity waves, while Rossby waves arise from non-constant rotation across latitude ($\beta \ne 0$).  Balanced flow exists for appropriate combinations of $F_0$ and $\omega_\nu$ values.  We will see that in the pseudo-spherical (\S\ref{sect:2d_beta}) and spherical (\S\ref{sect:spherical}) cases, the algebra is intractable for evaluating balanced flows and one must instead resort to solving for the steady state of an atmosphere ($\omega_{\rm R}=\omega_{\rm I}=0$).

\subsection{Magnetohydrodynamic (Vertical Background Field)}

We employ the same mathematical trick with one of the perturbed equations,
\begin{equation}
\begin{split}
\frac{\partial^2 v^\prime_{\rm x}}{\partial t \partial y} =& -g \frac{\partial^2 h^\prime}{\partial x \partial y} + \beta v^\prime_{\rm y} + f \frac{\partial v^\prime_{\rm y}}{\partial y} - \frac{\bar{B}_z}{4 \pi \rho H} \frac{\partial b^\prime_{\rm x}}{\partial y} \\
&+ \nu \left( \frac{\partial^3 v^\prime_{\rm x}}{\partial x^2 \partial y} + \frac{\partial^3 v^\prime_{\rm x}}{\partial y^3} \right) - \omega_{\rm drag} \frac{\partial v^\prime_{\rm x}}{\partial y}.
\end{split}
\end{equation}
Seeking wave solutions and constructing the $\hat{{\cal M}}$ matrix as before, we obtain
\[ \hat{{\cal M}} = \left( \begin{array}{ccc}
k_{\rm y} \omega_{\rm B_0} & - \left( \beta + i f k_{\rm y} \right) & -g k_{\rm x} k_{\rm y} \\
f & -i \omega_{\rm B_0} & i g k_{\rm y} \\
k_{\rm x} H & k_{\rm y} H & - \omega_{\rm F} \end{array} \right),\]
where we have defined
\begin{equation}
\begin{split}
\omega_\eta &\equiv \omega + i \eta k^2, \\
\omega_{\rm B} &\equiv \omega_\nu + \frac{i}{\omega_\eta} \left( \frac{v_{\rm A}}{H} \right)^2, \\
\omega_{\rm B_0} &\equiv \omega + i \omega_{\rm B}.
\end{split}
\end{equation}
Setting $\mbox{det}\hat{{\cal M}}=0$ yields
\begin{equation}
i k_{\rm y} \omega^2_{\rm B_0} \omega_{\rm F} - i g H k_{\rm y} \left( k^2 \omega_{\rm B_0} + k_{\rm x} \beta \right) - \beta \omega_{\rm F} f - i f^2 k_{\rm y} \omega_{\rm F} = 0.
\end{equation}
To proceed, we find it convenient to first write
\begin{equation}
\omega_{\rm B} = \zeta_{\rm R} + i \zeta_{\rm I},
\end{equation}
en route to obtaining a pair of equations for $\omega_{\rm R}$ and $\omega_{\rm I}$,
\begin{equation}
\begin{split}
&\omega_{\rm I}^3 - 3 \omega_{\rm R}^2 \omega_{\rm I} - \left( 2 \zeta_{\rm R} - F_0 \right) \left( \omega_{\rm R}^2 - \omega_{\rm I}^2 \right) + 4 \zeta_{\rm I} \omega_{\rm R} \omega_{\rm I} \\
&-\omega_{\rm I} \left[ \zeta_{\rm R} \left( 2 F_0 - \zeta_{\rm R} \right) + \zeta_{\rm I}^2 \right] - 2 \zeta_{\rm I} \left( F_0 - \zeta_{\rm R} \right) \omega_{\rm R} \\
&+ gHk^2 \left( \omega_{\rm I} + \zeta_{\rm R} \right) - \frac{\beta f \omega_{\rm R}}{k_{\rm y}} + f^2 \left( \omega_{\rm I} - F_0 \right) \\
&-\left( \zeta_{\rm R}^2 - \zeta_{\rm I}^2 \right) F_0 = 0, \\
&\omega_{\rm R}^3 - 3\omega_{\rm R} \omega_{\rm I}^2 - 2 \left( 2 \zeta_{\rm R} - F_0 \right) \omega_{\rm R} \omega_{\rm I} - 2 \zeta_{\rm I} \left( \omega_{\rm R}^2 - \omega_{\rm I}^2 \right) \\
&+ \omega_{\rm R} \left[ \zeta_{\rm R} \left( 2 F_0 - \zeta_{\rm R} \right) + \zeta_{\rm I}^2 \right]  - 2 \zeta_{\rm I} \left( F_0 - \zeta_{\rm R} \right) \omega_{\rm I} \\
&+ gHk^2 \left( \zeta_{\rm I} - \omega_{\rm R} \right) - f^2 \omega_{\rm R} + \frac{\beta f}{k_{\rm y}} \left( F_0 - \omega_{\rm I} \right) \\
&- 2 \zeta_{\rm R} \zeta_{\rm I} F_0 - gHk_{\rm x} \beta = 0.
\end{split}
\label{eq:2d_cartesian_mhd_vert_dispersion}
\end{equation}
It can then be shown, after the fact, that
\begin{equation}
\begin{split}
\zeta_{\rm R} &\equiv \omega_\nu + \left(\frac{v_{\rm A}}{H}\right)^2 \frac{\eta k^2 + \omega_{\rm I}}{\omega_{\rm R}^2 + \left( \eta k^2 + \omega_{\rm I} \right)^2}, \\
\zeta_{\rm I} &\equiv \left(\frac{v_{\rm A}}{H}\right)^2 \frac{\omega_{\rm R}}{\omega_{\rm R}^2 + \left( \eta k^2 + \omega_{\rm I} \right)^2},
\end{split}
\end{equation}
by substituting the expression for $\omega_\eta$ into $\omega_{\rm B}$ and separating out the real and imaginary components.

At this point, the algebra is becoming intractable, since the pair of expressions in (\ref{eq:2d_cartesian_mhd_vert_dispersion}) no longer takes the form of a cubic equation.

\subsection{Magnetohydrodynamic (Horizontal Background Field)}

For a horizontal background field, the equation for the velocity perturbation in the $x$-direction reads,
\begin{equation}
\begin{split}
\frac{\partial^2 v^\prime_{\rm x}}{\partial t \partial y} =& -g \frac{\partial^2 h^\prime}{\partial x \partial y} + \beta v^\prime_{\rm y} + f \frac{\partial v^\prime_{\rm y}}{\partial y} + \frac{\bar{B}_x}{4 \pi \rho} \frac{\partial^2 b^\prime_{\rm x}}{\partial x \partial y} + \frac{\bar{B}_y}{4 \pi \rho} \frac{\partial^2 b^\prime_{\rm x}}{\partial y^2} \\
&+ \nu \left( \frac{\partial^3 v^\prime_{\rm x}}{\partial x^2 \partial y} + \frac{\partial^3 v^\prime_{\rm x}}{\partial y^3} \right) - \omega_{\rm drag} \frac{\partial v^\prime_{\rm x}}{\partial y}.
\end{split}
\end{equation}
Seeking wave solutions, we find that the $\hat{{\cal M}}$ matrix is identical to the one derived for the vertical background field, except that
\begin{equation}
\begin{split}
\vec{v}_{\rm A} . \vec{k} &\equiv \frac{\bar{B}_{\rm x} k_{\rm x} + \bar{B}_{\rm y} k_{\rm y}}{2 \left( \pi \rho \right)^{1/2}}, \\
\omega_{\rm B} &\equiv \omega_\nu + \frac{i \left( \vec{v}_{\rm A} . \vec{k} \right)^2}{\omega_\eta}.
\end{split}
\end{equation}
The quantities $\omega_\eta$ and $\omega_{\rm B_0}$ have the same definitions as in the case of a vertical background field.  Again, we first write $\omega_{\rm B} = \zeta_{\rm R} + i \zeta_{\rm I}$ and evaluate $\mbox{det}\hat{{\cal M}}=0$.  Since $\hat{{\cal M}}$ is mathematically identical to the previous situation, we recover the pair of expressions in (\ref{eq:2d_cartesian_mhd_vert_dispersion}), although the definitions for $\zeta_{\rm R}$ and $\zeta_{\rm I}$ differ slightly,
\begin{equation}
\begin{split}
\zeta_{\rm R} &\equiv \omega_\nu + \frac{\left(\vec{v}_{\rm A} . \vec{k} \right)^2 \left(\eta k^2 + \omega_{\rm I}\right)}{\omega_{\rm R}^2 + \left( \eta k^2 + \omega_{\rm I} \right)^2}, \\
\zeta_{\rm I} &\equiv \frac{\omega_{\rm R} \left(\vec{v}_{\rm A} . \vec{k} \right)^2}{\omega_{\rm R}^2 + \left( \eta k^2 + \omega_{\rm I} \right)^2}.
\end{split}
\end{equation}

\section{2D Pseudo-Spherical Models (Equatorial $\beta$-Plane)}
\label{sect:2d_beta}

The first step towards considering the effects of sphericity is to allow for the physical quantities to have an arbitrary, rather than a sinusoidal, functional dependence on the latitude ($y$).  Specifically, we solve for the functional dependence of $X_0 \Psi_0$ in $y$ (see \S\ref{subsect:seek_wave}).  This approximation captures the essence of being near the equator on a sphere without actually working in full spherical coordinates.  We shall refer to this approach as the ``equatorial $\beta$-plane approximation".

In this section, we choose to non-dimensionalize our quantities by defining the following characteristic length ($L_0$) and time ($t_{\rm dyn}$) scales,
\begin{equation}
L_0 = \left( \frac{c_0}{\beta} \right)^{1/2}, ~t_{\rm dyn} = \left( c_0 \beta \right)^{-1/2},
\end{equation}
following \cite{matsuno66}.  The characteristic velocity is $c_0 \equiv \left( g H \right)^{1/2}$.  Furthermore, the shallow water height perturbation is normalized by $H$.  Near the equator, we have $f \approx \beta y$.  

Generally, we find that to obtain the dispersion relations, one needs to evaluate the square root of a complex quantity, $\zeta_{\rm R} + i \zeta_{\rm I}$, which requires the use of De Moivre's formula (e.g., \citealt{aw95}).  The expressions for $\zeta_{\rm R}$ and $\zeta_{\rm I}$, in terms of the other quantities, are generally tedious.  To avoid introducing more notation than is necessary, we will generally make use of the separation functions $\zeta_{\rm R}$ and $\zeta_{\rm I}$ when invoking De Moivre's formula, but we note that their exact definitions will vary between models.

\subsection{Relationship to the Quantum Harmonic Oscillator}

Consider the general equation for the function $F=F(z)$,
\begin{equation}
\frac{d^2F}{dz^2} - \left( {\cal A} z^2 - {\cal B} z - {\cal C} \right) F = 0,
\label{eq:general_beta_0}
\end{equation}
where the coefficients ${\cal A}$, ${\cal B}$ and ${\cal C}$ may be complex.  By completing the square, this equation may be written as
\begin{equation}
\frac{d^2F}{dz^2} - {\cal A} \left( z - \frac{{\cal B}}{2 {\cal A}} \right)^2 F + \left(\frac{{\cal B}^2}{4 {\cal A}} + {\cal C} \right) F = 0.
\label{eq:general_beta}
\end{equation}

The next step is to rescale $z$.  It turns out that the choice of rescaling is important.  If we pick
\begin{equation}
z \rightarrow 2^{1/2} {\cal A}^{1/4} \left( z - \frac{{\cal B}}{2 {\cal A}} \right),
\end{equation}
then equation (\ref{eq:general_beta}) becomes one of the Weber equations,
\begin{equation}
\frac{d^2F}{dz^2} - \frac{z^2 F}{4} + \frac{{\cal A}^\prime F}{2} = 0.
\end{equation}
But if we instead pick
\begin{equation}
z \rightarrow {\cal A}^{1/4} \left( z - \frac{{\cal B}}{2 {\cal A}} \right),
\end{equation}
then equation (\ref{eq:general_beta}) becomes the governing equation for the quantum harmonic oscillator,
\begin{equation}
\frac{d^2F}{dz^2} - z^2 F + {\cal A}^\prime F = 0,
\end{equation}
provided that
\begin{equation}
{\cal A}^\prime \equiv \left( \frac{{\cal B}^2}{4 {\cal A}^{3/2}} + \frac{{\cal C}}{{\cal A}^{1/2}} \right)
\end{equation}
is discretized/quantized.

The subtle difference between these two equations may be emphasized by considering the equation,
\begin{equation}
\frac{d^2F}{dz^2} + \left( 2n + 1 - {\cal D}_1 z^2 \right) F = 0,
\label{eq:hermite_demo}
\end{equation}
where $n$ is an integer.  Consider the proposed solution,
\begin{equation}
F \propto \exp{\left(- \frac{{\cal D}_2 z^2}{2} \right)} {\cal H}_n,
\label{eq:proposed_solution}
\end{equation}
where ${\cal H}_n = {\cal H}_n(z)$ is the Hermite polynomial.\footnote{These are the ``physicists' Hermite polynomials", where ${\cal H}_1 = 2z$ rather than $z$.}  The function $F$ is the parabolic cylinder function.  By using the recurrence relations for Hermite polynomials,
\begin{equation}
\begin{split}
{\cal H}_{n-1} &= \frac{1}{n} \left( z {\cal H}_n - \frac{{\cal H}_{n+1}}{2} \right), \\
\frac{d^2{\cal H}_n}{dz^2} &= 2n \left( 2 z {\cal H}_{n-1} - {\cal H}_n \right),
\end{split}
\end{equation}
we may reduce equation (\ref{eq:hermite_demo}) to
\begin{equation}
4 n z {\cal H}_{n-1} \left( 1 - {\cal D}_2 \right) + {\cal H}_n \left[ z^2 \left( {\cal D}_2 - {\cal D}_1 \right) - {\cal D}_2 + 1 \right] = 0.
\end{equation}
For $z \ne 0$, the equality holds only if ${\cal D}_1 = {\cal D}_2 = 1$.  Thus, the proposed solution in equation (\ref{eq:proposed_solution}) is true only if the governing equation follows the same form as that for the quantum harmonic oscillator.  But since all three equations are related by the appropriate transformations, one may always cast the problem in terms of a quantum harmonic oscillator and transform back to the appropriate form.

Generally, a shallow water model on the $\beta$-plane, with forcing and friction, is described by equation (\ref{eq:general_beta_0}) with ${\cal B}=0$.  For a free system, the coefficients ${\cal A}$ and ${\cal C}$ are real; for a forced or damped system, they are generally imaginary.  

The recognition that the meridional velocity obeys the same governing equation as a quantum harmonic oscillator, for a free system, is not novel \citep{matsuno66}.  However, the insight that the governing equation for the meridional velocity may always be transformed into the quantum harmonic oscillator equation, even in the presence of forcing, friction and magnetic fields for time-dependent systems, is novel and constitutes an improvement over the time-independent (stationary) analyses of \citep{matsuno66} and \citep{gill80}.  Furthermore, the use of De Moivre's formula to derive the dispersion relations is novel.

\subsection{Hydrodynamic}

\subsubsection{Free System}
\label{subsect:2d_beta_free}

For the most basic equatorial $\beta$-plane model (i.e., no molecular viscosity or Rayleigh drag), the dimensionless governing equations are
\begin{equation}
\begin{split}
\frac{\partial v_{\rm x}^\prime}{\partial t} &= - \frac{\partial h^\prime}{\partial x} + y v_{\rm y}^\prime, \\
\frac{\partial v_{\rm y}^\prime}{\partial t} &= - \frac{\partial h^\prime}{\partial y} - y v_{\rm x}^\prime, \\
\frac{\partial h^\prime}{\partial t} &= - \frac{\partial v_{\rm x}^\prime}{\partial x} - \frac{\partial v_{\rm y}^\prime}{\partial y}.
\end{split}
\end{equation}

By seeking wave solutions, one obtains three expressions for $v_{\rm x_0}$,
\begin{equation}
\begin{split}
v_{\rm x_0} &= \omega^{-1} \left( k_{\rm x} h_0 + i y v_{\rm y_0} \right), \\
v_{\rm x_0} &= y^{-1} \left( i \omega v_{\rm y_0} - \frac{\partial h_0}{\partial y} \right), \\
v_{\rm x_0} &= k_{\rm x}^{-1} \left( \omega h_0 + i \frac{\partial v_{\rm y_0}}{\partial y} \right).
\end{split}
\label{eq:beta_basic}
\end{equation}
Differentiating the third equation in (\ref{eq:beta_basic}) yields
\begin{equation}
\frac{\partial v_{\rm x_0}}{\partial y} = k_{\rm x}^{-1} \left( \omega \frac{\partial h_0}{\partial y} + i \frac{\partial^2 v_{\rm y_0}}{\partial y^2} \right). 
\end{equation}
Clearly, one needs expressions for $\partial v_{\rm x_0}/\partial y$ and $\partial h_0/\partial y$ in terms of $h_0$ and $v_{\rm y_0}$.  Differentiating the first equation in (\ref{eq:beta_basic}) gives $\partial v_{\rm x_0}/\partial y$.  Combining the first and third equations yields,
\begin{equation}
h_0 = i \left( \frac{1}{k_{\rm x}} \frac{\partial v_{\rm y_0}}{\partial y} - \frac{y v_{\rm y_0}}{\omega} \right) \left( \frac{k_{\rm x}}{\omega} - \frac{\omega}{k_{\rm x}} \right)^{-1}.
\label{eq:beta_h0}
\end{equation}
Combining the first and second equations gives
\begin{equation}
\begin{split}
\left( \frac{k_{\rm x}}{\omega} - \frac{\omega}{k_{\rm x}} \right) \frac{\partial h_0}{\partial y} =& i v_{\rm y_0} \left( \omega - \frac{y^2}{\omega} \right) \left( \frac{k_{\rm x}}{\omega} - \frac{\omega}{k_{\rm x}} \right)\\
&- \frac{iy}{\omega} \frac{\partial v_{\rm y_0}}{\partial y} + \frac{i y^2 k_{\rm x} v_{\rm y_0}}{\omega^2}.
\end{split}
\end{equation}
Putting it all together, we get
\begin{equation}
\frac{\partial^2 v_{\rm y_0}}{\partial y^2} + \left( \omega^2 - k^2_{\rm x} - \frac{k_{\rm x}}{\omega} - y^2 \right) v_{\rm y_0} = 0.
\label{eq:parabolic}
\end{equation}
Equation (\ref{eq:parabolic}) is exactly the equation for the quantum harmonic oscillator if the following combination of quantities is discretized,
\begin{equation}
\omega^2 - k^2_{\rm x} - \frac{k_{\rm x}}{\omega} = 2n + 1,
\end{equation}
where $n$ is the meridional wavenumber and takes on integer values.  Physically, $k_{\rm x}$ and $n$ behave like the $m$ and $l$ quantum numbers \citep{matsuno66}---an analogy we will formalize in \S\ref{sect:spherical}---except that $k_{\rm x}$ is free to take on non-integer values.  We note that equation (\ref{eq:parabolic}) differs from equation (6) of \cite{matsuno66}, because in that work free solutions with $\Psi \propto \exp{(i \omega t)}$, rather than $\Psi \propto \exp{(-i \omega t)}$, were sought.

While it is clear that the imaginary part of the wave frequency vanishes for a free system, we find it instructive to consider $\omega = \omega_{\rm R} + i \omega_{\rm I}$, which produces two expressions from the dispersion relation,
\begin{equation}
\begin{split}
&\omega_{\rm I} \left[ \omega_{\rm I}^2 - 3 \omega_{\rm R}^2 + \left( 2n + 1 + k^2_{\rm x} \right) \right] = 0, \\
&\omega_{\rm R}^3 - 3 \omega_{\rm R} \omega_{\rm I}^2 - \left( 2n + 1 + k^2_{\rm x} \right) \omega_{\rm R} - k_{\rm x} = 0. \\
\end{split}
\label{eq:beta_hydro_dispersions}
\end{equation}
One of the solutions of the first expression is $\omega_{\rm I}=0$, which corresponds to balanced flow in a free system.  The other two solutions for $\omega_{\rm I}$ are not physically meaningful, because they permit finite or imaginary values of $\omega_{\rm I}$.  The second expression yields three solutions for $\omega_{\rm R}$, which corresponds to the expected three types of waves that exist in free, hydrodynamic systems: eastward- and westward-propagating Poincar\'{e} waves and westward-propagating Rossby waves \citep{matsuno66,lh67,lh68,gill80}.  In other words, the first expression describes how balanced flow may be obtained, while the second expression describes how the waves oscillate.  In a balanced flow ($\omega_{\rm I}=0$), the dispersion relation becomes \citep{matsuno66}
\begin{equation}
\omega_{\rm R}^3 - \left( 2n + 1 + k^2_{\rm x} \right) \omega_{\rm R} - k_{\rm x} = 0.
\end{equation}

From solving equation (\ref{eq:parabolic}) and using equation (\ref{eq:beta_h0}) and the first expression in equation (\ref{eq:beta_basic}), we obtain the amplitudes of the perturbations,
\begin{equation}
\begin{split}
v_{\rm y_0} =& v_0 ~\exp{\left(-\frac{y^2}{2}\right)} ~{\cal H}_n, \\
h_0 =& i v_0 ~\exp{\left(-\frac{y^2}{2}\right)} \left( \frac{k_{\rm x}}{\omega_{\rm R}} - \frac{\omega_{\rm R}}{k_{\rm x}} \right)^{-1} \\
&\times \left[ \frac{2n {\cal H}_{n-1}}{k_{\rm x}} - y \left( \frac{1}{k_{\rm x}} + \frac{1}{\omega_{\rm R}} \right) {\cal H}_n \right], \\
v_{\rm x_0} =& \frac{i v_0}{k^2_{\rm x} - \omega_{\rm R}^2} ~\exp{\left(-\frac{y^2}{2}\right)} \\
&\times \left[ 2n k_{\rm x} {\cal H}_{n-1} - y \left( \omega_{\rm R} + k_{\rm x} \right) {\cal H}_n \right],
\end{split}
\label{eq:beta_hydro_amp}
\end{equation}
where $v_0$ is an arbitrary normalization factor.  As realized by \cite{matsuno66}, the $\omega_{\rm R}=k_{\rm x}$ solution is rejected when $n=0$.  The perturbations are obtained from taking the real parts of the wave solutions,
\begin{equation}
\begin{split}
v^\prime_{\rm y} =& v_0 ~\exp{\left(-\frac{y^2}{2}\right)} ~{\cal H}_n \cos{\left( k_{\rm x} x - \omega_{\rm R} t \right)}, \\
h^\prime =& v_0 ~\exp{\left(-\frac{y^2}{2}\right)} \left( \frac{k_{\rm x}}{\omega_{\rm R}} - \frac{\omega_{\rm R}}{k_{\rm x}} \right)^{-1} \sin{\left( k_{\rm x} x - \omega_{\rm R} t \right)} \\
&\times \left[ y \left( \frac{1}{k_{\rm x}} + \frac{1}{\omega_{\rm R}} \right) {\cal H}_n - \frac{2n {\cal H}_{n-1}}{k_{\rm x}}\right], \\
v^\prime_{\rm x} =& \frac{v_0}{k^2_{\rm x} - \omega_{\rm R}^2} ~\exp{\left(-\frac{y^2}{2}\right)} \sin{\left( k_{\rm x} x - \omega_{\rm R} t \right)} \\
&\times \left[ y \left( \omega_{\rm R} + k_{\rm x} \right) {\cal H}_n - 2n k_{\rm x} {\cal H}_{n-1} \right].
\end{split}
\label{eq:beta_hydro_solutions}
\end{equation}
In deriving equations (\ref{eq:beta_hydro_amp}) and (\ref{eq:beta_hydro_solutions}), we find that a useful recurrence relation to use is $d {\cal H}_n/dy = 2n {\cal H}_{n-1}$.

For a steady-state system ($\omega_{\rm R}=0$), we obtain
\begin{equation}
\begin{split}
v^\prime_{\rm y} =& v_0 ~\exp{\left(-\frac{y^2}{2}\right)} ~{\cal H}_n \cos{\left( k_{\rm x} x\right)}, \\
h^\prime =& \frac{v_0 y}{k_{\rm x}} ~\exp{\left(-\frac{y^2}{2}\right)} {\cal H}_n \sin{\left( k_{\rm x} x \right)}, \\
v^\prime_{\rm x} =& \frac{v_0}{k_{\rm x}} ~\exp{\left(-\frac{y^2}{2}\right)} \sin{\left( k_{\rm x} x \right)} \left[ y {\cal H}_n - 2n {\cal H}_{n-1} \right].
\end{split}
\label{eq:beta_hydro_steady}
\end{equation}

\subsubsection{Forcing with Hydrodynamic Friction}
\label{subsect:2d_beta_hydro}

\begin{figure*}
\begin{center}
\vspace{-0.2in}
\includegraphics[width=\columnwidth]{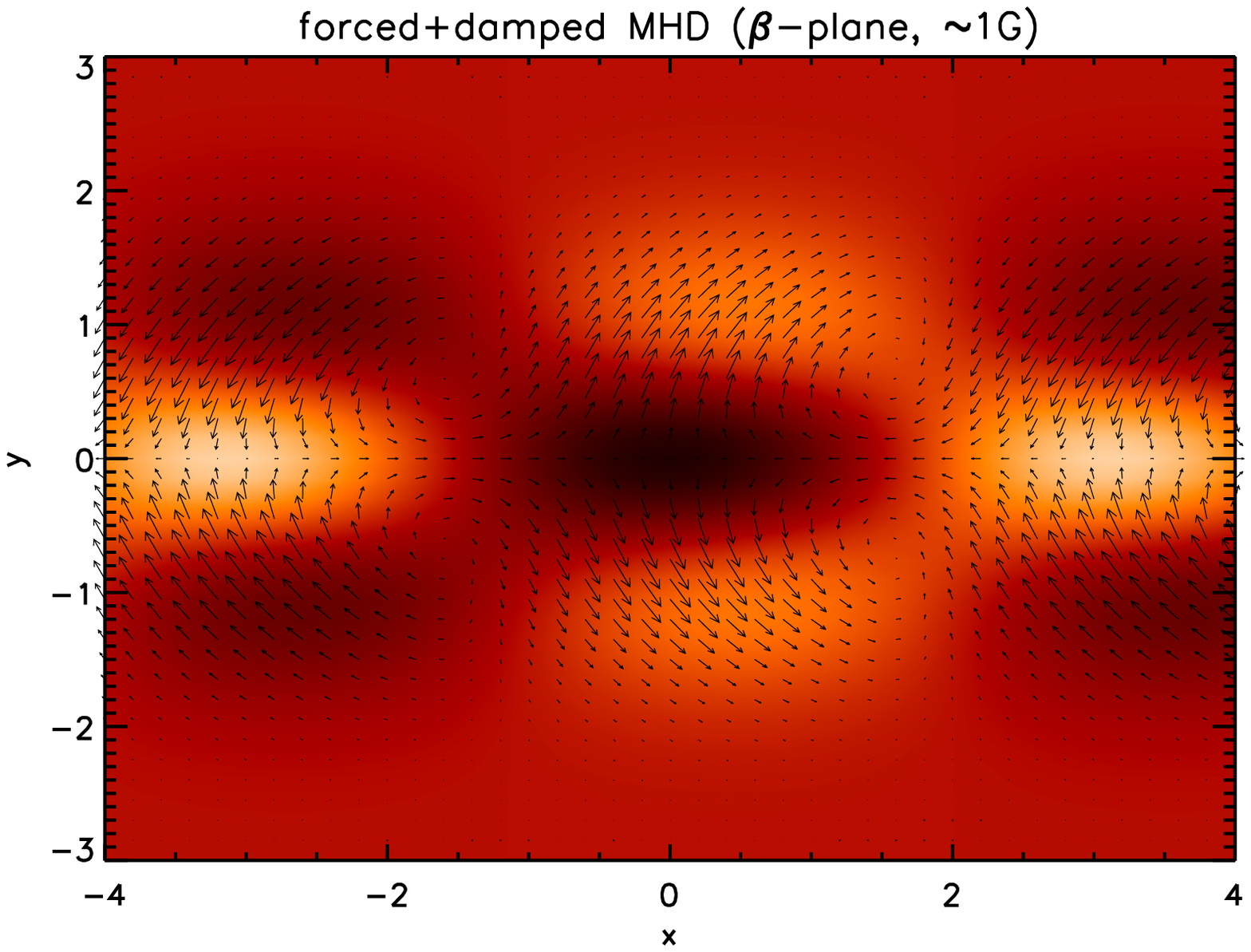}
\includegraphics[width=\columnwidth]{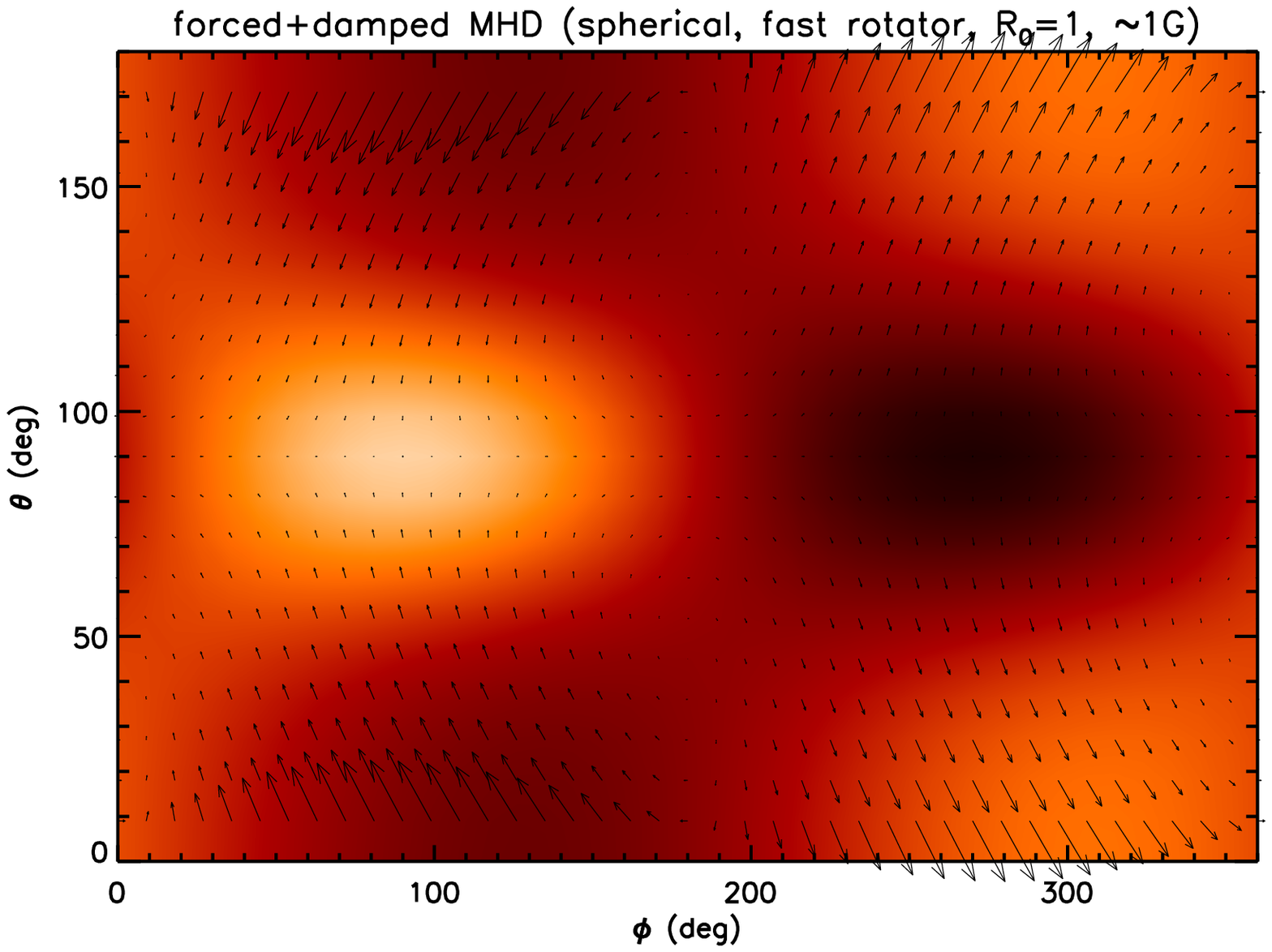}
\includegraphics[width=\columnwidth]{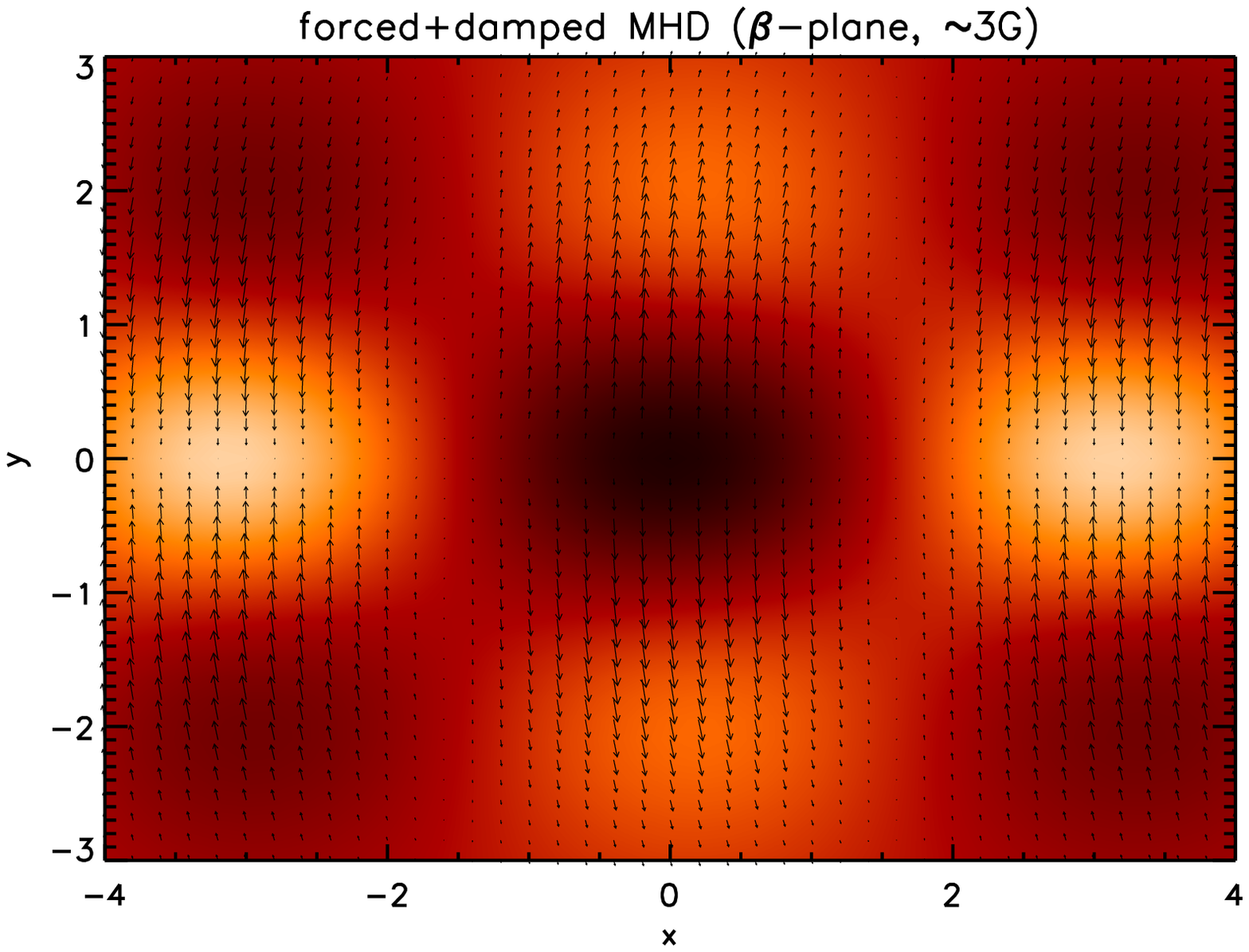}
\includegraphics[width=\columnwidth]{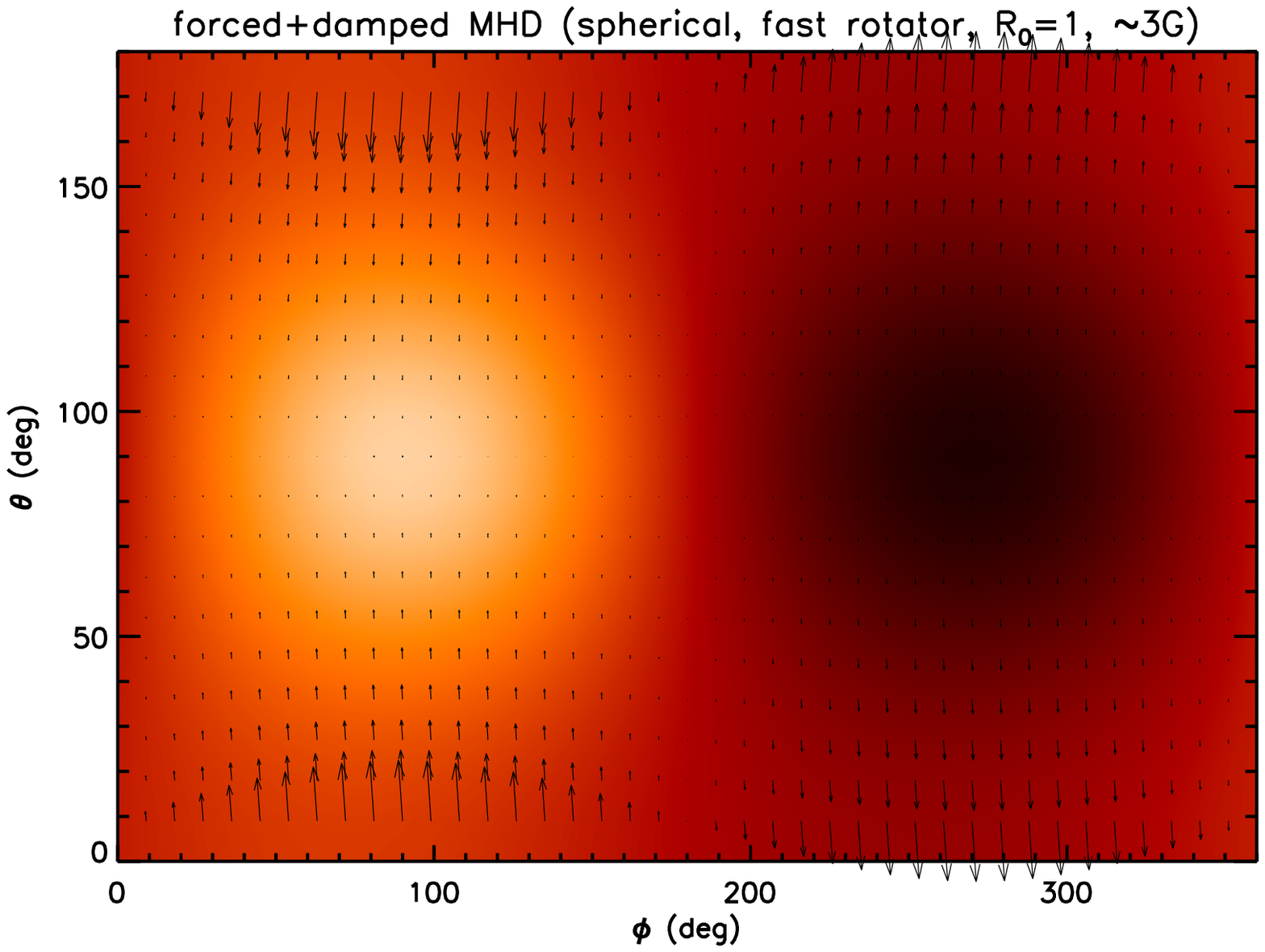}
\includegraphics[width=\columnwidth]{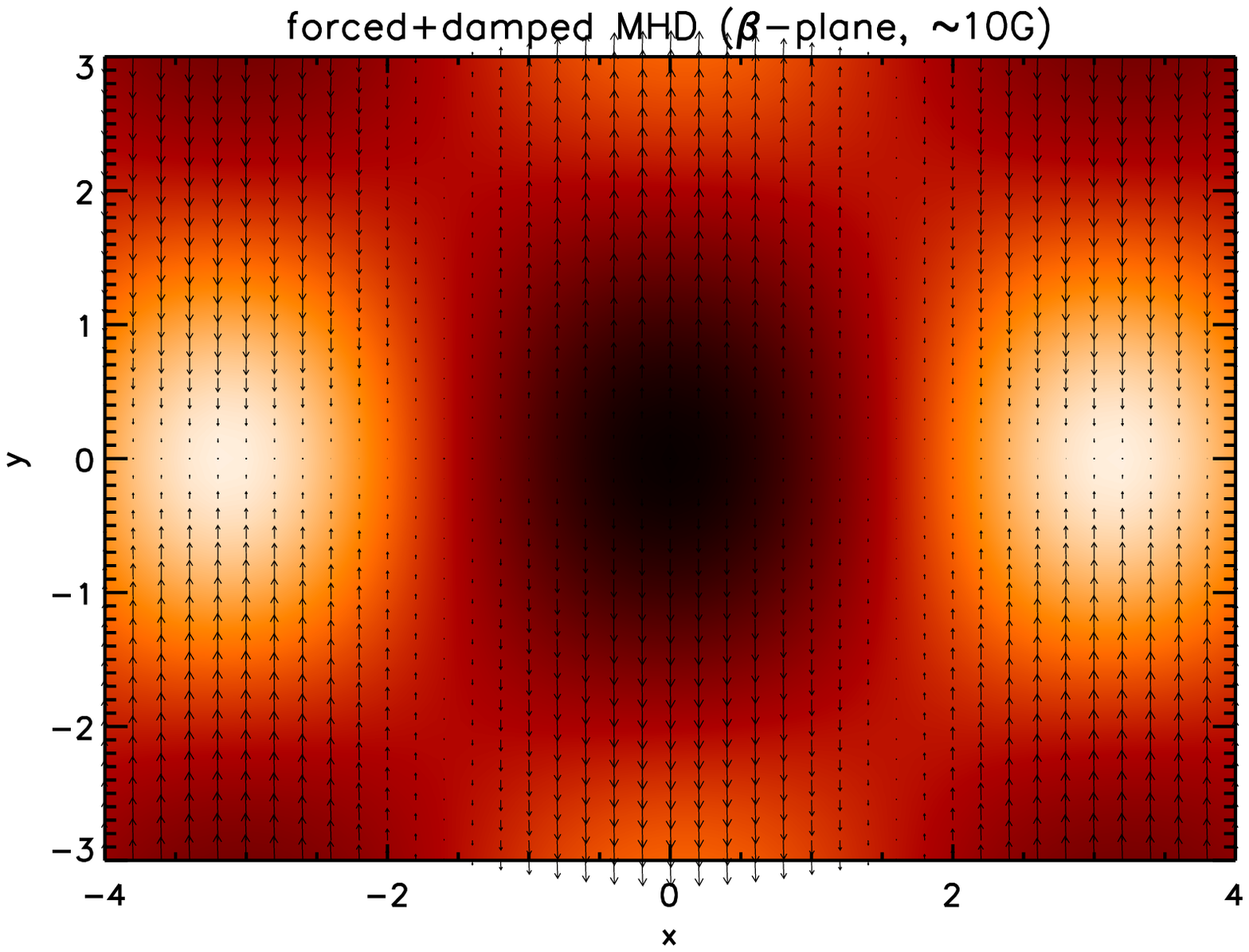}
\includegraphics[width=\columnwidth]{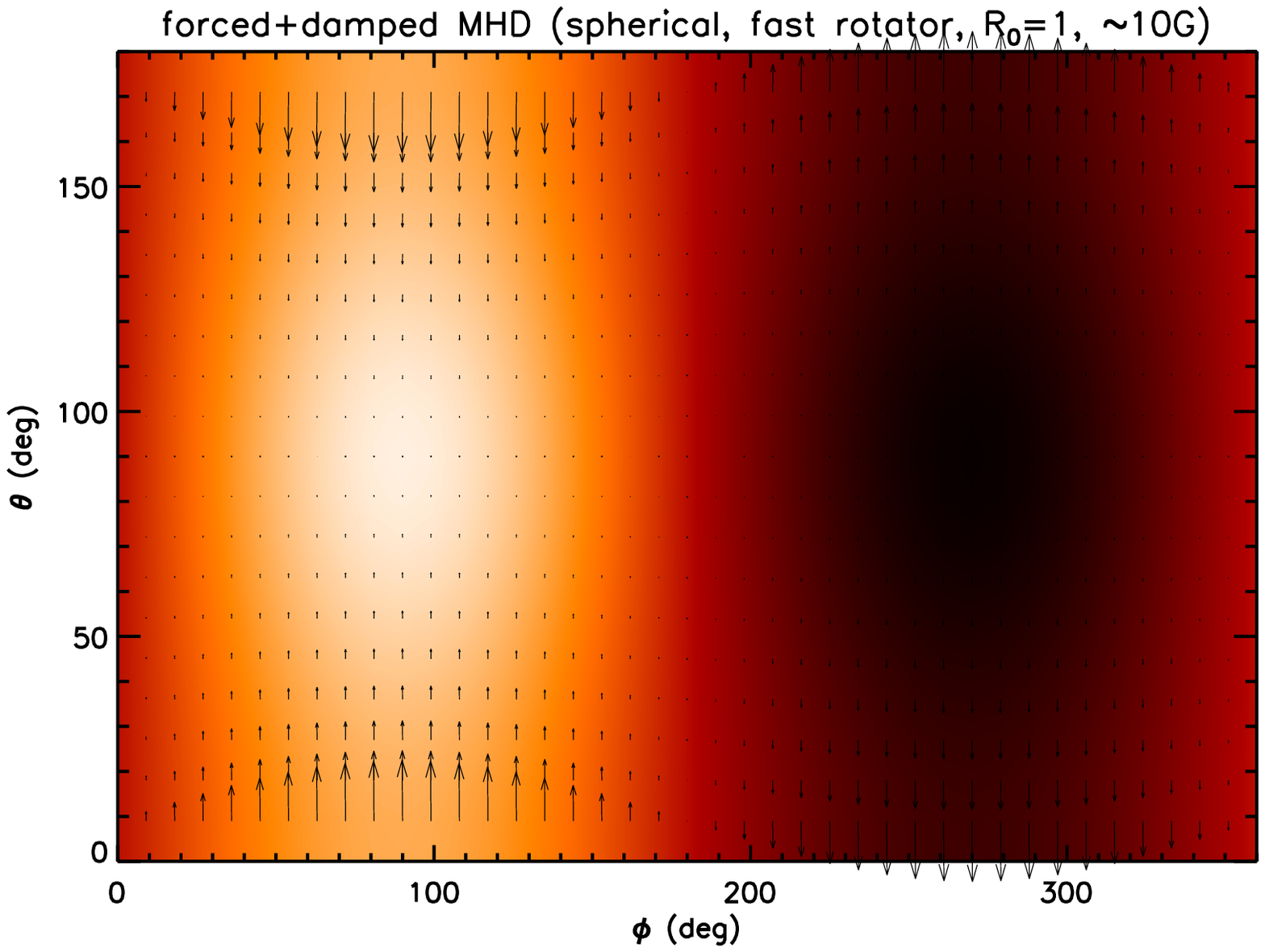}
\end{center}
\vspace{-0.2in}
\caption{Montage of plots of velocity (arrows) and water height perturbations ($\beta$-plane) or $h^\prime_{\rm v}$ (spherical), shown as contours, for steady-state, MHD systems in both the equatorial $\beta$-plane approximation (left column; for $n=k_{\rm x}=1$) and full spherical geometry (right column; for $l=m=1$) with purely vertical/radial background magnetic fields.  We adopt fixed values of the forcing ($\vert F_0 \vert =5$) and hydrodynamic friction ($\omega_\nu=0.1$).  The different panels are for $\bar{B}_{\rm z,r} \sim 1$ G, $\sim 3$ G and $\sim 10$ G.  All quantities are computed in terms of an arbitrary velocity normalization ($v_0$).  Bright and dark colors correspond to positive and negative height perturbations, respectively.}
\vspace{0.1in}
\label{fig:mhd_forced}
\end{figure*}

When forcing and friction (both molecular viscosity and Rayleigh drag) are included, the dimensionless governing equations are
\begin{equation}
\begin{split}
\frac{\partial v_{\rm x}^\prime}{\partial t} &= - \frac{\partial h^\prime}{\partial x} + y v_{\rm y}^\prime + \frac{1}{{\cal R}} \left( \frac{\partial^2 v_{\rm x}^\prime}{\partial x^2} + \frac{\partial^2 v_{\rm x}^\prime}{\partial y^2} \right) - \omega_{\rm drag} v_{\rm x}^\prime, \\
\frac{\partial v_{\rm y}^\prime}{\partial t} &= - \frac{\partial h^\prime}{\partial y} - y v_{\rm x}^\prime + \frac{1}{{\cal R}} \left( \frac{\partial^2 v_{\rm y}^\prime}{\partial x^2} + \frac{\partial^2 v_{\rm y}^\prime}{\partial y^2} \right) - \omega_{\rm drag} v_{\rm y}^\prime, \\
\frac{\partial h^\prime}{\partial t} &= - \frac{\partial v_{\rm x}^\prime}{\partial x} - \frac{\partial v_{\rm y}^\prime}{\partial y} + F_0 h^\prime,
\end{split}
\end{equation}
where $\omega_{\rm drag}$ and $F_0$ have been normalized by $t^{-1}_0$.  The Reynolds number is ${\cal R} = c_0^{3/2}/\nu \beta^{1/2} = c_0 L_0/ \nu$.  As before, we define
\begin{equation}
\omega_\nu \equiv \omega_{\rm drag} + \frac{k^2_{\rm x}}{{\cal R}}.
\end{equation}
We further define the dimensionless quantities,
\begin{equation}
\begin{split}
\omega_0 &\equiv \omega + i \omega_\nu, \\
\omega_{\rm F} &\equiv \omega - i F_0.
\end{split}
\end{equation}

The three expressions for $v_{\rm x_0}$ become
\begin{equation}
\begin{split}
v_{\rm x_0} &= \omega_0^{-1} \left( k_{\rm x} h_0 + i y v_{\rm y_0} + \frac{i}{{\cal R}} \frac{\partial^2 v_{\rm x_0}}{\partial y^2} \right), \\
v_{\rm x_0} &= y^{-1} \left( i \omega_0 v_{\rm y_0} - \frac{\partial h_0}{\partial y} + \frac{1}{{\cal R}} \frac{\partial^2 v_{\rm y_0}}{\partial y^2} \right), \\
v_{\rm x_0} &= k_{\rm x}^{-1} \left( h_0 \omega_{\rm F} + i \frac{\partial v_{\rm y_0}}{\partial y} \right).
\end{split}
\label{eq:beta_hydro}
\end{equation}
The expression for $h_0$ becomes
\begin{equation}
h_0 = i \left( \frac{1}{k_{\rm x}} \frac{\partial v_{\rm y_0}}{\partial y} - \frac{y v_{\rm y_0}}{\omega_0} - \frac{1}{\omega_0 {\cal R}} \frac{\partial^2 v_{\rm x_0}}{\partial y^2}\right) \left( \frac{k_{\rm x}}{\omega_0} - \frac{\omega_{\rm F}}{k_{\rm x}} \right)^{-1}.
\end{equation}
Furthermore, we have
\begin{equation}
\begin{split}
& i v_{\rm y_0} \left( \omega_0 - \frac{y^2}{\omega_0} \right) \left( \frac{k_{\rm x}}{\omega_0} - \frac{\omega_{\rm F}}{k_{\rm x}} \right) - \frac{iy}{\omega_0} \frac{\partial v_{\rm y_0}}{\partial y} + \frac{i y^2 k_{\rm x} v_{\rm y_0}}{\omega^2_0} \\
&+ \frac{1}{{\cal R}} \frac{\partial^2 v_{\rm y_0}}{\partial y^2} \left( \frac{k_{\rm x}}{\omega_0} - \frac{\omega_{\rm F}}{k_{\rm x}} \right) + \frac{i y \omega_{\rm F}}{\omega_0 {\cal R} k_{\rm x}} \frac{\partial^2 v_{\rm x_0}}{\partial y^2} \\
&=\left( \frac{k_{\rm x}}{\omega_0} - \frac{\omega_{\rm F}}{k_{\rm x}} \right) \frac{\partial h_0}{\partial y}.
\end{split}
\end{equation}
Putting it all together, we get
\begin{equation}
\begin{split}
&\frac{\partial^2 v_{\rm y_0}}{\partial y^2} \left[ 1 + \frac{i}{{\cal R}} \left( \frac{k^2_{\rm x}}{\omega_0} - \omega_{\rm F} \right) \right] \\
&+ \left( \omega_0 \omega_{\rm F} - k^2_{\rm x} - \frac{k_{\rm x}}{\omega_0} - \frac{y^2 \omega_{\rm F}}{\omega_0} \right) v_{\rm y_0} \\
&- \frac{1}{\omega_0 {\cal R}} \left( y \omega_{\rm F} \frac{\partial^2 v_{\rm x_0}}{\partial y^2} + k_{\rm x} \frac{\partial^3 v_{\rm x_0}}{\partial y^3} \right) = 0.
\end{split}
\label{eq:parabolic_hydro1}
\end{equation}
We see that the challenging terms are those explicitly involving ${\cal R}$ in equation (\ref{eq:parabolic_hydro1}), which prevent the equation from being cast in the form for a quantum harmonic oscillator.  To proceed analytically, we perform a trick: we take the limit as ${\cal R} \rightarrow \infty$, while allowing it to retain a finite value within $\omega_\nu$.  Physically, this amounts to assuming that molecular viscosity has a scale dependence across longitude, but not across latitude.  The governing equation for $v_{\rm y_0}$ becomes
\begin{equation}
\frac{\partial^2 v_{\rm y_0}}{\partial y^2} + \left( \omega_0 \omega_{\rm F} - k^2_{\rm x} - \frac{k_{\rm x}}{\omega_0} - \frac{y^2 \omega_{\rm F}}{\omega_0} \right) v_{\rm y_0} = 0.
\label{eq:parabolic_hydro2}
\end{equation}
Written in this form, it is easy to see that when forcing and friction are absent, we obtain $\omega = \omega_0 = \omega_{\rm F}$ and we recover equation (\ref{eq:parabolic}).  Since $\omega_0$ and $\omega_{\rm F}$ are both complex in general, it is easy to see that equation (\ref{eq:parabolic_hydro2}) can be cast in terms of a quantum harmonic oscillator equation with complex coefficients,
\begin{equation}
\frac{\partial^2 v_{\rm y_0}}{\partial \tilde{y}^2} + \left[ \left( \omega_0 \omega_{\rm F} - k^2_{\rm x} - \frac{k_{\rm x}}{\omega_0} \right) \left( \frac{\omega_0}{\omega_{\rm F}} \right)^{1/2} - \tilde{y}^2 \right] v_{\rm y_0} = 0,
\end{equation}
where the transformed latitude is
\begin{equation}
\begin{split}
\tilde{y} &\equiv \alpha y, \\
\alpha &\equiv \left( \frac{\omega_{\rm F}}{\omega_0} \right)^{1/4}.
\end{split}
\end{equation}

The dispersion relation follows from considering
\begin{equation}
\left( \omega_0 \omega_{\rm F} - k^2_{\rm x} - \frac{k_{\rm x}}{\omega_0} \right) \left( \frac{\omega_0}{\omega_{\rm F}} \right)^{1/2} = 2n + 1,
\end{equation}
from which two expressions again follow,
\begin{equation}
\begin{split}
&\omega^3_{\rm I} - 3 \omega_{\rm R}^2 \omega_{\rm I} + \left( \omega_{\rm R}^2 - \omega_{\rm I}^2 \right) \left( F_0 - 2 \omega_\nu\right) \\
&+ \omega_{\rm I} \left[ k_{\rm x}^2 + \omega_\nu\left( \omega_\nu - 2 F_0 \right) \right] + \omega_\nu \left( k^2_{\rm x} - \omega_\nu F_0 \right) \\
&+ \left( 2n + 1 \right) \left( \frac{\zeta- \zeta_{\rm R}}{2} \right)^{1/2} = 0, \\
&\omega^3_{\rm R} - 3 \omega_{\rm R} \omega_{\rm I}^2 + 2 \omega_{\rm R} \omega_{\rm I} \left( F_0 - 2 \omega_\nu\right) \\
&+\omega_{\rm R} \left[ \omega_\nu\left( 2 F_0 - \omega_\nu \right) - k_{\rm x}^2 \right] - k_{\rm x} \\
&- \left( 2n + 1 \right) \left( \frac{\zeta + \zeta_{\rm R}}{2} \right)^{1/2} = 0, \\
\end{split}
\label{eq:beta_hydro_forced_damped_dispersions}
\end{equation}
where we have defined the separation functions,
\begin{equation}
\begin{split}
\zeta_{\rm R} &\equiv \omega_{\rm R}^2 - \omega_{\rm I}^2 + \omega_{\rm I} \left( F_0 - \omega_\nu \right) + \omega_\nu F_0, \\
\zeta_{\rm I} &\equiv \omega_{\rm R} \left( 2 \omega_{\rm I} + \omega_\nu - F_0 \right), \\
\zeta &= \left( \zeta_{\rm R}^2 + \zeta_{\rm I}^2 \right)^{1/2}.
\end{split}
\end{equation}
The expressions in (\ref{eq:beta_hydro_forced_damped_dispersions}) make use of De Moivre's formula,
\begin{equation}
\left( \omega_0 \omega_{\rm F} \right)^{1/2} =  \left( \zeta_{\rm R} + i \zeta_{\rm I} \right)^{1/2} = \zeta^{1/2} \left( \cos{\frac{\psi}{2}} + i \sin{\frac{\psi}{2}} \right),
\end{equation}
where the cosine and sine terms may be expressed in terms of the separation functions,
\begin{equation}
\begin{split}
\sin{\frac{\psi}{2}} &= \left( \frac{1- \zeta_{\rm R}/\zeta}{2} \right)^{1/2}, \\
\cos{\frac{\psi}{2}} &= \left( \frac{1 + \zeta_{\rm R}/\zeta}{2} \right)^{1/2}.
\end{split}
\end{equation}
The complex quantity $(\omega_0 \omega_{\rm F})^{1/2}$ is generally double-valued; $\sin(\psi/2)$ and $\cos(\psi/2)$ are mathematically associated with $\pm$ signs.  However, we pick the positive sign on the physical ground that equation (\ref{eq:beta_hydro_forced_damped_dispersions}) needs to reduce to (\ref{eq:beta_hydro_dispersions}) in the limit of a free, hydrodynamic system.

In the limit of ${\cal R} \rightarrow \infty$, the amplitudes of the wave perturbations are
\begin{equation}
\begin{split}
&v_{\rm y_0} = v_0 ~\exp{\left( -\frac{\tilde{y}^2}{2} \right)} ~\tilde{{\cal H}}_n, \\
&h_0 = i v_0 ~\exp{\left( -\frac{\tilde{y}^2}{2} \right)} \left( \frac{k_{\rm x}}{\omega_0} - \frac{\omega_{\rm F}}{k_{\rm x}} \right)^{-1} \\
&\times \left\{ \frac{2n}{k_{\rm x}} \left( \frac{\omega_{\rm F}}{\omega_0} \right)^{1/4} \tilde{{\cal H}}_{n-1} - \tilde{{\cal H}}_n \left[ \frac{\tilde{y}}{k_{\rm x}} \left( \frac{\omega_{\rm F}}{\omega_0} \right)^{1/4} + \frac{y}{\omega_0} \right] \right\}, \\
&v_{\rm x_0} = \omega_0^{-1} \left( k_{\rm x} h_0 + i y v_{\rm y_0} \right),
\end{split}
\end{equation}
where $\tilde{{\cal H}}_{n-1} \equiv {\cal H}_{n-1}(\tilde{y})$ and $\tilde{{\cal H}}_n \equiv {\cal H}_n(\tilde{y})$.  Since $\tilde{y}$ is generally a complex quantity, one cannot easily write down the wave solutions in a closed form.  Instead, one has to multiply these amplitudes by $\Psi$ and take the real parts of the resulting expressions.

It is possible, however, to solve for the steady state of a forced, damped, hydrodynamic atmosphere in a closed form, as already demonstrated by \cite{matsuno66}, \cite{gill80} and \cite{sp11}.  We find that we arrive at equivalent solutions as these previous works only if we allow for $F_0$ to be negative,
\begin{equation}
F_0 = - \left \vert F_0 \right \vert.
\end{equation}
Physically, one expects $F_0 < 0$ because a positive height perturbation then leads to cooling, while a negative one leads to heating, just as one would expect for Newtonian cooling.  In a steady state, such a forcing allows for the key controlling parameter to assume a simple and real form,
\begin{equation}
\alpha \equiv \left( \frac{\left \vert F_0 \right \vert}{\omega_\nu} \right)^{1/4},
\end{equation}
and avoids the mathematical complication of having to evaluate $\tilde{y}$ using De Moivre's formula.  As realized by \cite{sp11}, the quantity $1/\alpha^4$ is the Prandtl number on the $\beta$-plane.  We will see later this is an incomplete description on the sphere as $\alpha$ generally involves the Rossby number as well.  It follows that the steady-state solutions are
\begin{equation}
\begin{split}
&v^\prime_{\rm y} = v_0 ~\exp{\left(-\frac{\alpha^2 y^2}{2}\right)} ~\tilde{{\cal H}}_n \cos{\left( k_{\rm x} x\right)}, \\
&h^\prime = - \frac{v_0 k_{\rm x} \omega_\nu}{k^2_{\rm x} + \omega_\nu \left \vert F_0 \right \vert} ~\exp{\left(-\frac{\alpha^2 y^2}{2}\right)} \\
&\times \left[ \frac{\alpha}{k_{\rm x}} \left( 2 n \tilde{{\cal H}}_{n-1} - \alpha y \tilde{{\cal H}}_n \right) \cos{\left( k_{\rm x} x\right)}  - \frac{y \tilde{{\cal H}}_n}{\omega_\nu} \sin{\left( k_{\rm x} x \right)} \right], \\
&v^\prime_{\rm x} = - \frac{v_0}{k^2_{\rm x} + \omega_\nu \left \vert F_0 \right \vert} ~\exp{\left(-\frac{\alpha^2 y^2}{2}\right)} \\
&\times [ - y \left \vert F_0 \right \vert \tilde{{\cal H}}_n \cos{\left( k_{\rm x} x\right)} \\
&+ k_{\rm x} \alpha \left( 2 n \tilde{{\cal H}}_{n-1} - \alpha y \tilde{{\cal H}}_n \right) \sin{\left( k_{\rm x} x \right)} ]. \\
\end{split}
\label{eq:beta_hydro_forced_steady}
\end{equation}

\subsection{Magnetohydrodynamic (Vertical Background Field)}

MHD shallow water systems on the $\beta$-plane, with a vertical background magnetic field, are generally described by five dimensionless parameters: the reciprocal of the aspect ratio ($\epsilon$), the strength of hydrodynamic versus magnetic wave propagation ($\Gamma$), the forcing ($F_0$), the hydrodynamic friction ($\omega_\nu$) and the magnetic Reynolds number (${\cal R}_{\rm B}$).  However, since $\epsilon$ and $\Gamma$ always appear as products of each other in the dispersion relations and wave solutions, only four parameters are independent.

\subsubsection{Free System, Ideal MHD}
\label{subsect:2d_beta_mhd_free}

Free MHD systems with vertical/radial background fields have previously been considered by \cite{hs09}, both in the equatorial $\beta$-plane and spherical geometries.  We expand upon their analysis and also cast the problem in terms of more intuitive notation.  For a purely vertical background magnetic field, the dimensionless governing equations are
\begin{equation}
\begin{split}
\frac{\partial v_{\rm x}^\prime}{\partial t} &= - \frac{\partial h^\prime}{\partial x} + y v_{\rm y}^\prime - \Gamma b^\prime_{\rm x}, \\
\frac{\partial v_{\rm y}^\prime}{\partial t} &= - \frac{\partial h^\prime}{\partial y} - y v_{\rm x}^\prime - \Gamma b^\prime_{\rm y}, \\
\frac{\partial h^\prime}{\partial t} &= - \frac{\partial v_{\rm x}^\prime}{\partial x} - \frac{\partial v_{\rm y}^\prime}{\partial y}, \\
\frac{\partial b^\prime_{\rm x,y}}{\partial t} &= \epsilon v_{\rm x,y}^\prime.
\end{split}
\end{equation}
The reciprocal of the aspect ratio is
\begin{equation}
\epsilon \equiv \frac{g^{1/4}}{\beta^{1/2} H^{3/4}} = \frac{L_0}{H},
\end{equation}
while the other dimensionless quantity is
\begin{equation}
\Gamma \equiv \frac{\bar{B}^2_{\rm z}}{4 \pi \rho \beta^{1/2} g^{3/4} H^{7/4}} = \epsilon \left( \frac{v_{\rm A}}{c_0} \right)^2,
\end{equation}
such that their product is the square of the ratio of dynamical to Alfv\'{e}n time scales,
\begin{equation}
\epsilon \Gamma = \left( \frac{t_{\rm dyn}}{t_{\rm A}} \right)^2,
\end{equation}
if we write $t_{\rm A} = H/v_{\rm A}$.

Stating the three expressions for $v_{\rm x_0}$ requires defining an additional dimensionless frequency,
\begin{equation}
\omega_{\rm B_0} \equiv \omega - \frac{\epsilon \Gamma}{\omega} = \omega - \frac{1}{\omega} \left( \frac{t_{\rm dyn}}{t_{\rm A}} \right)^2.
\label{eq:omega_b0_free}
\end{equation}
It is important to recognize that $\omega_{\rm B_0}$ is real in the free MHD limit.  The expressions for $v_{\rm x_0}$ are very similar to the basic, hydrodynamic situation,
\begin{equation}
\begin{split}
v_{\rm x_0} &= \omega_{\rm B_0}^{-1} \left( k_{\rm x} h_0 + i y v_{\rm y_0} \right), \\
v_{\rm x_0} &= y^{-1} \left( i \omega_{\rm B_0} v_{\rm y_0} - \frac{\partial h_0}{\partial y} \right), \\
v_{\rm x_0} &= k_{\rm x}^{-1} \left( h_0 \omega + i \frac{\partial v_{\rm y_0}}{\partial y} \right).
\end{split}
\label{eq:beta_mhd_free_3vx}
\end{equation}
The expression for $h_0$ becomes
\begin{equation}
h_0 = i \left( \frac{1}{k_{\rm x}} \frac{\partial v_{\rm y_0}}{\partial y} - \frac{y v_{\rm y_0}}{\omega_{\rm B_0}} \right) \left( \frac{k_{\rm x}}{\omega_{\rm B_0}} - \frac{\omega}{k_{\rm x}} \right)^{-1}.
\end{equation}
The expression for $\partial h_0/\partial y$ becomes
\begin{equation}
\begin{split}
\left( \frac{k_{\rm x}}{\omega_{\rm B_0}} - \frac{\omega}{k_{\rm x}} \right) \frac{\partial h_0}{\partial y} =& i v_{\rm y_0} \left( \omega_{\rm B_0} - \frac{y^2}{\omega_{\rm B_0}} \right) \left( \frac{k_{\rm x}}{\omega_{\rm B_0}} - \frac{\omega}{k_{\rm x}} \right)\\
&- \frac{iy}{\omega_{\rm B_0}} \frac{\partial v_{\rm y_0}}{\partial y} + \frac{i y^2 k_{\rm x} v_{\rm y_0}}{\omega^2_{\rm B_0}}.
\end{split}
\end{equation}
Assembling all of the different parts yields
\begin{equation}
\frac{\partial^2 v_{\rm y_0}}{\partial y^2} + \left( \omega \omega_{\rm B_0} - k^2_{\rm x} - \frac{k_{\rm x}}{\omega_{\rm B_0}} - \frac{y^2 \omega}{\omega_{\rm B_0}} \right) v_{\rm y_0} = 0,
\label{eq:parabolic_mhd1}
\end{equation}
which is identical to equation (\ref{eq:parabolic}) when $\omega=\omega_{\rm B_0}$.  The coefficients in the preceding equation remain real.

As in the hydrodynamic case, we find it instructive to first consider both $\omega_{\rm R}$ and $\omega_{\rm I}$ to be non-vanishing.  From discretizing the following expression,
\begin{equation}
\left( \omega_{\rm B_0} \omega - k^2_{\rm x} - \frac{k_{\rm x}}{\omega_{\rm B_0}} \right) \left( \frac{\omega_{\rm B_0}}{\omega} \right)^{1/2} = 2n + 1,
\label{eq:beta_mhd_vert_discretized}
\end{equation}
we obtain
\begin{equation}
\begin{split}
&\omega_{\rm I} \left[ \zeta_- \left( \zeta_- + 2 \zeta_+ \right) \omega_{\rm R}^2 - \zeta_+^2 \omega_{\rm I}^2 \right] - \zeta_+ k^2_{\rm x} \omega_{\rm I} \\
&- \left( 2n+1 \right) \left( \frac{\zeta - \zeta_{\rm R}}{2} \right)^{1/2} = 0, \\
&\omega_{\rm R} \left[ \zeta_-^2 \omega_{\rm R}^2 - \zeta_+ \left( 2\zeta_- + \zeta_+ \right) \omega_{\rm I}^2 \right] - \zeta_- k^2_{\rm x} \omega_{\rm R} - k \\
&- \left( 2n+1 \right) \left( \frac{\zeta + \zeta_{\rm R}}{2} \right)^{1/2} = 0, \\
\end{split}
\label{eq:beta_mhd_vert_dispersions}
\end{equation}
where we have defined the separation functions,
\begin{equation}
\begin{split}
\zeta_\pm &\equiv 1 \pm \frac{\epsilon \Gamma}{\omega_{\rm R}^2 + \omega_{\rm I}^2}, \\
\zeta_{\rm R} &\equiv \zeta_- \omega_{\rm R}^2 - \zeta_+ \omega_{\rm I}^2, \\
\zeta_{\rm I} &\equiv \left( \zeta_- + \zeta_+ \right) \omega_{\rm R} \omega_{\rm I}, \\
\zeta &= \left( \zeta_{\rm R}^2 + \zeta_{\rm I}^2 \right)^{1/2}.
\end{split}
\end{equation}
For ease of evaluating equation (\ref{eq:beta_mhd_vert_discretized}), we have written
\begin{equation}
\omega_{\rm B_0} = \zeta_- \omega_{\rm R} + i \zeta_+ \omega_{\rm I}.
\end{equation}
The expression $(\omega_{\rm B_0} \omega_{\rm F})^{1/2}$ is again evaluated using De Moivre's formula; the double-valued nature of this quantity is eliminated by ensuring that equation (\ref{eq:beta_hydro_dispersions}) is obtained in the free, hydrodynamic limit ($\zeta_\pm=1$).  In this limit, we previously showed that $\omega_{\rm I}=0$ is a solution of the growth/decay dispersion relation.  We are unable to show that this is generally true for the first expression in equation (\ref{eq:beta_mhd_vert_dispersions}).  Nevertheless, we expect $\omega_{\rm I}=0$ on physical grounds, since we are dealing with a free MHD system, which yields the dispersion relation,
\begin{equation}
\zeta_-^2 \omega_{\rm R}^3 - \zeta_- k^2_{\rm x} \omega_{\rm R} - \zeta_-^{1/2} \left( 2n + 1 \right) \omega_{\rm R} - k_{\rm x} = 0.
\end{equation}

Physically, the quantity $\zeta_-$ controls the effect of ``magnetic pinching" \citep{hs09}.  It is contained within the transformed latitude,
\begin{equation}
\begin{split}
\tilde{y} &\equiv \alpha y, \\
\alpha &\equiv \frac{1}{\zeta^{1/4}_-}.
\end{split}
\end{equation}
Since $1/\zeta_- = \omega_{\rm R}^2/(\omega_{\rm R}^2 - \epsilon \Gamma) \ge 1$, we have $\alpha \ge 1$ and the transformed latitude is always \emph{longer} than the actual latitude (i.e., $\tilde{y} \ge y$) when a magnetic field is present, implying that the term $\exp{(-\tilde{y}^2/2)}$ becomes smaller.  This has the effect that waves will be more concentrated across latitude; it becomes more pronounced as the magnetic field strength increases (larger $\Gamma$).  In terms of the transformed latitude, the equation for $v_{\rm y_0}$ is again that of a quantum harmonic oscillator and yields the following solutions for the wave amplitudes,
\begin{equation}
\begin{split}
v_{\rm y_0} =& v_0 ~\exp{\left(-\frac{\tilde{y}^2}{2}\right)} ~\tilde{{\cal H}}_n, \\
h_0 =& i v_0 ~\exp{\left(-\frac{\tilde{y}^2}{2}\right)} \left( \frac{k_{\rm x}}{\zeta_- \omega_{\rm R}} - \frac{\omega_{\rm R}}{k_{\rm x}} \right)^{-1} \\
&\times \left[ \frac{2n \tilde{{\cal H}}_{n-1}}{\zeta_-^{1/4} k_{\rm x}} - \left( \frac{\tilde{y}}{\zeta_-^{1/4} k_{\rm x}} + \frac{y}{\zeta_- \omega_{\rm R}} \right) \tilde{{\cal H}}_n \right], \\
v_{\rm x_0} =& \frac{i v_0}{k^2_{\rm x} - \zeta_- \omega_{\rm R}^2} ~\exp{\left(-\frac{\tilde{y}^2}{2}\right)} \\
&\times \left[ \frac{2n k_{\rm x} \tilde{{\cal H}}_{n-1}}{\zeta_-^{1/4}} - \left( y \omega_{\rm R} + \frac{ \tilde{y} k_{\rm x} }{\zeta^{1/4}_-} \right) \tilde{{\cal H}}_n \right],
\end{split}
\label{eq:beta_mhd_vert_amp}
\end{equation}
which reduce to equation (\ref{eq:beta_hydro_amp}) in the free, hydrodynamic limit.  Again, it is possible to state the general wave solutions for a free system,
\begin{equation}
\begin{split}
v^\prime_{\rm y} =& v_0 ~\exp{\left(-\frac{\tilde{y}^2}{2}\right)} ~\tilde{{\cal H}}_n \cos{\left( k_{\rm x} x - \omega_{\rm R} t \right)}, \\
h^\prime =& v_0 ~\exp{\left(-\frac{\tilde{y}^2}{2}\right)} \left( \frac{k_{\rm x}}{\zeta_- \omega_{\rm R}} - \frac{\omega_{\rm R}}{k_{\rm x}} \right)^{-1} \sin{\left( k_{\rm x} x - \omega_{\rm R} t \right)} \\
&\times \left[ \left( \frac{\tilde{y}}{\zeta_-^{1/4} k_{\rm x}} + \frac{y}{\zeta_- \omega_{\rm R}} \right) \tilde{{\cal H}}_n - \frac{2n \tilde{{\cal H}}_{n-1}}{\zeta^{1/4}_- k_{\rm x}}\right], \\
v^\prime_{\rm x} =& \frac{v_0}{k^2_{\rm x} - \zeta_- \omega_{\rm R}^2} ~\exp{\left(-\frac{\tilde{y}^2}{2}\right)} \sin{\left( k_{\rm x} x - \omega_{\rm R} t \right)} \\
&\times \left[ \left( y \omega_{\rm R} + \frac{\tilde{y} k_{\rm x} }{\zeta_-^{1/4}} \right) \tilde{{\cal H}}_n - \frac{2n k_{\rm x} \tilde{{\cal H}}_{n-1}}{\zeta_-^{1/4}} \right].
\end{split}
\label{eq:beta_mhd_vert_solutions}
\end{equation}

\subsubsection{Forcing with Hydrodynamic Friction}
\label{subsect:2d_beta_mhd_hydrofric}

Some insight is gained by considering a situation of intermediate complexity, namely that involving forcing and only hydrodynamic friction (molecular viscosity and Rayleigh drag).  Magnetic drag is omitted for now.  The dimensionless governing equations are
\begin{equation}
\begin{split}
\frac{\partial v_{\rm x}^\prime}{\partial t} =& - \frac{\partial h^\prime}{\partial x} + y v_{\rm y}^\prime + \frac{1}{{\cal R}} \left( \frac{\partial^2 v_{\rm x}^\prime}{\partial x^2} + \frac{\partial^2 v_{\rm x}^\prime}{\partial y^2} \right) \\
&- \omega_{\rm drag} v_{\rm x}^\prime - \Gamma b^\prime_{\rm x}, \\
\frac{\partial v_{\rm y}^\prime}{\partial t} =& - \frac{\partial h^\prime}{\partial y} - y v_{\rm x}^\prime + \frac{1}{{\cal R}} \left( \frac{\partial^2 v_{\rm y}^\prime}{\partial x^2} + \frac{\partial^2 v_{\rm y}^\prime}{\partial y^2} \right) \\
&- \omega_{\rm drag} v_{\rm y}^\prime - \Gamma b^\prime_{\rm y}, \\
\frac{\partial h^\prime}{\partial t} =& - \frac{\partial v_{\rm x}^\prime}{\partial x} - \frac{\partial v_{\rm y}^\prime}{\partial y} + F_0 h^\prime,\\
\frac{\partial b^\prime_{\rm x,y}}{\partial t} =& \epsilon v_{\rm x,y}^\prime.
\end{split}
\end{equation}
In the three expressions for $v_{\rm x_0}$,
\begin{equation}
\begin{split}
v_{\rm x_0} &= \omega_{\rm B_0}^{-1} \left( k_{\rm x} h_0 + i y v_{\rm y_0} + \frac{i}{{\cal R}} \frac{\partial^2 v_{\rm x_0}}{\partial y^2} \right), \\
v_{\rm x_0} &= y^{-1} \left( i \omega_{\rm B_0} v_{\rm y_0} - \frac{\partial h_0}{\partial y} + \frac{1}{{\cal R}} \frac{\partial^2 v_{\rm y_0}}{\partial y^2} \right), \\
v_{\rm x_0} &= k_{\rm x}^{-1} \left( h_0 \omega_{\rm F} + i \frac{\partial v_{\rm y_0}}{\partial y} \right),
\end{split}
\end{equation}
the expression for $\omega_{\rm B_0}$ is generalized to
\begin{equation}
\omega_{\rm B_0} \equiv \omega + i \omega_{\rm B},
\label{eq:omega_b0}
\end{equation}
where an additional dimensionless frequency is defined,
\begin{equation}
\omega_{\rm B} \equiv \omega_\nu + \frac{i \epsilon \Gamma}{\omega}.
\end{equation}
One may verify that equation (\ref{eq:omega_b0}) reduces to equation (\ref{eq:omega_b0_free}) in the absence of hydrodynamic friction ($\omega_\nu=0$).

The expression for $h_0$ is structurally identical to the hydrodynamic situation with forcing and friction, except that $\omega_0$ is replaced by $\omega_{\rm B_0}$,
\begin{equation}
h_0 = i \left( \frac{1}{k_{\rm x}} \frac{\partial v_{\rm y_0}}{\partial y} - \frac{y v_{\rm y_0}}{\omega_{\rm B_0}} - \frac{1}{\omega_{\rm B_0} {\cal R}} \frac{\partial^2 v_{\rm x_0}}{\partial y^2}\right) \left( \frac{k_{\rm x}}{\omega_{\rm B_0}} - \frac{\omega_{\rm F}}{k_{\rm x}} \right)^{-1}.
\end{equation}
The same applies for the expression for $\partial h_0/\partial y$,
\begin{equation}
\begin{split}
& i v_{\rm y_0} \left( \omega_{\rm B_0} - \frac{y^2}{\omega_{\rm B_0}} \right) \left( \frac{k_{\rm x}}{\omega_{\rm B_0}} - \frac{\omega_{\rm F}}{k_{\rm x}} \right) - \frac{iy}{\omega_{\rm B_0}} \frac{\partial v_{\rm y_0}}{\partial y} + \frac{i y^2 k_{\rm x} v_{\rm y_0}}{\omega^2_{\rm B_0}} \\
&+ \frac{1}{{\cal R}} \frac{\partial^2 v_{\rm y_0}}{\partial y^2} \left( \frac{k_{\rm x}}{\omega_{\rm B_0}} - \frac{\omega_{\rm F}}{k_{\rm x}} \right) + \frac{i y \omega_{\rm F}}{\omega_{\rm B_0} {\cal R} k_{\rm x}} \frac{\partial^2 v_{\rm x_0}}{\partial y^2} \\
&=\left( \frac{k_{\rm x}}{\omega_{\rm B_0}} - \frac{\omega_{\rm F}}{k_{\rm x}} \right) \frac{\partial h_0}{\partial y}.
\end{split}
\end{equation}

Putting it all together, the general governing equation for $v_{\rm y_0}$ is
\begin{equation}
\begin{split}
&\frac{\partial^2 v_{\rm y_0}}{\partial y^2} \left[ 1 + \frac{i}{{\cal R}} \left( \frac{k^2_{\rm x}}{\omega_{\rm B_0}} - \omega_{\rm F} \right) \right] \\
&+ \left( \omega_{\rm B_0} \omega_{\rm F} - k^2_{\rm x} - \frac{k_{\rm x}}{\omega_{\rm B_0}} - \frac{y^2 \omega_{\rm F}}{\omega_{\rm B_0}} \right) v_{\rm y_0} \\
&- \frac{1}{\omega_{\rm B_0} {\cal R}} \left( y \omega_{\rm F} \frac{\partial^2 v_{\rm x_0}}{\partial y^2} + k_{\rm x} \frac{\partial^3 v_{\rm x_0}}{\partial y^3} \right) = 0.
\end{split}
\label{eq:parabolic_mhd_hydrofric1}
\end{equation}
Again, if we make the approximation that molecular viscosity has a scale dependence across longitude but not latitude, the governing equation becomes amenable to analytical solution,
\begin{equation}
\frac{\partial^2 v_{\rm y_0}}{\partial y^2} + \left( \omega_{\rm B_0} \omega_{\rm F} - k^2_{\rm x} - \frac{k_{\rm x}}{\omega_{\rm B_0}} - \frac{y^2 \omega_{\rm F}}{\omega_{\rm B_0}} \right) v_{\rm y_0} = 0.
\label{eq:parabolic_mhd_hydrofric2}
\end{equation}

The key point is that when a vertical, background magnetic field is added to a system with forcing and hydrodynamic friction, one only needs to replace $\omega_0$ in the equations and solutions by its magnetic counterpart ($\omega_{\rm B_0}$).  Both dimensionless frequencies are generally complex.  This intermediate case also shows the progressive generalization of $\omega_{\rm B}$ and $\omega_{\rm B_0}$.

\subsubsection{Forcing with Friction}
\label{subsect:2d_beta_mhd_fric}

When magnetic drag is added to a system with forcing, hydrodynamic friction and magnetic fields, the dimensionless induction equations become
\begin{equation}
\frac{\partial b^\prime_{\rm x,y}}{\partial t} = \epsilon v^\prime_{\rm x,y} + \frac{1}{{\cal R}_{\rm B}} \left( \frac{\partial^2 b^\prime_{\rm x,y}}{\partial x^2} + \frac{\partial^2 b^\prime_{\rm x,y}}{\partial y^2} \right),
\end{equation}
where the magnetic Reynolds number is defined as
\begin{equation}
{\cal R}_{\rm B} \equiv \frac{c_0^{3/2}}{\beta^{1/2} \eta} = \frac{c_0 L_0}{\eta}.
\end{equation}

A final, additional dimensionless frequency is needed,
\begin{equation}
\omega_\eta \equiv \omega + \frac{i k^2_{\rm x}}{{\cal R}_{\rm B}}.
\end{equation}
The definition for $\omega_{\rm B}$ is generalized,
\begin{equation}
\omega_{\rm B} \equiv \omega_\nu + \frac{i \epsilon \Gamma}{\omega_\eta},
\end{equation}
while still retaining $\omega_{\rm B_0} \equiv \omega + i \omega_{\rm B}$.  The expressions for $v_{\rm x_0}$ pick up extra contributions involving ${\cal R}_{\rm B}$ and $\omega_\eta$,
\begin{equation}
\begin{split}
v_{\rm x_0} &= \omega_{\rm B_0}^{-1} \left( k_{\rm x} h_0 + i y v_{\rm y_0} + \frac{i}{{\cal R}} \frac{\partial^2 v_{\rm x_0}}{\partial y^2} + \frac{\Gamma}{\omega_\eta {\cal R}_{\rm B}} \frac{\partial^2 b^\prime_{\rm x_0}}{\partial y^2} \right), \\
v_{\rm x_0} &= y^{-1} \left( i \omega_{\rm B_0} v_{\rm y_0} - \frac{\partial h_0}{\partial y} + \frac{1}{{\cal R}} \frac{\partial^2 v_{\rm y_0}}{\partial y^2} - \frac{i \Gamma}{\omega_\eta {\cal R}_{\rm B}} \frac{\partial^2 b^\prime_{\rm y_0}}{\partial y^2} \right), \\
v_{\rm x_0} &= k_{\rm x}^{-1} \left( h_0 \omega_{\rm F} + i \frac{\partial v_{\rm y_0}}{\partial y} \right).
\end{split}
\end{equation}
The expressions for $h_0$,
\begin{equation}
\begin{split}
h_0 =& i \left( \frac{1}{k_{\rm x}} \frac{\partial v_{\rm y_0}}{\partial y} - \frac{y v_{\rm y_0}}{\omega_{\rm B_0}} - \frac{1}{\omega_{\rm B_0} {\cal R}} \frac{\partial^2 v_{\rm x_0}}{\partial y^2} + \frac{i \Gamma}{\omega_{\rm B_0} \omega_\eta {\cal R}_{\rm B}} \frac{\partial^2 b^\prime_{\rm x_0}}{\partial y^2} \right) \\
&\times \left( \frac{k_{\rm x}}{\omega_{\rm B_0}} - \frac{\omega_{\rm F}}{k_{\rm x}} \right)^{-1},
\end{split}
\end{equation}
and its derivative,
\begin{equation}
\begin{split}
&i v_{\rm y_0} \left( \omega_{\rm B_0} - \frac{y^2}{\omega_{\rm B_0}} \right) \left( \frac{k_{\rm x}}{\omega_{\rm B_0}} - \frac{\omega_{\rm F}}{k_{\rm x}} \right) - \frac{iy}{\omega_{\rm B_0}} \frac{\partial v_{\rm y_0}}{\partial y} \\
&+ \frac{i y^2 k_{\rm x} v_{\rm y_0}}{\omega^2_{\rm B_0}} + \frac{1}{{\cal R}} \frac{\partial^2 v_{\rm y_0}}{\partial y^2} \left( \frac{k_{\rm x}}{\omega_{\rm B_0}} - \frac{\omega_{\rm F}}{k_{\rm x}} \right) \\
&+ \frac{i y \omega_{\rm F}}{\omega_{\rm B_0} {\cal R} k_{\rm x}} \frac{\partial^2 v_{\rm x_0}}{\partial y^2} + \frac{\Gamma k_{\rm x} y}{\omega^2_{\rm B_0} \omega_\eta {\cal R}_{\rm B}} \frac{\partial^2 b^\prime_{\rm x_0}}{\partial y^2} \\
&- \frac{\Gamma}{\omega_\eta {\cal R}_{\rm B}} \left( \frac{k_{\rm x}}{\omega_{\rm B_0}} - \frac{\omega_{\rm F}}{k_{\rm x}} \right) \left( \frac{y}{\omega_{\rm B_0}} \frac{\partial^2 b^\prime_{\rm x_0}}{\partial y^2} + i \frac{\partial^2 b^\prime_{\rm y_0}}{\partial y^2} \right) \\
&= \left( \frac{k_{\rm x}}{\omega_{\rm B_0}} - \frac{\omega_{\rm F}}{k_{\rm x}} \right) \frac{\partial h_0}{\partial y},
\end{split}
\end{equation}
also pick up extra contributions.

In the $\beta$-plane approximation, the most general governing equation for $v_{\rm y_0}$ is
\begin{equation}
\begin{split}
&\frac{\partial^2 v_{\rm y_0}}{\partial y^2} \left[ 1 + \frac{i}{{\cal R}} \left( \frac{k^2_{\rm x}}{\omega_{\rm B_0}} - \omega_{\rm F} \right) \right] \\
&+ \left( \omega_{\rm B_0} \omega_{\rm F} - k^2_{\rm x} - \frac{k_{\rm x}}{\omega_{\rm B_0}} - \frac{y^2 \omega_{\rm F}}{\omega_{\rm B_0}} \right) v_{\rm y_0} \\
&- \frac{1}{\omega_{\rm B_0} {\cal R}} \left( y \omega_{\rm F} \frac{\partial^2 v_{\rm x_0}}{\partial y^2} + k_{\rm x} \frac{\partial^3 v_{\rm x_0}}{\partial y^3} \right) \\
&-\frac{\Gamma}{\omega_\eta {\cal R}_{\rm B}} \left( \frac{k_{\rm x}^2}{\omega_{\rm B_0}} - \omega_{\rm F} \right) \left( \frac{i y}{\omega_{\rm B_0}} \frac{\partial^2 b^\prime_{\rm x_0}}{\partial y^2} - \frac{\partial^2 b^\prime_{\rm y_0}}{\partial y^2} \right) \\
&+ \frac{i \Gamma k_{\rm x}}{\omega_{\rm B_0} \omega_\eta {\cal R}_{\rm B}} \left( \frac{k_{\rm x} y}{\omega_{\rm B_0}} \frac{\partial^2 b^\prime_{\rm x_0}}{\partial y^2} + \frac{\partial^3 b^\prime_{\rm x_0}}{\partial y^3} \right) = 0.
\end{split}
\label{eq:parabolic_mhd_fric1}
\end{equation}
To proceed analytically, both molecular viscosity and magnetic drag are assumed to have scale dependences only across longitude.  Mathematically, we set ${\cal R}, {\cal R}_{\rm B} \rightarrow \infty$ wherever they appear explicitly in equation (\ref{eq:parabolic_mhd_fric1}), while allowing them to retain finite values within $\omega_\nu$ and $\omega_\eta$.  In this limit, one ends up with equation (\ref{eq:parabolic_mhd_hydrofric2}), except with a more general definition of $\omega_{\rm B_0}$. 

The dispersion relations are obtained from discretizing the following expression,
\begin{equation}
\left( \omega_{\rm B_0} \omega_{\rm F} - k^2_{\rm x} - \frac{k_{\rm x}}{\omega_{\rm B_0}} \right) \left( \frac{\omega_{\rm B_0}}{\omega_{\rm F}} \right)^{1/2} = 2n + 1.
\end{equation}
We again find the algebra to be more tractable if we write
\begin{equation}
\begin{split}
\left( \omega_{\rm B_0} \omega_{\rm F} \right)^{1/2} &= \left( \zeta_{\rm R} + i \zeta_{\rm I} \right)^{1/2} \\
&= \left( \frac{\zeta + \zeta_{\rm R}}{2} \right)^{1/2} + i \left( \frac{\zeta - \zeta_{\rm R}}{2} \right)^{1/2},
\end{split}
\end{equation}
and 
\begin{equation}
\omega_{\rm B_0} = \zeta_- \omega_{\rm R} + i \left( \zeta_0 + \zeta_+ \omega_{\rm I} \right),
\end{equation}
where we have defined the separation functions,
\begin{equation}
\begin{split}
\zeta_\pm &\equiv 1 \pm \frac{\epsilon \Gamma}{\omega_{\rm R}^2 + \left( k^2_{\rm x}/{\cal R}_{\rm B} + \omega_{\rm I}\right)^2}, \\
\zeta_0 &\equiv \omega_\nu + \frac{\epsilon \Gamma k^2_{\rm x}}{{\cal R}_{\rm B} \left[ \omega_{\rm R}^2 + \left( k^2_{\rm x}/{\cal R}_{\rm B} + \omega_{\rm I}\right)^2 \right]}, \\
\zeta_{\rm R} &\equiv \zeta_- \omega_{\rm R}^2 - \left( \omega_{\rm I} - F_0 \right) \left( \zeta_0 + \zeta_+ \omega_{\rm I} \right), \\
\zeta_{\rm I} &\equiv \omega_{\rm R} \left[ \zeta_0 + \zeta_+ \omega_{\rm I} + \zeta_- \left( \omega_{\rm I} - F_0 \right) \right], \\
\zeta &= \left( \zeta_{\rm R}^2 + \zeta_{\rm I}^2 \right)^{1/2}.
\end{split}
\end{equation}
It follows that
\begin{equation}
\begin{split}
&\zeta^2_- \omega_{\rm R}^3 - \omega_{\rm R} \left( \zeta_0 + \zeta_+ \omega_{\rm I} \right)^2 - 2 \zeta_- \omega_{\rm R} \left( \omega_{\rm I} - F_0 \right) \left( \zeta_0 + \zeta_+ \omega_{\rm I} \right) \\
&- k^2_{\rm x} \zeta_- \omega_{\rm R} - k_{\rm x} - \left( 2n + 1 \right) \left( \frac{\zeta + \zeta_{\rm R}}{2} \right)^{1/2} = 0, \\
&2 \zeta_- \left( \zeta_0 + \zeta_+ \omega_{\rm I} \right) \omega_{\rm R}^2 + \zeta_-^2 \omega_{\rm R}^2 \left( \omega_{\rm I} - F_0 \right) \\
&- \left( \omega_{\rm I} - F_0 \right) \left( \zeta_0 + \zeta_+ \omega_{\rm I} \right)^2 - k^2_{\rm x} \left( \zeta_0 + \zeta_+ \omega_{\rm I} \right) \\
&- \left( 2n + 1 \right) \left( \frac{\zeta - \zeta_{\rm R}}{2} \right)^{1/2} = 0.
\end{split}
\end{equation}
In the preceding expressions, we have again used De Moivre's formula and picked the positive root, such that the dispersion relations reduce to equation (\ref{eq:beta_hydro_dispersions}) in the free, hydrodynamic limit.

In the limit of ${\cal R}, {\cal R}_{\rm B} \rightarrow \infty$, the amplitudes of the wave perturbations are
\begin{equation}
\begin{split}
&v_{\rm y_0} = v_0 ~\exp{\left( -\frac{\tilde{y}^2}{2} \right)} ~\tilde{{\cal H}}_n, \\
&h_0 = i v_0 ~\exp{\left( -\frac{\tilde{y}^2}{2} \right)} \left( \frac{k_{\rm x}}{\omega_{\rm B_0}} - \frac{\omega_{\rm F}}{k_{\rm x}} \right)^{-1} \\
&\times \left\{ \frac{2n}{k_{\rm x}} \left( \frac{\omega_{\rm F}}{\omega_{\rm B_0}} \right)^{1/4} \tilde{{\cal H}}_{n-1} - \tilde{{\cal H}}_n \left[ \frac{\tilde{y}}{k_{\rm x}} \left( \frac{\omega_{\rm F}}{\omega_{\rm B_0}} \right)^{1/4} + \frac{y}{\omega_{\rm B_0}} \right] \right\}, \\
&v_{\rm x_0} = \omega_{\rm B_0}^{-1} \left( k_{\rm x} h_0 + i y v_{\rm y_0} \right),
\end{split}
\label{eq:beta_mhd_forced_amp}
\end{equation}
where we have
\begin{equation}
\begin{split}
\tilde{y} &\equiv \alpha y, \\
\alpha &\equiv \left( \frac{\omega_{\rm F}}{\omega_{\rm B_0}} \right)^{1/4}.
\end{split}
\end{equation}
Again, it is easier to write down the steady-state solutions (with $F_0 = -\vert F_0 \vert$),
\begin{equation}
\begin{split}
&v^\prime_{\rm y} = v_0 ~\exp{\left(-\frac{\alpha^2 y^2}{2}\right)} ~\tilde{{\cal H}}_n \cos{\left( k_{\rm x} x\right)}, \\
&h^\prime = - \frac{v_0 k_{\rm x} \zeta_0}{k^2_{\rm x} + \zeta_0  \left \vert F_0 \right \vert} ~\exp{\left(-\frac{\alpha^2 y^2}{2}\right)} \\
&\times \left[ \frac{\alpha}{k_{\rm x}} \left( 2 n \tilde{{\cal H}}_{n-1} - \alpha y \tilde{{\cal H}}_n \right) \cos{\left( k_{\rm x} x\right)}  - \frac{y \tilde{{\cal H}}_n}{\zeta_0} \sin{\left( k_{\rm x} x \right)} \right], \\
&v^\prime_{\rm x} = - \frac{v_0}{k^2_{\rm x} + \zeta_0  \left \vert F_0 \right \vert} ~\exp{\left(-\frac{\alpha^2 y^2}{2}\right)} \\
&\times [ - y  \left \vert F_0 \right \vert \tilde{{\cal H}}_n \cos{\left( k_{\rm x} x\right)} \\
&+ k_{\rm x} \alpha \left( 2 n \tilde{{\cal H}}_{n-1} - \alpha y \tilde{{\cal H}}_n \right) \sin{\left( k_{\rm x} x \right)} ], \\
\end{split}
\label{eq:beta_mhd_forced_steady}
\end{equation}
where in the steady-state limit we have
\begin{equation}
\begin{split}
\alpha &= \left( \frac{ \left \vert F_0 \right \vert }{\zeta_0} \right)^{1/4}, \\
\zeta_0 &\equiv \omega_{\rm drag} + \frac{k_{\rm x}^2}{{\cal R}} + \frac{\epsilon \Gamma {\cal R}_{\rm B}}{k^2_{\rm x}}.\\
\end{split}
\end{equation}
The ``generalized friction" $\zeta_0$ contains all of the physical effects which act like sources of friction, including magnetic tension.  It has the property that molecular viscosity acts predominantly on small scales, while magnetic tension and magnetic drag act collectively and preferentially on large scales.

As is evident, the parameters $\epsilon$ and $\Gamma$ are degenerate and appear only as products of each other.  The purely hydrodynamic case is recovered when $\epsilon \Gamma = 0$.  Magnetic drag vanishes in the limit of ${\cal R}_{\rm B} \rightarrow \infty$.

\subsection{Magnetohydrodynamic (Horizontal Background Field)}

\subsubsection{Free System, Ideal MHD}
\label{subsect:2d_beta_mhd_free_hori}

For a purely horizontal background magnetic field, the dimensionless governing equations are
\begin{equation}
\begin{split}
\frac{\partial v_{\rm x}^\prime}{\partial t} &= - \frac{\partial h^\prime}{\partial x} + y v_{\rm y}^\prime + \Gamma_{\rm x} \frac{\partial b^\prime_{\rm x}}{\partial x} + \left( \Gamma_{\rm x} \Gamma_{\rm y} \right)^{1/2} \frac{\partial b^\prime_{\rm x}}{\partial y}, \\
\frac{\partial v_{\rm y}^\prime}{\partial t} &= - \frac{\partial h^\prime}{\partial y} - y v_{\rm x}^\prime + \left( \Gamma_{\rm x} \Gamma_{\rm y} \right)^{1/2} \frac{\partial b^\prime_{\rm y}}{\partial x} + \Gamma_{\rm y} \frac{\partial b^\prime_{\rm y}}{\partial y}, \\
\frac{\partial h^\prime}{\partial t} &= - \frac{\partial v_{\rm x}^\prime}{\partial x} - \frac{\partial v_{\rm y}^\prime}{\partial y}, \\
\frac{\partial b^\prime_{\rm x}}{\partial t} &= \frac{\partial v^\prime_{\rm x}}{\partial x} + \epsilon \frac{\partial v_{\rm x}^\prime}{\partial y}, \\
\frac{\partial b^\prime_{\rm y}}{\partial t} &= \frac{1}{\epsilon} \frac{\partial v^\prime_{\rm y}}{\partial x} + \frac{\partial v_{\rm y}^\prime}{\partial y},
\end{split}
\end{equation}
where the definitions for $\epsilon$ and $\Gamma$ (now separated into two components) have changed.  The former is now the ratio of magnetic field strengths,
\begin{equation}
\epsilon \equiv \frac{\bar{B}_{\rm y}}{\bar{B}_{\rm x}},
\end{equation}
while the latter is the ratio of Alfv\'{e}n to dynamical velocities,
\begin{equation}
\Gamma_{\rm x,y} \equiv \left(\frac{v_{\rm A_{x,y}}}{c_0}\right)^2 = \left(\frac{t_{\rm dyn}}{t_{\rm A}}\right)^2,
\end{equation}
where $v_{\rm A_{x,y}} \equiv \bar{B}_{\rm x,y}/2\sqrt{\pi \rho}$.

The three expressions for the velocity amplitude in the $x$-direction are
\begin{equation}
\begin{split}
v_{\rm x_0} =& \omega_{\rm B_0}^{-1} \left( k_{\rm x} h_0 + i y v_{\rm y_0} \right) \\
&- \frac{i k_{\rm x}}{\omega \omega_{\rm B_0}} \left[ \epsilon \Gamma_{\rm x} + \left( \Gamma_{\rm x} \Gamma_{\rm y} \right)^{1/2} \right] \frac{\partial v_{\rm x_0}}{\partial y} \\
&- \frac{\epsilon \left( \Gamma_{\rm x} \Gamma_{\rm y} \right)^{1/2}}{\omega \omega_{\rm B_0}} \frac{\partial^2 v_{\rm x_0}}{\partial y^2}, \\
v_{\rm x_0} =& \frac{i \omega v_{\rm y_0}}{y} \left[ 1 - \frac{k^2_{\rm x} \left( \Gamma_{\rm x} \Gamma_{\rm y} \right)^{1/2}}{\epsilon \omega^2} \right] - \frac{1}{y} \frac{\partial h_0}{\partial y} \\
&- \frac{k_{\rm x}}{\omega y} \left[ \left( \Gamma_{\rm x} \Gamma_{\rm y} \right)^{1/2} + \frac{\Gamma_{\rm y}}{\epsilon} \right] \frac{\partial v_{\rm y_0}}{\partial y} + \frac{i \Gamma_{\rm y}}{\omega y} \frac{\partial^2 v_{\rm y_0}}{\partial y^2}, \\
v_{\rm x_0} =& k_{\rm x}^{-1} \left( h_0 \omega + i \frac{\partial v_{\rm y_0}}{\partial y} \right),
\end{split}
\end{equation}
where the definition of $\omega_{\rm B_0}$ has changed,
\begin{equation}
\omega_{\rm B_0} \equiv \omega - \frac{k_{\rm x}^2 \Gamma_{\rm x}}{\omega}.
\end{equation}
It is apparent that the preceding set of equations cannot be solved in the usual way.  However, if we demand that the poloidal background field vanishes ($\bar{B}_{\rm y}=0$) and do not consider any perturbation of the poloidal magnetic field ($b^\prime_{\rm y}=0$),\footnote{This step needs to be done right from the beginning, before performing non-dimensionalization.} the algebra becomes tractable, since we end up with
\begin{equation}
\begin{split}
v_{\rm x_0} =& \omega_{\rm B_0}^{-1} \left( k_{\rm x} h_0 + i y v_{\rm y_0} \right), \\
v_{\rm x_0} =& y^{-1} \left( i \omega v_{\rm y_0} - \frac{\partial h_0}{\partial y} \right), \\
v_{\rm x_0} =& k_{\rm x}^{-1} \left( h_0 \omega + i \frac{\partial v_{\rm y_0}}{\partial y} \right).
\end{split}
\end{equation}
Notice that these expressions are almost identical to the free MHD case with a vertical background field (equation [\ref{eq:beta_mhd_free_3vx}]), except that the second expression has $\omega$ instead of $\omega_{\rm B_0}$.  Employing the usual mathematical machinery, we obtain
\begin{equation}
\frac{\partial^2 v_{\rm y_0}}{\partial \tilde{y}^2} + \left[ \left( \omega^2 - \frac{k^2_{\rm x} \omega}{\omega_{\rm B_0}} - \frac{k_{\rm x}}{\omega_{\rm B_0}} \right) \left( \frac{\omega_{\rm B_0}}{\omega} \right)^{1/2} - \tilde{y}^2 \right] v_{\rm y_0} = 0,
\end{equation}
where we have defined
\begin{equation}
\begin{split}
\tilde{y} &\equiv \alpha y, \\
\alpha &\equiv \left( \frac{\omega}{\omega_{\rm B_0}} \right)^{1/4}.
\end{split}
\end{equation}

To obtain the dispersion relations, we again find it convenient to write $\omega_{\rm B_0} = \zeta_- \omega_{\rm R} + i \zeta_+ \omega_{\rm I}$, from which it follows that
\begin{equation}
\begin{split}
&\omega_{\rm R} \left( \zeta_- \omega_{\rm R}^2 - \zeta_+ \omega_{\rm I}^2 \right) - \omega_{\rm I}^2 \omega_{\rm R} \left( \zeta_- + \zeta_+ \right) \\
&-k_{\rm x}^2 \omega_{\rm R} - k_{\rm x} - \left( 2n + 1 \right) \left( \frac{\zeta + \zeta_{\rm R}}{2} \right)^{1/2} = 0, \\
&\omega_{\rm R}^2 \omega_{\rm I} \left( \zeta_- + \zeta_+ \right) + \omega_{\rm I} \left( \zeta_- \omega_{\rm R}^2 - \zeta_+ \omega_{\rm I}^2 \right) \\
&- k^2_{\rm x} \omega_{\rm I} - \left( 2n + 1 \right) \left( \frac{\zeta - \zeta_{\rm R}}{2} \right)^{1/2} = 0, \\
\end{split}
\end{equation}
where we have defined
\begin{equation}
\begin{split}
\zeta_\pm &\equiv 1 \pm \frac{k^2_{\rm x} \Gamma_{\rm x}}{\omega_{\rm R}^2 + \omega_{\rm I}^2}, \\
\zeta_{\rm R} &\equiv \zeta_- \omega_{\rm R}^2 - \zeta_+ \omega_{\rm I}^2, \\
\zeta_{\rm I} &\equiv \left( \zeta_- + \zeta_+ \right) \omega_{\rm R} \omega_{\rm I}, \\
\zeta &= \left( \zeta_{\rm R}^2 + \zeta_{\rm I}^2 \right)^{1/2}.
\end{split}
\end{equation}
It turns out that the wave amplitudes and solutions are identical to those previously stated in (\ref{eq:beta_mhd_vert_amp}) and (\ref{eq:beta_mhd_vert_solutions}), respectively, except with different definitions of $\omega_{\rm B_0}$ and $\zeta_\pm$.

\subsubsection{Forcing with Friction}
\label{subsect:2d_beta_mhd_fric_horizontal}

When all sources of friction are added, the dimensionless governing equations are
\begin{equation}
\begin{split}
\frac{\partial v_{\rm x}^\prime}{\partial t} =& - \frac{\partial h^\prime}{\partial x} + y v_{\rm y}^\prime + \Gamma_{\rm x} \frac{\partial b^\prime_{\rm x}}{\partial x} + \left( \Gamma_{\rm x} \Gamma_{\rm y} \right)^{1/2} \frac{\partial b^\prime_{\rm x}}{\partial y} \\
&+\frac{1}{{\cal R}} \left( \frac{\partial^2 v_{\rm x}^\prime}{\partial x^2} + \frac{\partial^2 v_{\rm x}^\prime}{\partial y^2} \right) - \omega_{\rm drag} v_{\rm x}^\prime, \\
\frac{\partial v_{\rm y}^\prime}{\partial t} =& - \frac{\partial h^\prime}{\partial y} - y v_{\rm x}^\prime + \left( \Gamma_{\rm x} \Gamma_{\rm y} \right)^{1/2} \frac{\partial b^\prime_{\rm y}}{\partial x} + \Gamma_{\rm y} \frac{\partial b^\prime_{\rm y}}{\partial y} \\
&+\frac{1}{{\cal R}} \left( \frac{\partial^2 v_{\rm y}^\prime}{\partial x^2} + \frac{\partial^2 v_{\rm y}^\prime}{\partial y^2} \right) - \omega_{\rm drag} v_{\rm y}^\prime, \\
\frac{\partial h^\prime}{\partial t} &= - \frac{\partial v_{\rm x}^\prime}{\partial x} - \frac{\partial v_{\rm y}^\prime}{\partial y} + F_0 h^\prime, \\
\frac{\partial b^\prime_{\rm x}}{\partial t} &= \frac{\partial v^\prime_{\rm x}}{\partial x} + \epsilon \frac{\partial v_{\rm x}^\prime}{\partial y} + \frac{1}{{\cal R}_{\rm B}} \left( \frac{\partial^2 b^\prime_{\rm x}}{\partial x^2} + \frac{\partial^2 b^\prime_{\rm x}}{\partial y^2} \right), \\
\frac{\partial b^\prime_{\rm y}}{\partial t} &= \frac{1}{\epsilon} \frac{\partial v^\prime_{\rm y}}{\partial x} + \frac{\partial v_{\rm y}^\prime}{\partial y} + \frac{1}{{\cal R}_{\rm B}} \left( \frac{\partial^2 b^\prime_{\rm y}}{\partial x^2} + \frac{\partial^2 b^\prime_{\rm y}}{\partial y^2} \right),
\end{split}
\end{equation}
from which it follows that
\begin{equation}
\begin{split}
v_{\rm x_0} =& \omega_{\rm B_0}^{-1} \left( k_{\rm x} h_0 + i y v_{\rm y_0} \right) \\
&- \frac{i k_{\rm x}}{\omega_{\rm B_0} \omega_\eta} \left[ \epsilon \Gamma_{\rm x} + \left( \Gamma_{\rm x} \Gamma_{\rm y} \right)^{1/2} \right] \frac{\partial v_{\rm x_0}}{\partial y} \\
&+ \frac{1}{\omega_{\rm B_0}} \left[ \frac{i}{{\cal R}} - \frac{\epsilon \left( \Gamma_{\rm x} \Gamma_{\rm y} \right)^{1/2}}{\omega_
\eta} \right] \frac{\partial^2 v_{\rm x_0}}{\partial y^2} \\
&- \frac{1}{\omega_{\rm B_0} \omega_\eta {\cal R}_{\rm B}} \left[ i k_{\rm x} \Gamma_{\rm x} \frac{\partial^2 b_{\rm x_0}}{\partial y^2} + \left( \Gamma_{\rm x} \Gamma_{\rm y} \right)^{1/2} \frac{\partial^3 b_{\rm x_0}}{\partial y^3} \right], \\
v_{\rm x_0} =& \frac{i \omega_0 v_{\rm y_0}}{y} - \frac{i k^2_{\rm x} \left( \Gamma_{\rm x} \Gamma_{\rm y} \right)^{1/2} v_{\rm y_0}}{\epsilon \omega_\eta y} - \frac{1}{y} \frac{\partial h_0}{\partial y} \\
&- \frac{k_{\rm x}}{\omega_\eta y} \left[ \left( \Gamma_{\rm x} \Gamma_{\rm y} \right)^{1/2} + \frac{\Gamma_{\rm y}}{\epsilon} \right] \frac{\partial v_{\rm y_0}}{\partial y} \\
&+ \frac{1}{y} \left( \frac{1}{{\cal R}} + \frac{i \Gamma_{\rm y}}{\omega_\eta} \right) \frac{\partial^2 v_{\rm y_0}}{\partial y^2} \\
&- \frac{1}{\omega_\eta {\cal R}_{\rm B} y} \left[ k_{\rm x} \left( \Gamma_{\rm x} \Gamma_{\rm y} \right)^{1/2} \frac{\partial^2 b_{\rm y_0}}{\partial y^2} - i \Gamma_{\rm y} \frac{\partial^3 b_{\rm y_0}}{\partial y^3} \right], \\
v_{\rm x_0} =& k_{\rm x}^{-1} \left( h_0 \omega_{\rm F} + i \frac{\partial v_{\rm y_0}}{\partial y} \right).
\end{split}
\end{equation}
Again, to make the algebra tractable, we assume $\bar{B}_{\rm y} = b^\prime_{\rm y} = 0$ and take the ${\cal R}, {\cal R}_{\rm B} \rightarrow \infty$ limit, which yields
\begin{equation}
\begin{split}
v_{\rm x_0} =& \omega_{\rm B_0}^{-1} \left( k_{\rm x} h_0 + i y v_{\rm y_0} \right), \\
v_{\rm x_0} =& y^{-1} \left( i \omega_0 v_{\rm y_0} - \frac{\partial h_0}{\partial y} \right), \\
v_{\rm x_0} =& k_{\rm x}^{-1} \left( h_0 \omega_{\rm F} + i \frac{\partial v_{\rm y_0}}{\partial y} \right),
\end{split}
\end{equation}
where the definitions for some of the dimensionless frequencies have changed,
\begin{equation}
\begin{split}
\omega_\eta &\equiv \omega + \frac{i k^2_{\rm x}}{{\cal R}_{\rm B}}, \\
\omega_{\rm B} &\equiv \omega_\nu + \frac{i k^2_{\rm x} \Gamma_{\rm x}}{\omega_\eta},
\end{split}
\end{equation}
whilst retaining $\omega_\nu \equiv \omega_{\rm drag} + k^2_{\rm x}/{\cal R}$, $\omega_0 \equiv \omega + i \omega_\nu$ and $\omega_{\rm B_0} \equiv \omega + i \omega_{\rm B}$ as in the case of a vertical background magnetic field.  Performing the same mathematical procedure, we obtain
\begin{equation}
\frac{\partial^2 v_{\rm y_0}}{\partial \tilde{y}^2} + \left[ \left( \omega_{\rm F} \omega_0 - \frac{k^2_{\rm x} \omega_0}{\omega_{\rm B_0}} - \frac{k_{\rm x}}{\omega_{\rm B_0}} \right) \left( \frac{\omega_{\rm B_0}}{\omega_{\rm F}} \right)^{1/2} - \tilde{y}^2 \right] v_{\rm y_0} = 0,
\end{equation}
where we have $\tilde{y} \equiv \alpha y$ and $\alpha \equiv ( \omega_{\rm F} / \omega_{\rm B_0} )^{1/4}$.  

In deriving the dispersion relations, we again find it useful to first write $\omega_{\rm B_0} = \zeta_- \omega_{\rm R} + i (\zeta_0 + \zeta_+ \omega_{\rm I} )$ and evaluate $(\omega_{\rm B_0} \omega_{\rm F})^{1/2} = (\zeta_{\rm R} + i \zeta_{\rm I})^{1/2}$ using De Moivre's formula.  It follows that
\begin{equation}
\begin{split}
&\zeta_- \omega_{\rm R} \left( \omega_{\rm R}^2 - \omega_{\rm I}^2 \right) - 2 \omega_{\rm I} \omega_{\rm R} \left( \zeta_0 + \zeta_+ \omega_{\rm I} \right) \\
&- \zeta_- \left( \omega_\nu - F_0 \right) \omega_{\rm I} \omega_{\rm R} - \omega_{\rm R} \left( \omega_\nu - F_0 \right) \left( \zeta_0 + \zeta_+ \omega_{\rm I} \right) \\
&+ F_0 \omega_\nu \zeta_- \omega_{\rm R} - k^2_{\rm x} \omega_{\rm R} - k_{\rm x} - \left( 2n + 1 \right) \left( \frac{\zeta + \zeta_{\rm R}}{2} \right)^{1/2} = 0, \\
&2 \zeta_- \omega_{\rm I} \omega_{\rm R}^2 + \left( \zeta_0 + \zeta_+ \omega_{\rm I} \right) \left( \omega_{\rm R}^2 - \omega_{\rm I}^2 \right) + \zeta_- \left( \omega_\nu - F_0 \right) \omega_{\rm R}^2 \\
&- \left( \omega_\nu - F_0 \right) \left( \zeta_0 + \zeta_+ \omega_{\rm I} \right) \omega_{\rm I} + \left( \zeta_0 + \zeta_+ \omega_{\rm I} \right) F_0 \omega_\nu \\
&-k^2_{\rm x} \left( \omega_\nu + \omega_{\rm I} \right) - \left( 2n + 1 \right) \left( \frac{\zeta - \zeta_{\rm R}}{2} \right)^{1/2} = 0, \\
\end{split}
\end{equation}
where we have defined the separation functions,
\begin{equation}
\begin{split}
\zeta_\pm &\equiv 1 \pm \frac{\Gamma_{\rm x} k_{\rm x}^2}{\omega_{\rm R}^2 + \left( k^2_{\rm x}/{\cal R}_{\rm B} + \omega_{\rm I}\right)^2}, \\
\zeta_0 &\equiv \omega_\nu + \frac{\Gamma_{\rm x} k^4_{\rm x}}{{\cal R}_{\rm B} \left[ \omega_{\rm R}^2 + \left( k^2_{\rm x}/{\cal R}_{\rm B} + \omega_{\rm I}\right)^2 \right]}, \\
\zeta_{\rm R} &\equiv \zeta_- \omega_{\rm R}^2 - \left( \omega_{\rm I} - F_0 \right) \left( \zeta_0 + \zeta_+ \omega_{\rm I} \right), \\
\zeta_{\rm I} &\equiv \omega_{\rm R} \left[ \zeta_0 + \zeta_+ \omega_{\rm I} + \zeta_- \left( \omega_{\rm I} - F_0 \right) \right], \\
\zeta &= \left( \zeta_{\rm R}^2 + \zeta_{\rm I}^2 \right)^{1/2}.
\end{split}
\end{equation}

The wave amplitudes and steady-state solutions are identical to the vertical-field situation, as given in equations (\ref{eq:beta_mhd_forced_amp}) and (\ref{eq:beta_mhd_forced_steady}) respectively, except that the definition of $\zeta_0$ has changed,
\begin{equation}
\zeta_0 \equiv \omega_{\rm drag} + \frac{k^2_{\rm x}}{{\cal R}} + \Gamma_{\rm x} {\cal R}_{\rm B}.
\end{equation}
We retain $\alpha = ( \vert F_0 \vert / \zeta_0 )^{1/4}$.  The major difference is that magnetic tension and magnetic drag operate equally on all scales in a collective manner.

\section{2D Models (Spherical)}
\label{sect:spherical}

Generally, a shallow water model on a sphere may be described by five parameters: the forcing ($F_0$), the hydrodynamic friction ($\omega_\nu$), the magnetic Reynolds number (${\cal R}_{\rm B}$), the Rossby number (${\cal R}_0$) and the ratio of dynamical to Alfv\'{e}n timescales ($t_{\rm dyn}/t_{\rm A}$).  One also needs to specify the characteristic length scales of interest via the zonal ($m$) and meridional ($l$) wavenumbers, analogous to the quantum numbers for the quantum harmonic oscillator.

\subsection{Hydrodynamic}

\begin{figure*}
\begin{center}
\vspace{-0.2in}
\includegraphics[width=\columnwidth]{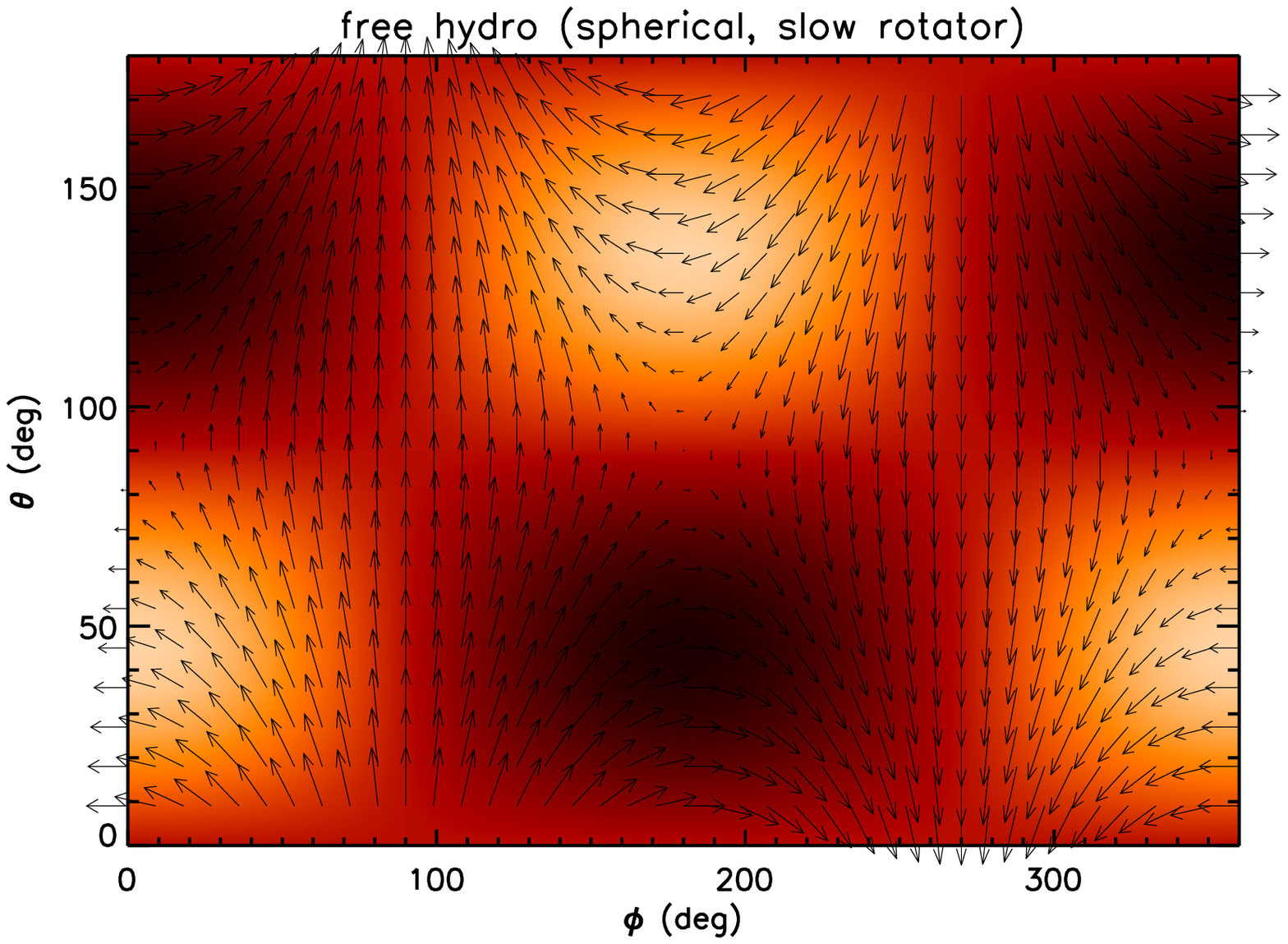}
\includegraphics[width=\columnwidth]{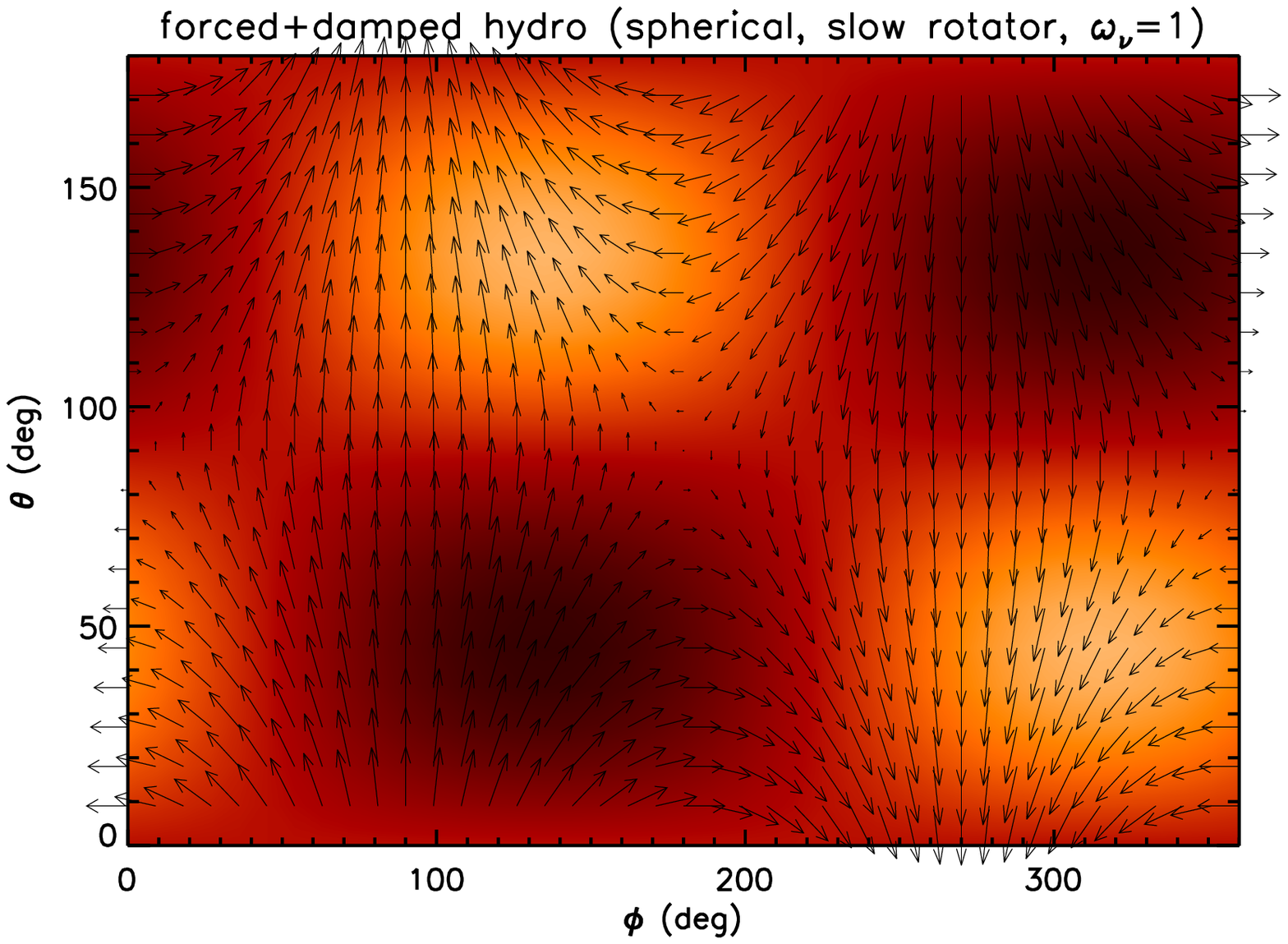}
\includegraphics[width=\columnwidth]{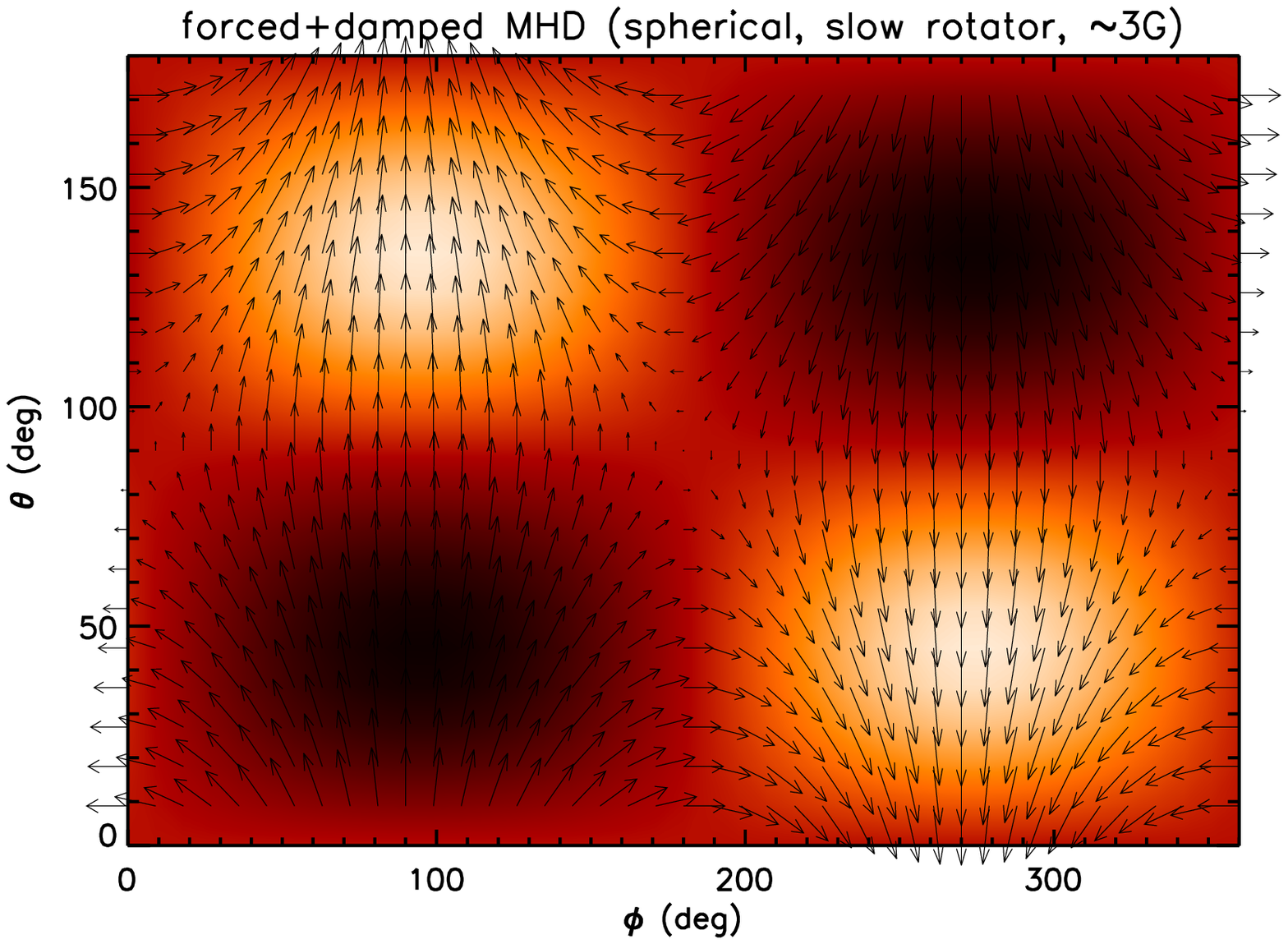}
\includegraphics[width=\columnwidth]{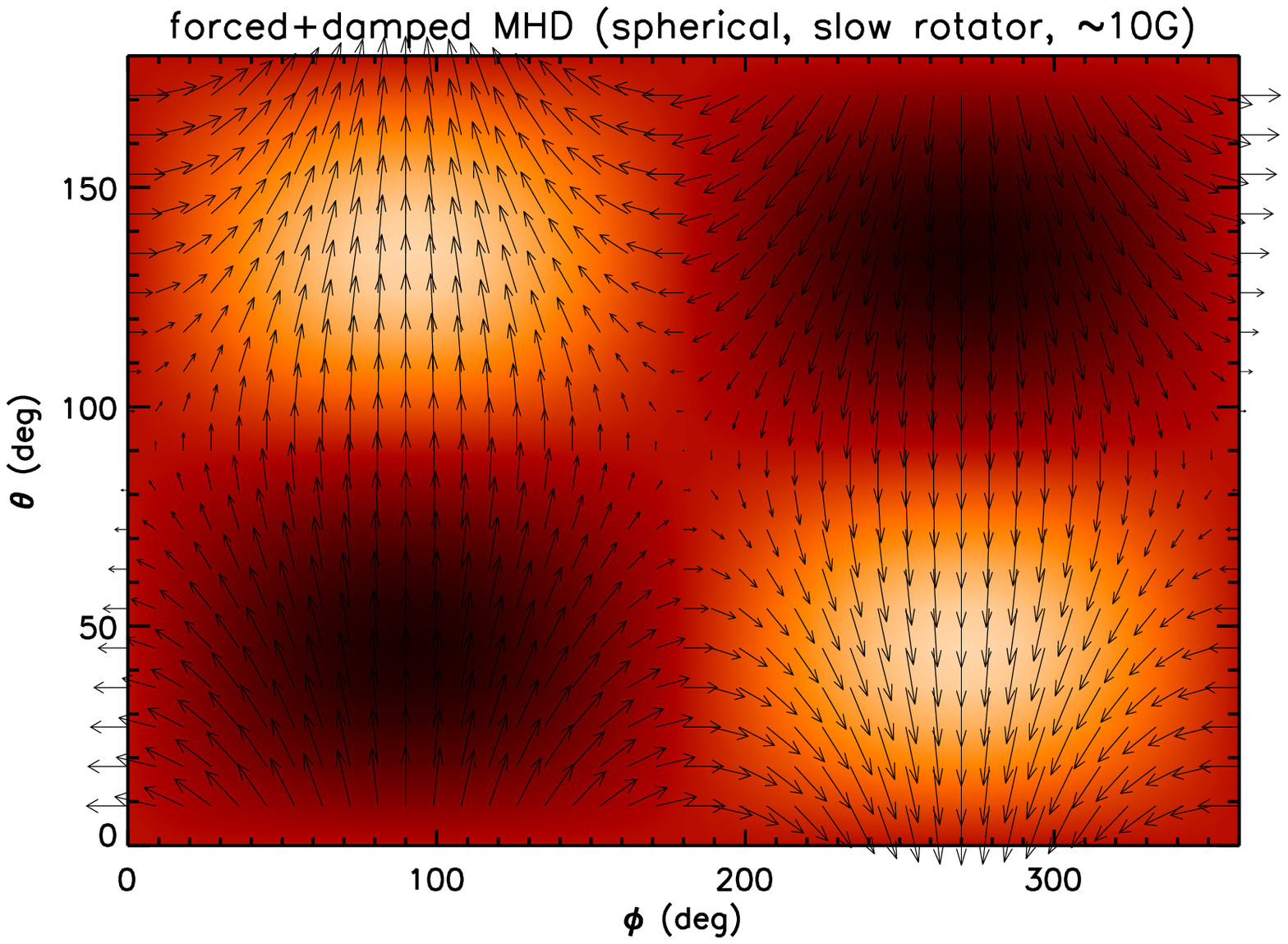}
\end{center}
\vspace{-0.2in}
\caption{Montage of plots of velocity perturbations (arrows) and $h^\prime_{\rm v}$ (contours) for steady-state, hydrodynamic and MHD systems in full spherical geometry for $l=m=1$.  In the slowly-rotating limit, forcing has no effect on the atmospheric structure, both for hydrodynamic and magnetized systems.  All quantities are computed in terms of an arbitrary velocity normalization ($v_0$).  Bright and dark colors correspond to positive and negative height perturbations, respectively.}
\vspace{0.1in}
\label{fig:sphere_slow}
\end{figure*}

\subsubsection{Free System}
\label{subsect:spherical_hydro_free}

We revisit the classical work of \cite{lh68}, who studied free, hydrodynamic shallow water systems on a sphere, as a basis for generalizing to forced, damped, magnetized systems in spherical geometry.  We note that \cite{hs09} has previously rederived the work of \cite{lh68} in a condensed form, but we still find it useful to introduce self-consistent notation and also reconcile some differences between the present work and that of \cite{lh68}.

In the free, hydrodynamic limit, the governing equations are
\begin{equation}
\begin{split}
&\frac{\partial v^\prime_\theta}{\partial t} = -\frac{g}{R} \frac{\partial h^\prime}{\partial \theta} + 2 \Omega v^\prime_\phi \cos\theta, \\
&\frac{\partial v^\prime_\phi}{\partial t} = -\frac{g}{R \sin\theta} \frac{\partial h^\prime}{\partial \phi} - 2 \Omega v^\prime_\theta \cos\theta, \\
&\frac{\partial h^\prime}{\partial t} + \frac{H}{R \sin\theta} \left[ \frac{\partial}{\partial \theta} \left( v^\prime_\theta \sin\theta \right) + \frac{\partial v_\phi^\prime}{\partial \phi} \right] = 0.
\end{split}
\end{equation}
These equations differ from those of \cite{lh68}, presented in equations (2.1)--(2.3) of that paper, for a simple reason: \cite{lh68} calls his polar coordinate the ``co-latitude"; if we denote this by $\theta_{\rm LH}$, then it is related to our co-latitude by $\theta = \pi - \theta_{\rm LH}$.  If we assume a unit sphere ($R=1$) and substitute $\sin\theta = \sin\theta_{\rm LH}$, $\cos\theta = -\cos\theta_{\rm LH}$ and $\partial/\partial \theta = -\partial/\partial \theta_{\rm LH}$ into the preceding equations, we (nearly) recover equations (2.1)--(2.3) of \cite{lh68}.  We note a typographical error in equation (2.2) of that work, where the $2 \Omega v^\prime_\phi \sin\theta_{\rm LH}$ term should instead be $2 \Omega v^\prime_\phi \cos\theta_{\rm LH}$, otherwise one does not recover equation (7.2) of the same work when seeking wave solutions.\footnote{An easier way to see this is to note that ``cosines stay cosines and sines stay sines" in the $\theta \rightarrow \theta_{\rm LH}$ transformation.  This rule of thumb applies to the other four sinusoidal terms in equations (2.1)--(2.3) of \cite{lh68}, so it should naturally also apply to the fifth.}  This error does not propagate into the rest of the results in \cite{lh68}.\footnote{One may already infer this by examining equation (2.4) of \cite{lh68}.}

To proceed, we first need to recast the equations using the following transformations \citep{margules,lh68}:
\begin{equation}
\begin{split}
v^{\prime\prime}_{\theta,\phi} &\equiv v^\prime_{\theta,\phi} \sin\theta, \\
h^\prime_{\rm v} &\equiv \frac{g h^\prime}{2 \Omega R}, \\
\mu &\equiv \cos\theta, \\
\hat{D} &\equiv -\sin\theta \frac{\partial}{\partial \theta} = \left( 1 - \mu^2 \right) \frac{\partial}{\partial \mu}, \\
t_0 &\equiv 2 \Omega t. \\
\end{split}
\end{equation}
Several aspects of these transformations are worth emphasizing.  Unlike for the $\beta$-plane models, the velocities retain their dimensional form.  The quantity $h^\prime_{\rm v}$ has the physical units of velocity, rather than length.  In defining the dimensionless time ($t_0$), we automatically allow for the wave frequency ($\omega$) to be cast in dimensionless units.  By taking the characteristic length scale to be the radius of the exoplanet, we also define the Rossby number,
\begin{equation}
{\cal R}_0 \equiv \frac{c_0}{2 \Omega R},
\end{equation}
and write Lamb's parameter as \citep{lh68}
\begin{equation}
\xi \equiv \frac{1}{{\cal R}_0^2}.
\end{equation}

With these transformations, the governing equations become
\begin{equation}
\begin{split}
&-i \frac{\partial}{\partial t_0} \left( i v^{\prime\prime}_\theta \right) - \mu v^{\prime\prime}_\phi - \hat{D}h^\prime_{\rm v} = 0, \\
&i \mu \left( i v^{\prime\prime}_\theta \right) - \frac{\partial v^{\prime\prime}_\phi}{\partial t_0} - \frac{\partial h^\prime_{\rm v}}{\partial \phi} = 0, \\
&\left( 1 - \mu^2 \right) \frac{\partial h^\prime_{\rm v}}{\partial t_0} + {\cal R}_0^2 \left[ i \hat{D}\left(i v^{\prime\prime}_\theta \right) + \frac{\partial v_\phi^{\prime\prime}}{\partial \phi} \right] = 0.
\end{split}
\end{equation}
This somewhat peculiar way of writing the governing equations comes from the desire to seek solutions for the following quantities,
\begin{equation}
\begin{split}
i v^{\prime\prime}_\theta &= v_{\theta_0} \Psi, \\
v^{\prime\prime}_\phi &= v_{\phi_0} \Psi, \\
h^\prime_{\rm v} &= h_{\rm v_0} \Psi, \\
\end{split}
\end{equation}
where $\Psi \equiv \exp{[i(m\phi - \omega t_0)]}$ and $m$ is the zonal wavenumber.

From seeking wave solutions, one obtains the equations for the wave amplitudes,
\begin{equation}
\begin{split}
&\omega v_{\theta_0} + \mu v_{\phi_0} + \hat{D} h_{\rm v_0} = 0, \\
&\mu v_{\theta_0} + \omega v_{\phi_0} - m h_{\rm v_0} = 0, \\
&\hat{D} v_{\theta_0} + m v_{\phi_0} - \omega \xi \left( 1 -\mu^2 \right) h_{\rm v_0} = 0,
\end{split}
\label{eq:spherical_hydro_amp}
\end{equation}
where we have used Lamb's parameter, instead of the Rossby number, in order to write the preceding expressions in a more compact form.  These expressions differ from those in equation (7.2) of \cite{lh68} due to the $\theta \rightarrow \theta_{\rm LH}$ transformation previously described.

We next employ a series of mathematical steps first described in \cite{lh68}.  From the second equation in (\ref{eq:spherical_hydro_amp}), we obtain
\begin{equation}
v_{\phi_0} = \omega^{-1} \left( m h_{\rm v_0} - \mu v_{\theta_0} \right).
\label{eq:lh_trick0}
\end{equation}
Substituting this expression into the first equation in (\ref{eq:spherical_hydro_amp}) yields
\begin{equation}
\left( \omega \hat{D} + \mu m \right) h_{\rm v_0} = \left( \mu^2 - \omega^2 \right) v_{\theta_0}.
\label{eq:lh_trick1}
\end{equation}
Substituting the same expression into the third equation in (\ref{eq:spherical_hydro_amp}) gives
\begin{equation}
\left( \omega \hat{D} - \mu m \right) v_{\theta_0} = \left[ \omega^2 \xi \left( 1 - \mu^2 \right) - m^2 \right] h_{\rm v_0}.
\label{eq:lh_trick2}
\end{equation}
Combining equations (\ref{eq:lh_trick1}) and (\ref{eq:lh_trick2}) to eliminate $h_{\rm v_0}$ yields
\begin{equation}
\left( \omega \hat{D} + \mu m \right) \left[ \frac{\left( \omega \hat{D} - \mu m\right) v_{\theta_0}}{\omega^2 \xi \left( 1 - \mu^2 \right) - m^2} \right] + \left( \omega^2 - \mu^2 \right) v_{\theta_0} = 0,
\label{eq:spherical_hydro_provisional}
\end{equation}
the provisional governing equation for $v_{\theta_0}$.  Equation (\ref{eq:spherical_hydro_provisional}) is in agreement with equation (7.5) of \cite{lh68} because equations (\ref{eq:lh_trick1}) and (\ref{eq:lh_trick2}) each possess a sign flip resulting from the $\theta \rightarrow \theta_{\rm LH}$ transformation, rendering the provisional governing equation invariant to it.

To proceed, we need to generalize an identity described in \cite{hs09}.  For an arbitrary function ${\cal G}$, one may show that 
\begin{equation}
\begin{split}
&\left( \omega_1 \hat{D} + \mu m \right) \left[ {\cal G} \left( \omega_2 \hat{D} - \mu m \right) \right] v_{\theta_0} \\
&= {\cal G} \left( \omega_1 \hat{D} + \mu m \right) \left( \omega_2 \hat{D} - \mu m \right) v_{\theta_0} \\
&+ \omega_1 \left( \hat{D} {\cal G} \right) \left( \omega_2 \hat{D} - \mu m \right) v_{\theta_0},
\end{split}
\label{eq:identity}
\end{equation}
where $\omega_1$ and $\omega_2$ are arbitrary constants.  In the free hydrodynamic limit, we use the identity with $\omega_1 = \omega_2 = \omega$.  In our case, we have
\begin{equation}
{\cal G} \equiv \frac{1}{\omega^2 \xi \left( 1 - \mu^2 \right) - m^2},
\end{equation}
whence using the identity in equation (\ref{eq:spherical_hydro_provisional}) yields
\begin{equation}
\begin{split}
&{\cal G} \left( \omega \hat{D} + \mu m \right) \left( \omega \hat{D} - \mu m \right) v_{\theta_0} + \omega \left( \hat{D} {\cal G} \right) \left( \omega \hat{D} - \mu m \right) v_{\theta_0} \\
&+ \left( \omega^2 - \mu^2 \right) v_{\theta_0} = 0.
\end{split}
\end{equation}
It is useful to note that
\begin{equation}
\hat{D} {\cal G} = \frac{2 \omega^2 \xi \mu \left( 1 - \mu^2 \right) {\cal G}}{\omega^2 \xi \left( 1 - \mu^2 \right) - m^2} .
\end{equation}
Following through on the algebra, one obtains the general governing equation for the meridional velocity amplitude,
\begin{equation}
\begin{split}
&\frac{\partial}{\partial \mu} \left[ \left( 1 - \mu^2 \right) \frac{\partial}{\partial \mu} \right] v_{\theta_0} - \frac{m v_{\theta_0}}{\omega} - \frac{m^2 v_{\theta_0}}{1 - \mu^2} \\
&+ \xi \left( \omega^2 - \mu^2 \right) v_{\theta_0}  + \frac{2 \omega \xi \mu}{\omega^2 \xi \left( 1- \mu^2 \right) - m^2} \left( \omega \hat{D} - \mu m \right) v_{\theta_0} \\
&= 0,
\end{split}
\label{eq:sphere_hydro_general_governing}
\end{equation}
in agreement with equation (7.8) of \cite{lh68} and equation (B3) of \cite{hs09}.  This general equation is amenable to analytical solution only in the limits of slow or fast rotation.

In the limit of slow rotation ($\xi \rightarrow 0$), the governing equation for $v_{\theta_0}$ reduces to the associated Legendre equation \citep{abram,aw95},
\begin{equation}
\left( 1 - \mu^2 \right) \frac{\partial^2 v_{\theta_0}}{\partial \mu^2} - 2\mu \frac{\partial v_{\theta_0}}{\partial \mu} - m \left[ \frac{1}{\omega} + \frac{m}{\left( 1 - \mu^2 \right)} \right] v_{\theta_0} = 0,
\label{eq:legendre_hydro_free}
\end{equation}
if the following quantity is discretized in terms of an integer $l$,
\begin{equation}
-\frac{m}{\omega} = l \left( l + 1 \right).
\end{equation}
Here, $l$ is the meridional wavenumber.  The dispersion relations follow immediately,
\begin{equation}
\begin{split}
\omega_{\rm R} &= - \frac{m}{l \left( l + 1 \right)}, \\
\omega_{\rm I} &= 0,
\end{split}
\end{equation}
and they describe the slow, undamped Rossby waves in the system, as expected.\footnote{In Appendix B of \cite{hs09}, it was erroneously stated that eastward-propagating Poincar\'{e} waves exist exclusively in slowly-rotating shallow water systems.}

The solution to equation (\ref{eq:legendre_hydro_free}) is
\begin{equation}
v_{\theta_0} = v_0 {\cal P}^m_l,
\end{equation}
where ${\cal P}^m_l$ is the associated Legendre function.  We then use equation (\ref{eq:lh_trick2}) and the following recurrence relation for associated Legendre functions,
\begin{equation}
\hat{D} {\cal P}^m_l = \left( l+1 \right) \mu {\cal P}^m_l - \left( l - m + 1 \right) {\cal P}^m_{l+1},
\end{equation}
from which we obtain
\begin{equation}
h_{\rm v_0} = \frac{v_0}{m^2} \left\{ \omega_{\rm R} \left( l - m + 1 \right) {\cal P}^m_{l+1} - \mu \left[ \omega_{\rm R} \left( l + 1 \right) - m \right] {\cal P}^m_l \right\}.
\end{equation}
Finally, by using equation (\ref{eq:lh_trick0}), we obtain
\begin{equation}
v_{\phi_0} = \frac{v_0}{m} \left[ \left( l - m + 1 \right) {\cal P}^m_{l+1} - \mu \left( l + 1 \right) {\cal P}^m_l \right].
\end{equation}
It is worth noting that, short of a normalization factor, ${\cal P}^m_l \exp{(im\phi)}$ are spherical harmonics.

The wave solutions are
\begin{equation}
\begin{split}
v^\prime_\theta =& \frac{v_0 {\cal P}^m_l}{\sin\theta} \sin{\left( m \phi - \omega_{\rm R} t_0\right)}, \\
h^\prime_{\rm v} =& \frac{v_0}{m^2} \left\{ \omega_{\rm R} \left( l - m + 1 \right) {\cal P}^m_{l+1} - \mu \left[ \omega_{\rm R} \left( l + 1 \right) - m \right] {\cal P}^m_l \right\} \\
&\times \cos{\left( m \phi - \omega_{\rm R} t_0 \right)}, \\
v^\prime_\phi =& \frac{v_0}{m \sin\theta} \left[ \left( l - m + 1 \right) {\cal P}^m_{l+1} - \mu \left( l + 1 \right) {\cal P}^m_l \right] \\
&\times \cos{\left( m \phi - \omega_{\rm R} t_0 \right)},
\end{split}
\end{equation}
from which the steady-state solutions naturally follow,
\begin{equation}
\begin{split}
v^\prime_\theta =& \frac{v_0 {\cal P}^m_l}{\sin\theta} \sin{\left( m \phi \right)}, \\
h^\prime_{\rm v} =& \frac{v_0 \mu {\cal P}^m_l}{m} \cos{\left( m \phi \right)}, \\
v^\prime_\phi =& \frac{v_0}{m \sin\theta} \left[ \left( l - m + 1 \right) {\cal P}^m_{l+1} - \mu \left( l + 1 \right) {\cal P}^m_l \right] \cos{\left( m \phi \right)}.
\end{split}
\end{equation}
At first sight, one may be concerned that the solutions blow up when $\theta=0^\circ$, since $\sin\theta=0$.  However, we also have ${\cal P}^m_l=0$ when $\theta=0^\circ$ ($\mu=1$).  Using the following recurrence relation,
\begin{equation}
\frac{{\cal P}^m_l}{\left( 1 - \mu^2 \right)^{1/2}} = \frac{{\cal P}^{m+1}_l + \left[ l \left( l + 1 \right) - m \left( m-1 \right) \right] {\cal P}^{m-1}_l}{2 m \mu},
\end{equation}
one realises that ${\cal P}^m_l/\sin\theta = 0$ when $\theta=0^\circ$ (for $m \ne 0$), so the solutions for $v^\prime_\theta$ and $v^\prime_\phi$ simply vanish at the poles.  These free, hydrodynamic solutions on a sphere, in the slowly-rotating limit, have previously been described by \cite{margules} and \cite{hough}.

In the rapidly-rotating limit ($\xi \rightarrow \infty$), \cite{lh68} has previously shown that for the solutions to satisfy the boundary conditions at $\mu \rightarrow \pm 1$ (i.e., be finite), the first four terms in equation (\ref{eq:sphere_hydro_general_governing}) need to be retained.  However, \cite{lh68} then made the approximation that $(1-\mu^2) \approx 1$ (near-equator solutions), which causes these solutions to formally diverge at $\mu = \pm 1$, as we will see.  Nevertheless, we follow this approach to obtain approximate solutions on a sphere in the rapidly-rotating limit, while being aware that our solutions will break down near the poles.  Much of the mathematical machinery for rapid rotators on a sphere has already been constructed during our $\beta$-plane analysis.  By making the transformation,
\begin{equation}
\begin{split}
\tilde{\mu} &= \alpha \mu, \\
\alpha &\equiv \xi^{1/4},
\end{split}
\end{equation}
we obtain the governing equation for $v_{\theta_0}$ in the rapidly-rotating limit,
\begin{equation}
\frac{\partial^2 v_{\theta_0}}{\partial \tilde{\mu}^2} + \left[ \left( \xi \omega^2 - m^2 - \frac{m}{\omega} \right) \xi^{-1/2} - \tilde{\mu}^2 \right] v_{\theta_0} = 0,
\end{equation}
which is again the quantum harmonic oscillator equation if the following quantity is quantized,
\begin{equation}
\left( \xi \omega^2 - m^2 - \frac{m}{\omega} \right) \xi^{-1/2} = 2l + 1.
\end{equation}
Unlike on the $\beta$-plane, we already have $\alpha \ne 1$ for a free hydrodynamic system on the sphere, because $\xi \ne 1$.  It follows that the dispersion relations are:
\begin{equation}
\begin{split}
&\omega_{\rm I} \left[ \xi \omega_{\rm I}^2 - 3 \xi \omega_{\rm R}^2 + m^2 + \left( 2l + 1 \right) \xi^{1/2} \right] = 0, \\
&\xi \omega_{\rm R}^3 - 3 \xi \omega_{\rm R} \omega_{\rm I}^2 - \omega_{\rm R} \left[ m^2 + \left( 2l + 1 \right) \xi^{1/2} \right] - m = 0.
\end{split}
\end{equation}
As before (on the $\beta$-plane), $\omega_{\rm I}=0$ is a solution of the dispersion relations.  The equations for the wave amplitudes are
\begin{equation}
\begin{split}
v_{\theta_0} =& v_0 \exp{\left(-\frac{\alpha^2 \mu^2}{2} \right)} \tilde{{\cal H}}_l, \\
h_{\rm v_0} =& \frac{v_0}{\omega_{\rm R}^2 \xi - m^2} \exp{\left(-\frac{\alpha^2 \mu^2}{2} \right)} \\
&\times \left[ 2 l \omega_{\rm R} \alpha \tilde{{\cal H}}_{l-1} - \tilde{{\cal H}}_l \left( \omega_{\rm R} \alpha \tilde{\mu} + \mu m \right) \right], \\
v_{\phi_0} =& \frac{v_0}{\omega_{\rm R}^2 \xi - m^2} \exp{\left(-\frac{\alpha^2 \mu^2}{2} \right)} \\
&\times \left[ 2 m l \alpha \tilde{{\cal H}}_{l-1} - \tilde{{\cal H}}_l \left( \tilde{\mu} m \alpha + \mu \omega_{\rm R} \xi \right) \right],
\end{split}
\end{equation}
where $\tilde{{\cal H}}_l \equiv {\cal H}_l(\tilde{\mu})$.  The wave solutions are
\begin{equation}
\begin{split}
v^\prime_\theta =& \frac{v_0}{\sin\theta} \exp{\left(-\frac{\alpha^2 \mu^2}{2} \right)} \tilde{{\cal H}}_l \sin{\left( m \phi - \omega_{\rm R} t_0 \right)}, \\
h^\prime_{\rm v} =& \frac{v_0}{\omega_{\rm R}^2 \xi - m^2} \exp{\left(-\frac{\alpha^2 \mu^2}{2} \right)} \cos{\left( m \phi - \omega_{\rm R} t_0 \right)} \\
&\times \left[ 2 l \omega_{\rm R} \alpha \tilde{{\cal H}}_{l-1} - \tilde{{\cal H}}_l \left( \omega_{\rm R} \alpha \tilde{\mu} + \mu m \right) \right], \\
v^\prime_\phi =& \frac{v_0}{\left( \omega_{\rm R}^2 \xi - m^2 \right) \sin\theta} \exp{\left(-\frac{\alpha^2 \mu^2}{2} \right)} \cos{\left( m \phi - \omega_{\rm R} t_0 \right)} \\
&\times \left[ 2 m l \alpha \tilde{{\cal H}}_{l-1} - \tilde{{\cal H}}_l \left( \tilde{\mu} m \alpha + \mu \omega_{\rm R} \xi \right) \right].
\end{split}
\end{equation}
The steady-state solutions follow naturally,
\begin{equation}
\begin{split}
v^\prime_\theta =& \frac{v_0}{\sin\theta} \exp{\left(-\frac{\alpha^2 \mu^2}{2} \right)} \tilde{{\cal H}}_l \sin{\left( m \phi \right)}, \\
h^\prime_{\rm v} =& \frac{v_0 \mu}{m} \exp{\left(-\frac{\alpha^2 \mu^2}{2} \right)} \tilde{{\cal H}}_l \cos{\left( m \phi \right)} \\
v^\prime_\phi =& \frac{v_0 \alpha}{m \sin\theta} \exp{\left(-\frac{\alpha^2 \mu^2}{2} \right)} \left( \tilde{\mu} \tilde{{\cal H}}_l - 2 l \tilde{{\cal H}}_{l-1} \right) \cos{\left( m \phi \right)}.
\end{split}
\end{equation}
It is worth noting that the Hermite polynomials do not vanish as $\tilde{\mu} \rightarrow \pm 1$, unlike for the associated Legendre functions, implying that the solutions for the velocities do blow up at the poles.  This is an artifact of the $(1-\mu^2) \approx 1$ approximation made.

\subsubsection{Forcing with Hydrodynamic Friction}
\label{subsect:spherical_hydro_forced}

When hydrodynamic friction is added, we find that the problem is analytically tractable only when the $\nu \partial/\partial \theta$ and $\nu \partial^2/\partial \theta^2$ terms are neglected, which is equivalent to assuming that molecular viscosity acts uniformly across latitude.  Note that this assumption does not apply to Rayleigh drag.  Additionally, we ignore the geometric terms associated with $\nu$ to render the algebra tractable.  The governing equations become
\begin{equation}
\begin{split}
&-i \frac{\partial}{\partial t_0} \left( i v^{\prime\prime}_\theta \right) - \mu v^{\prime\prime}_\phi - \hat{D}h^\prime_{\rm v} - i \omega_{\rm drag} \left( i v^{\prime\prime}_\theta \right) \\
&+ \frac{i}{{\cal R} \left( 1 - \mu^2 \right)} \frac{\partial^2}{\partial \phi^2} \left( i v^{\prime\prime}_\theta \right) = 0, \\
&i \mu \left( i v^{\prime\prime}_\theta \right) - \frac{\partial v^{\prime\prime}_\phi}{\partial t_0} - \frac{\partial h^\prime_{\rm v}}{\partial \phi} - \omega_{\rm drag} v^{\prime\prime}_\phi + \frac{1}{{\cal R} \left( 1 - \mu^2 \right)} \frac{\partial^2 v^{\prime\prime}_\phi}{\partial \phi^2} = 0, \\
&\left( 1 - \mu^2 \right) \left( \frac{\partial h^\prime_{\rm v}}{\partial t_0} - F_0 h^\prime_{\rm v} \right) + {\cal R}_0^2 \left[ i \hat{D}\left(i v^{\prime\prime}_\theta \right) + \frac{\partial v_\phi^{\prime\prime}}{\partial \phi} \right] = 0,
\end{split}
\end{equation}
where $\omega_{\rm drag}$ and $F_0$ have been cast in dimensionless units (normalized by $2\Omega$).  We have defined the Reynolds number as
\begin{equation}
{\cal R} \equiv \frac{2\Omega R^2}{\nu},
\end{equation}
where we have assumed that the characteristic velocity is $\Omega R$.  Note that this implies that more rapidly rotating exoplanets have faster wind speeds, which is not necessarily the case.  When modeling a specific object, there is no confusion as long as one specifies $\Omega$, $R$ and $\nu$ and then use it to construct ${\cal R}$.

The wave amplitudes are
\begin{equation}
\begin{split}
&\omega_0 v_{\theta_0} + \mu v_{\phi_0} + \hat{D} h_{\rm v_0} = 0, \\
&\mu v_{\theta_0} + \omega_0 v_{\phi_0} - m h_{\rm v_0} = 0, \\
&\hat{D} v_{\theta_0} + m v_{\phi_0} - \omega_{\rm F} \xi \left( 1 -\mu^2 \right) h_{\rm v_0} = 0,
\end{split}
\label{eq:spherical_hydro_forced_amp}
\end{equation}
where $\omega_0 \equiv \omega + i \omega_\nu$, $\omega_{\rm F} \equiv \omega - i F_0$ and
\begin{equation}
\omega_\nu \equiv \omega_{\rm drag} + \frac{m^2}{{\cal R} \left( 1 - \mu^2 \right)}.
\label{eq:omega_nu_spherical}
\end{equation}

Using the same mathematical machinery as outlined in \S\ref{subsect:spherical_hydro_free}, we derive
\begin{equation}
\begin{split}
&\left( \omega_0 \hat{D} + \mu m \right) h_{\rm v_0} = \left( \mu^2 - \omega^2_0 \right) v_{\theta_0}, \\
&\left( \omega_0 \hat{D} - \mu m \right) v_{\theta_0} = \left[ \omega_0 \omega_{\rm F} \xi \left( 1 - \mu^2 \right) - m^2 \right] h_{\rm v_0},
\end{split}
\end{equation}
from which we obtain
\begin{equation}
\left( \omega_0 \hat{D} + \mu m \right) \left[ \frac{\left( \omega_0 \hat{D} - \mu m\right) v_{\theta_0}}{\omega_0 \omega_{\rm F} \xi \left( 1 - \mu^2 \right) - m^2} \right] + \left( \omega^2_0 - \mu^2 \right) v_{\theta_0} = 0.
\end{equation}
To utilize the identity in equation (\ref{eq:identity}), we require that
\begin{equation}
\omega_\nu \approx \omega_{\rm drag} + \frac{m^2}{{\cal R}}.
\label{eq:omega_nu_approx}
\end{equation}
Physically, we are seeking near-equator solutions.  Using equation (\ref{eq:identity}) and noting that
\begin{equation}
\hat{D} {\cal G} = \frac{2 \omega_0 \omega_{\rm F} \xi \mu \left( 1 - \mu^2 \right) {\cal G}}{\omega_0 \omega_{\rm F} \xi \left( 1 - \mu^2 \right) - m^2},
\end{equation}
we obtain
\begin{equation}
\begin{split}
&\frac{\partial}{\partial \mu} \left[ \left( 1 - \mu^2 \right) \frac{\partial}{\partial \mu} \right] v_{\theta_0} - \frac{m v_{\theta_0}}{\omega_0} - \frac{m^2 v_{\theta_0}}{1 - \mu^2} \\
&+ \xi \omega_{\rm F} \left( \omega_0 - \frac{\mu^2}{\omega_0} \right) v_{\theta_0} \\
&+ \frac{2 \omega_{\rm F} \xi \mu}{\omega_0 \omega_{\rm F} \xi \left( 1- \mu^2 \right) - m^2} \left( \omega_0 \hat{D} - \mu m \right) v_{\theta_0} = 0.
\end{split}
\end{equation}

In the slowly-rotating limit ($\xi \rightarrow 0$), the governing equation for $v_{\theta_0}$ is again the associated Legendre equation,
\begin{equation}
\left( 1 - \mu^2 \right) \frac{\partial^2 v_{\theta_0}}{\partial \mu^2} - 2\mu \frac{\partial v_{\theta_0}}{\partial \mu} - m \left[ \frac{1}{\omega_0} + \frac{m}{\left( 1 - \mu^2 \right)} \right] v_{\theta_0} = 0.
\label{eq:legendre_hydro_forced}
\end{equation}
The dispersion relations are
\begin{equation}
\begin{split}
\omega_{\rm R} &= -\frac{m}{l \left(l+1\right)}, \\
\omega_{\rm I} &= -\omega_\nu. \\
\end{split}
\end{equation}
An immediate, curious inference is that forcing does not appear to affect the wave solutions.  A forced, damped, hydrodynamic atmosphere behaves like a purely damped one in the slowly-rotating limit.  We will confirm this finding by explicitly deriving $v^\prime_\theta$, $h^\prime_{\rm v}$ and $v^\prime_\phi$.

Using the same procedure described in \S\ref{subsect:spherical_hydro_free}, we obtain the solutions for the wave amplitudes,
\begin{equation}
\begin{split}
v_{\theta_0} =& v_0 {\cal P}^m_l, \\
h_{\rm v_0} =& \frac{v_0}{m^2} \left\{ \omega_0 \left( l - m + 1 \right) {\cal P}^m_{l+1} - \mu \left[ \omega_0 \left( l + 1 \right) - m \right] {\cal P}^m_l \right\}, \\
v_{\phi_0} =& \frac{v_0}{m} \left[ \left( l - m + 1 \right) {\cal P}^m_{l+1} - \mu \left( l + 1 \right) {\cal P}^m_l \right].
\end{split}
\end{equation}
The steady-state solutions are
\begin{equation}
\begin{split}
v^\prime_\theta =& \frac{v_0 {\cal P}^m_l}{\sin\theta} \sin{\left( m \phi \right)}, \\
h^\prime_{\rm v} =& \frac{v_0 \mu {\cal P}^m_l}{m} \cos{\left( m \phi \right)} \\
&- \frac{\omega_\nu v_0}{m^2} \left[ \left( l - m + 1 \right) {\cal P}^m_{l+1} - \mu \left( l + 1 \right) {\cal P}^m_l \right] \sin{\left( m \phi \right)}, \\
v^\prime_\phi =& \frac{v_0}{m \sin\theta} \left[ \left( l - m + 1 \right) {\cal P}^m_{l+1} - \mu \left( l + 1 \right) {\cal P}^m_l \right] \cos{\left( m \phi \right)}.
\end{split}
\end{equation}
One immediately sees that the solutions for $v^\prime_\theta$ and $v^\prime_\phi$ are exactly the same as in the hydrodynamic limit.  Hydrodynamic friction only affects $h^\prime_{\rm v}$ and introduces an out-of-phase component to the solution.  Forcing is completely absent from these solutions.

In the rapidly-rotating limit ($\xi \rightarrow \infty$), the governing equation for $v_{\theta_0}$ is 
\begin{equation}
\frac{\partial^2 v_{\theta_0}}{\partial \tilde{\mu}^2} + \left[ \left( \xi \omega_0 \omega_{\rm F} - m^2 - \frac{m}{\omega_0} \right) \left( \frac{\omega_0}{\xi \omega_{\rm F}} \right)^{1/2} - \tilde{\mu}^2 \right] v_{\theta_0} = 0,
\end{equation}
where we have defined the transformed (cosine of the) co-latitude as
\begin{equation}
\begin{split}
\tilde{\mu} &\equiv \alpha \mu, \\
\alpha &\equiv \left( \frac{\xi \omega_{\rm F}}{\omega_0} \right)^{1/4}.
\end{split}
\end{equation}
Via the usual use of De Moivre's formula, we obtain the dispersion relations,
\begin{equation}
\begin{split}
&\xi \omega^3_{\rm R} - 3 \xi \omega_{\rm R} \omega_{\rm I}^2 + 2 \xi \omega_{\rm R} \omega_{\rm I} \left( F_0 - 2 \omega_\nu \right) \\
&+ \omega_{\rm R} \left[ \xi \omega_\nu \left( 2 F_0 - \omega_\nu \right) - m^2 \right] - m \\
&- \left( 2 l + 1 \right) \left( \frac{\zeta + \zeta_{\rm R}}{2} \right)^{1/2} = 0, \\
&\xi \omega_{\rm I}^3 - 3 \xi \omega_{\rm R}^2 \omega_{\rm I} + \xi \left( F_0 - 2 \omega_\nu \right) \left( \omega_{\rm R}^2 - \omega_{\rm I}^2 \right) \\
&- \xi \omega_\nu \omega_{\rm I} \left( 2 F_0 - \omega_\nu \right) + m^2 \left( \omega_{\rm I} + \omega_\nu \right) - \omega_\nu^2 F_0 \xi \\
&+ \left( 2 l + 1 \right) \left( \frac{\zeta - \zeta_{\rm R}}{2} \right)^{1/2} = 0,
\end{split}
\end{equation}
where the separation functions are
\begin{equation}
\begin{split}
\zeta_{\rm R} \equiv& \xi \left[ \omega^2_{\rm R} - \omega^2_{\rm I} - \omega_{\rm I} \left( \omega_\nu - F_0 \right) + \omega_\nu F_0 \right], \\
\zeta_{\rm I} \equiv& \xi \omega_{\rm R} \left( 2 \omega_{\rm I} + \omega_\nu - F_0 \right), \\
\zeta =& \left( \zeta_{\rm R}^2 + \zeta_{\rm I}^2 \right)^{1/2}.
\end{split}
\end{equation}
The wave amplitudes are
\begin{equation}
\begin{split}
v_{\theta_0} =& v_0 \exp{\left(-\frac{\alpha^2 \mu^2}{2} \right)} \tilde{{\cal H}}_l, \\
h_{\rm v_0} =& \frac{v_0}{\omega_0 \omega_{\rm F} \xi - m^2} \exp{\left(-\frac{\alpha^2 \mu^2}{2} \right)} \\
&\times \left[ 2 l \omega_0 \alpha \tilde{{\cal H}}_{l-1} - \tilde{{\cal H}}_l \left( \omega_0 \alpha \tilde{\mu} + \mu m \right) \right], \\
v_{\phi_0} =& \frac{v_0}{\omega_0 \omega_{\rm F} \xi - m^2} \exp{\left(-\frac{\alpha^2 \mu^2}{2} \right)} \\
&\times \left[ 2 m l \alpha \tilde{{\cal H}}_{l-1} - \tilde{{\cal H}}_l \left( \tilde{\mu} m \alpha + \mu \omega_{\rm F} \xi \right) \right].
\end{split}
\end{equation}
In the steady-state limit, we set $F_0 = - \vert F_0 \vert$ and obtain
\begin{equation}
\alpha = \left( \frac{\xi \left \vert F_0 \right \vert}{\omega_\nu} \right)^{1/4}.
\end{equation}
Unlike in the $\beta$-plane treatment, $\alpha$ contains an extra factor of $\xi$, implying that $\alpha$ is related to both the Prandtl and Rossby numbers in the forced, damped hydrodynamic limit.  The steady-state solutions are
\begin{equation}
\begin{split}
&v^\prime_\theta = \frac{v_0}{\sin\theta} \exp{\left(-\frac{\alpha^2 \mu^2}{2} \right)} \tilde{{\cal H}}_l \sin{\left( m \phi \right)}, \\
&h^\prime_{\rm v} = \frac{v_0}{ \omega_\nu \left \vert F_0 \right \vert \xi + m^2 } \exp{\left(-\frac{\alpha^2 \mu^2}{2} \right)} \\
&\times \left[ \mu m \tilde{{\cal H}}_l \cos{\left( m \phi \right)} + \alpha \omega_\nu \left( 2 l \tilde{{\cal H}}_{l-1} - \tilde{\mu} \tilde{{\cal H}}_l \right) \sin{\left( m \phi \right)} \right], \\
&v^\prime_\phi = -\frac{v_0}{\left( \omega_\nu \left \vert F_0 \right \vert \xi + m^2 \right) \sin\theta} \exp{\left(-\frac{\alpha^2 \mu^2}{2} \right)} \\
&\times \left[ m \alpha \left( 2 l \tilde{{\cal H}}_{l-1} - \tilde{\mu} \tilde{{\cal H}}_l \right) \cos{\left( m \phi \right)} + \mu \left \vert F_0 \right \vert \xi \tilde{{\cal H}}_l \sin{\left( m \phi \right)} \right].
\end{split}
\end{equation}

\subsection{Magnetohydrodynamic (Radial Background Field)}

We consider forcing, magnetic tension and all forms of friction.  For a forced, dragged, shallow water system in spherical geometry with a purely radial background magnetic field, we find that the problem is analytically tractable only when the $\eta \partial/\partial \theta$ and $\eta \partial^2/\partial \theta^2$ terms are neglected.  We apply the same reasoning to the molecular viscosity.  We ignore the geometric terms associated with both molecular viscosity and magnetic drag.  With these simplifications, the governing equations become
\begin{equation}
\begin{split}
&-i \frac{\partial}{\partial t_0} \left( i v^{\prime\prime}_\theta \right) - \mu v^{\prime\prime}_\phi - \hat{D}h^\prime_{\rm v} - i \omega_{\rm drag} \left( i v^{\prime\prime}_\theta \right) \\
&+ \frac{i}{{\cal R} \left( 1 - \mu^2 \right)} \frac{\partial^2}{\partial \phi^2} \left( i v^{\prime\prime}_\theta \right) + \frac{\bar{B}_r b^{\prime\prime}_\theta}{8 \pi \rho H \Omega} = 0, \\
&i \mu \left( i v^{\prime\prime}_\theta \right) - \frac{\partial v^{\prime\prime}_\phi}{\partial t_0} - \frac{\partial h^\prime_{\rm v}}{\partial \phi} - \omega_{\rm drag} v^{\prime\prime}_\phi + \frac{1}{{\cal R} \left( 1 - \mu^2 \right)} \frac{\partial^2 v^{\prime\prime}_\phi}{\partial \phi^2} \\
&- \frac{\bar{B}_r b^{\prime\prime}_\phi}{8 \pi \rho H \Omega} = 0, \\
&\left( 1 - \mu^2 \right) \left( \frac{\partial h^\prime_{\rm v}}{\partial t_0} - F_0 h^\prime_{\rm v} \right) + {\cal R}_0^2 \left[ i \hat{D}\left(i v^{\prime\prime}_\theta \right) + \frac{\partial v_\phi^{\prime\prime}}{\partial \phi} \right] = 0, \\
&\frac{\partial b^{\prime\prime}_\theta}{\partial t_0} + \frac{i \bar{B}_r \left( i v^{\prime\prime}_\theta \right)}{2 \Omega H} - \frac{1}{{\cal R}_{\rm B} \left( 1 - \mu^2 \right)} \frac{\partial^2 b^{\prime\prime}_\theta}{\partial \phi^2} = 0, \\
&\frac{\partial b^{\prime\prime}_\phi}{\partial t_0} - \frac{\bar{B}_r v^{\prime\prime}_\phi}{2 \Omega H} - \frac{1}{{\cal R}_{\rm B} \left( 1 - \mu^2 \right)} \frac{\partial^2 b^{\prime\prime}_\phi}{\partial \phi^2} = 0, \\
\end{split}
\end{equation}
where we again have ${\cal R} \equiv 2 \Omega R^2 / \nu$ and the magnetic Reynolds number is
\begin{equation}
{\cal R}_{\rm B} \equiv \frac{2 \Omega R^2}{\eta}.
\end{equation}
We have also defined
\begin{equation}
b^{\prime\prime}_{\theta,\phi} \equiv b^\prime_{\theta,\phi} \sin\theta.
\end{equation}

From seeking wave solutions, the wave amplitudes are
\begin{equation}
\begin{split}
&\omega_{\rm B_0} v_{\theta_0} + \mu v_{\phi_0} + \hat{D} h_{\rm v_0} = 0, \\
&\mu v_{\theta_0} + \omega_{\rm B_0} v_{\phi_0} - m h_{\rm v_0} = 0, \\
&\hat{D} v_{\theta_0} + m v_{\phi_0} - \omega_{\rm F} \xi \left( 1 -\mu^2 \right) h_{\rm v_0} = 0,
\end{split}
\label{eq:spherical_mhd_amp}
\end{equation}
where we additionally define
\begin{equation}
\begin{split}
\omega_\eta &\equiv \omega + \frac{i m^2}{R_{\rm B} \left( 1 - \mu^2 \right)} \approx \omega + \frac{i m^2}{R_{\rm B}}, \\
\omega_{\rm B} &\equiv \omega_\nu + \frac{i}{\omega_\eta} \left( \frac{v_{\rm A}}{2 \Omega H} \right)^2 = \omega_\nu + \frac{i}{\omega_\eta} \left( \frac{t_{\rm dyn}}{t_{\rm A}} \right)^2, \\
\end{split}
\end{equation}
and $t_{\rm dyn} \equiv 1/2\Omega$.  The quantity $\omega_\nu$ is the same as defined in equation (\ref{eq:omega_nu_spherical}).  We again have $\omega_{\rm B_0} \equiv \omega + i \omega_{\rm B}$.  All of these generalized frequencies are dimensionless, as is $\omega$.  The approximation associated with $\omega_\eta$ is made such that the identity in equation (\ref{eq:identity}) can again be applied; we do the same for $\omega_\nu$ via equation (\ref{eq:omega_nu_approx}).  Physically, we are seeking near-equator solutions.

Since the expressions in equation (\ref{eq:spherical_mhd_amp}) are identical to those in (\ref{eq:spherical_hydro_forced_amp}), except for the substitution $\omega_0 \rightarrow \omega_{\rm B_0}$, we may immediately write down the general governing equation for $v_{\theta_0}$,
\begin{equation}
\begin{split}
&\frac{\partial}{\partial \mu} \left[ \left( 1 - \mu^2 \right) \frac{\partial}{\partial \mu} \right] v_{\theta_0} - \frac{m v_{\theta_0}}{\omega_{\rm B_0}} - \frac{m^2 v_{\theta_0}}{1 - \mu^2} \\
&+ \xi \omega_{\rm F} \left( \omega_{\rm B_0} - \frac{\mu^2}{\omega_{\rm B_0}} \right) v_{\theta_0} \\
&+ \frac{2 \omega_{\rm F} \xi \mu}{\omega_{\rm B_0} \omega_{\rm F} \xi \left( 1- \mu^2 \right) - m^2} \left( \omega_{\rm B_0} \hat{D} - \mu m \right) v_{\theta_0} = 0.
\end{split}
\end{equation}

In the slowly-rotating limit ($\xi \rightarrow 0$), the governing equation for $v_{\theta_0}$ is the same as equation (\ref{eq:legendre_hydro_forced}), except that $\omega_0$ is replaced by $\omega_{\rm B_0}$.  Making use of the separation functions and writing $\omega_{\rm B_0} = \zeta_- \omega_{\rm R} + i ( \zeta_0 + \zeta_+ \omega_{\rm I})$ as in the case of the $\beta$-plane treatment, we derive the dispersion relations,
\begin{equation}
\begin{split}
\omega_{\rm R} &= - \frac{m}{l \left( l + 1 \right) \zeta_-}, \\
\omega_{\rm I} &= - \frac{\zeta_0}{\zeta_+}, \\
\end{split}
\label{eq:dispersion_sphere_mhd_slow}
\end{equation}
where we have defined
\begin{equation}
\begin{split}
\zeta_\pm \equiv& 1 \pm \left( \frac{v_{\rm A}}{2 \Omega H} \right)^2 \frac{1}{\omega_{\rm R}^2 + \left( \omega_{\rm I} + m^2/{\cal R}_{\rm B} \right)^2}, \\
\zeta_0 \equiv& \omega_\nu + \left( \frac{v_{\rm A}}{2 \Omega H} \right)^2 \frac{m^2}{{\cal R}_{\rm B} \left[ \omega_{\rm R}^2 + \left( \omega_{\rm I} + m^2/{\cal R}_{\rm B} \right)^2 \right]}.
\end{split}
\end{equation}

The equations for the wave amplitudes are
\begin{equation}
\begin{split}
v_{\theta_0} =& v_0 {\cal P}^m_l, \\
h_{\rm v_0} =& \frac{v_0}{m^2} \left\{ \omega_{\rm B_0} \left( l - m + 1 \right) {\cal P}^m_{l+1} - \mu \left[ \omega_{\rm B_0} \left( l + 1 \right) - m \right] {\cal P}^m_l \right\}, \\
v_{\phi_0} =& \frac{v_0}{m} \left[ \left( l - m + 1 \right) {\cal P}^m_{l+1} - \mu \left( l + 1 \right) {\cal P}^m_l \right].
\end{split}
\end{equation}
In the steady-state limit, we have $\omega_{\rm B_0} = i \zeta_0$ and the generalized friction becomes
\begin{equation}
\zeta_0 = \omega_\nu + \left( \frac{v_{\rm A}}{2 \Omega H} \right)^2 \frac{{\cal R}_{\rm B}}{m^2}.
\end{equation}
The steady-state solutions are
\begin{equation}
\begin{split}
v^\prime_\theta =& \frac{v_0 {\cal P}^m_l}{\sin\theta} \sin{\left( m \phi \right)}, \\
h^\prime_{\rm v} =& \frac{v_0 \mu {\cal P}^m_l}{m} \cos{\left( m \phi \right)} \\
&- \frac{\zeta_0 v_0}{m^2} \left[ \left( l - m + 1 \right) {\cal P}^m_{l+1} - \mu \left( l + 1 \right) {\cal P}^m_l \right] \sin{\left( m \phi \right)}, \\
v^\prime_\phi =& \frac{v_0}{m \sin\theta} \left[ \left( l - m + 1 \right) {\cal P}^m_{l+1} - \mu \left( l + 1 \right) {\cal P}^m_l \right] \cos{\left( m \phi \right)}.
\end{split}
\label{eq:sphere_mhd_steady_slow}
\end{equation}
One can immediately see the justification for calling $\zeta_0$ the ``generalized friction": it replaces $\omega_\nu$ in the equation for $h^\prime_{\rm v}$ and includes the effects of hydrodynamic friction, magnetic tension and magnetic drag.  The velocities are unaffected by the generalized friction; forcing is again absent from the solutions in the slowly-rotating limit.

In the rapidly-rotating limit ($\xi \rightarrow \infty$), the governing equation for $v_{\theta_0}$ is 
\begin{equation}
\frac{\partial^2 v_{\theta_0}}{\partial \tilde{\mu}^2} + \left[ \left( \xi \omega_{\rm B_0} \omega_{\rm F} - m^2 - \frac{m}{\omega_{\rm B_0}} \right) \left( \frac{\omega_{\rm B_0}}{\xi \omega_{\rm F}} \right)^{1/2} - \tilde{\mu}^2 \right] v_{\theta_0} = 0,
\end{equation}
where we have defined
\begin{equation}
\begin{split}
\tilde{\mu} &\equiv \alpha \mu, \\
\alpha &\equiv \left( \frac{\xi \omega_{\rm F}}{\omega_{\rm B_0}} \right)^{1/4}.
\end{split}
\end{equation}
The dispersion relations are
\begin{equation}
\begin{split}
&\xi \omega_{\rm R}^3 \zeta_-^2 - \xi \omega_{\rm R} \left( \zeta_0 + \omega_{\rm I} \zeta_+ \right)^2 \\
&- 2 \xi \zeta_- \omega_{\rm R} \left( \omega_{\rm I} - F_0 \right) \left( \zeta_0 + \omega_{\rm I} \zeta_+ \right) - m^2 \zeta_- \omega_{\rm R} \\
&- m - \left( 2 l + 1 \right) \left( \frac{\zeta + \zeta_{\rm R}}{2} \right)^{1/2} = 0, \\
&2 \xi \zeta_- \omega_{\rm R}^2 \left( \zeta_0 + \omega_{\rm I} \zeta_+ \right) + \xi \zeta_-^2 \omega_{\rm R}^2 \left( \omega_{\rm I} - F_0 \right) \\
&- \xi \left( \omega_{\rm I} - F_0 \right) \left( \zeta_0 + \omega_{\rm I} \zeta_+ \right)^2 - m^2 \left( \zeta_0 + \omega_{\rm I} \zeta_+ \right) \\
&- \left( 2 l + 1 \right) \left( \frac{\zeta - \zeta_{\rm R}}{2} \right)^{1/2} = 0,\\
\end{split}
\end{equation}
where we have defined
\begin{equation}
\begin{split}
\zeta_{\rm R} \equiv& \xi \left[ \omega^2_{\rm R} \zeta_- - \left( \omega_{\rm I} - F_0 \right) \left( \zeta_0 + \omega_{\rm I} \zeta_+ \right) \right], \\
\zeta_{\rm I} \equiv& \xi \omega_{\rm R} \left[ \zeta_- \left( \omega_{\rm I} - F_0 \right) + \zeta_0 + \omega_{\rm I} \zeta_+ \right], \\
\zeta =& \left( \zeta_{\rm R}^2 + \zeta_{\rm I}^2 \right)^{1/2}.
\end{split}
\end{equation}
The wave amplitudes follow directly from the forced, damped hydrodynamic case with a $\omega_0 \rightarrow \omega_{\rm B_0}$ transformation,
\begin{equation}
\begin{split}
v_{\theta_0} =& v_0 \exp{\left(-\frac{\alpha^2 \mu^2}{2} \right)} \tilde{{\cal H}}_l, \\
h_{\rm v_0} =& \frac{v_0}{\omega_{\rm B_0} \omega_{\rm F} \xi - m^2} \exp{\left(-\frac{\alpha^2 \mu^2}{2} \right)} \\
&\times \left[ 2 l \omega_{\rm B_0} \alpha \tilde{{\cal H}}_{l-1} - \tilde{{\cal H}}_l \left( \omega_{\rm B_0} \alpha \tilde{\mu} + \mu m \right) \right], \\
v_{\phi_0} =& \frac{v_0}{\omega_{\rm B_0} \omega_{\rm F} \xi - m^2} \exp{\left(-\frac{\alpha^2 \mu^2}{2} \right)} \\
&\times \left[ 2 m l \alpha \tilde{{\cal H}}_{l-1} - \tilde{{\cal H}}_l \left( \tilde{\mu} m \alpha + \mu \omega_{\rm F} \xi \right) \right].
\end{split}
\label{eq:sphere_mhd_amp}
\end{equation}
The steady-state solutions follow,
\begin{equation}
\begin{split}
&v^\prime_\theta = \frac{v_0}{\sin\theta} \exp{\left(-\frac{\alpha^2 \mu^2}{2} \right)} \tilde{{\cal H}}_l \sin{\left( m \phi \right)}, \\
&h^\prime_{\rm v} = \frac{v_0}{ \zeta_0 \left \vert F_0 \right \vert \xi + m^2 } \exp{\left(-\frac{\alpha^2 \mu^2}{2} \right)} \\
&\times \left[ \mu m \tilde{{\cal H}}_l \cos{\left( m \phi \right)} + \alpha \zeta_0 \left( 2 l \tilde{{\cal H}}_{l-1} - \tilde{\mu} \tilde{{\cal H}}_l \right) \sin{\left( m \phi \right)} \right], \\
&v^\prime_\phi = -\frac{v_0}{\left( \zeta_0 \left \vert F_0 \right \vert \xi + m^2 \right) \sin\theta} \exp{\left(-\frac{\alpha^2 \mu^2}{2} \right)} \\
&\times \left[ m \alpha \left( 2 l \tilde{{\cal H}}_{l-1} - \tilde{\mu} \tilde{{\cal H}}_l \right) \cos{\left( m \phi \right)} + \mu \left \vert F_0 \right \vert \xi \tilde{{\cal H}}_l \sin{\left( m \phi \right)} \right],
\end{split}
\label{eq:sphere_mhd_steady}
\end{equation}
where we have $\alpha = ( \xi \vert F_0 \vert / \zeta_0 )^{1/4}$.  In the rapidly-rotating limit, these steady-state solutions are identical to the forced, damped hydrodynamic case, except for a $\omega_\nu \rightarrow \zeta_0$ transformation, once again illustrating why $\zeta_0$ is termed the ``generalized friction".

\subsection{Magnetohydrodynamic (Toroidal Background Field)}
\label{subsect:spherical_beta_mhd_toroidal}

\begin{figure*}
\begin{center}
\vspace{-0.2in}
\includegraphics[width=\columnwidth]{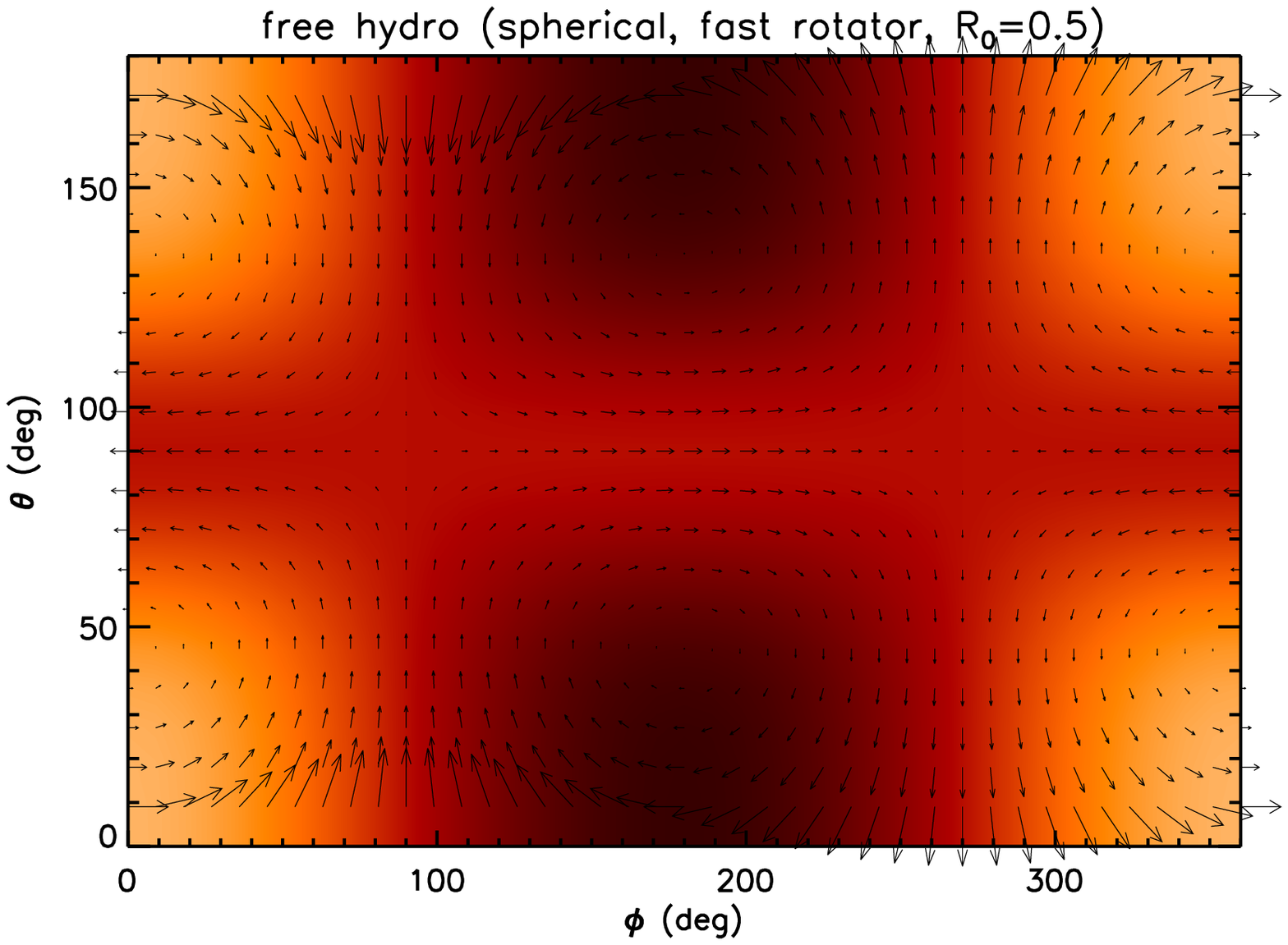}
\includegraphics[width=\columnwidth]{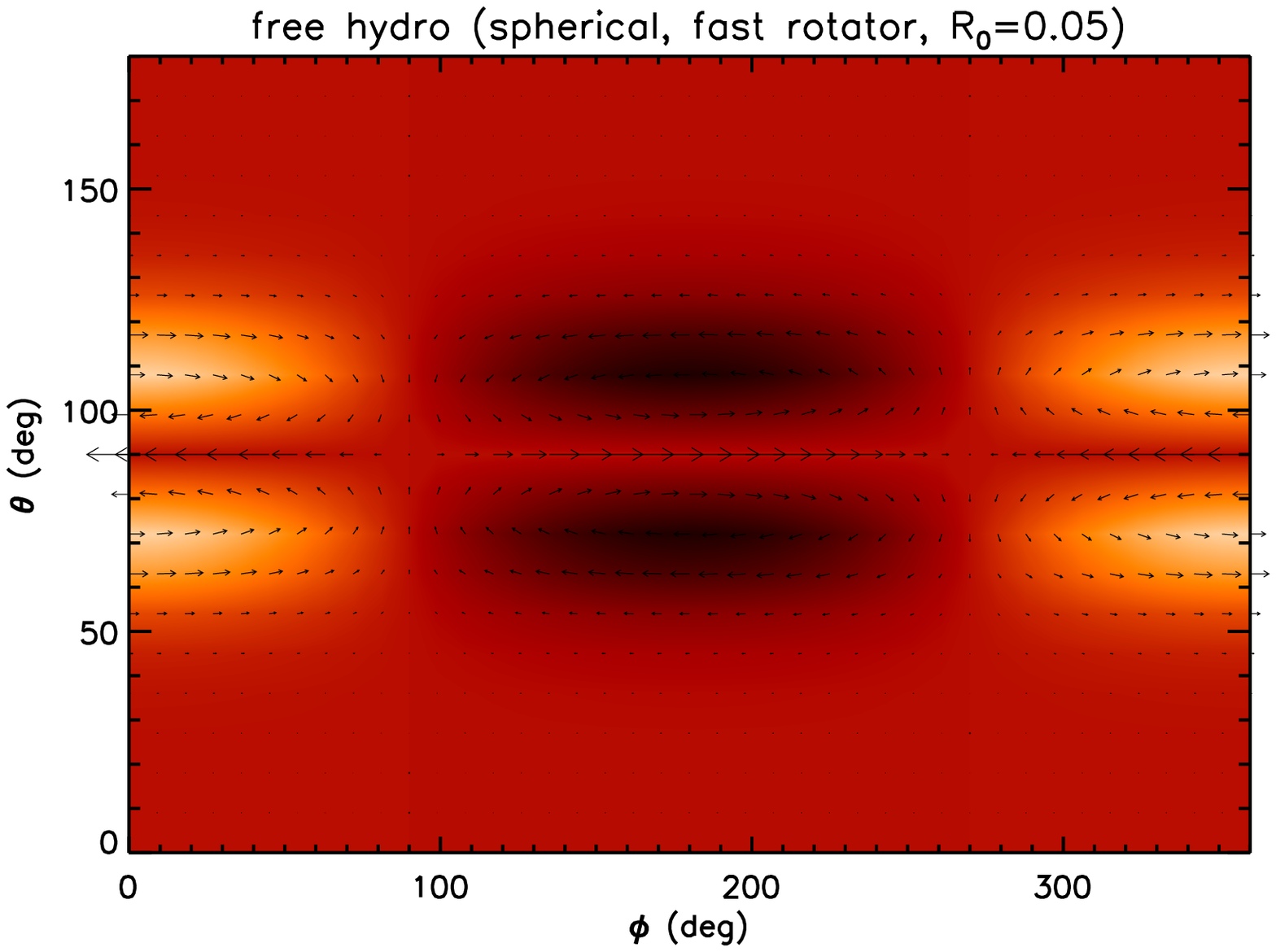}
\end{center}
\vspace{-0.2in}
\caption{Plots of velocity perturbations (arrows) and $h^\prime_{\rm v}$ (contours) for steady-state, hydrodynamic systems in full spherical geometry for $l=m=1$ in the rapidly-rotating limit.  All quantities are computed in terms of an arbitrary velocity normalization ($v_0$).  Bright and dark colors correspond to positive and negative height perturbations, respectively.}
\vspace{0.1in}
\label{fig:sphere_hydro_fast}
\end{figure*}

In evaluating a forced, damped, shallow water system with a horizontal background magnetic field on a sphere, we encounter an obstacle already elucidated on the $\beta$-plane: we need to set $\bar{B}_\theta = b^\prime_\theta = 0$ to proceed.  We ignore all geometric terms.  It follows that
\begin{equation}
\begin{split}
&-i \frac{\partial}{\partial t_0} \left( i v^{\prime\prime}_\theta \right) - \mu v^{\prime\prime}_\phi - \hat{D}h^\prime_{\rm v} - i \omega_{\rm drag} \left( i v^{\prime\prime}_\theta \right) \\
&+ \frac{i}{{\cal R} \left( 1 - \mu^2 \right)} \frac{\partial^2}{\partial \phi^2} \left( i v^{\prime\prime}_\theta \right) = 0, \\
&i \mu \left( i v^{\prime\prime}_\theta \right) - \frac{\partial v^{\prime\prime}_\phi}{\partial t_0} - \frac{\partial h^\prime_{\rm v}}{\partial \phi} - \omega_{\rm drag} v^{\prime\prime}_\phi + \frac{1}{{\cal R} \left( 1 - \mu^2 \right)} \frac{\partial^2 v^{\prime\prime}_\phi}{\partial \phi^2} \\
&+ \frac{\bar{B}_\phi}{8 \pi \rho R \Omega \sin\theta} \frac{\partial b^{\prime\prime}_\phi}{\partial \phi} = 0, \\
&\left( 1 - \mu^2 \right) \left( \frac{\partial h^\prime_{\rm v}}{\partial t_0} - F_0 h^\prime_{\rm v} \right) + {\cal R}_0^2 \left[ i \hat{D}\left(i v^{\prime\prime}_\theta \right) + \frac{\partial v_\phi^{\prime\prime}}{\partial \phi} \right] = 0, \\
&\frac{\partial b^{\prime\prime}_\phi}{\partial t_0} - \frac{\bar{B}_\phi}{2 \Omega R \sin\theta} \frac{\partial v^{\prime\prime}_\phi}{\partial \phi} - \frac{1}{{\cal R}_{\rm B} \left( 1 - \mu^2 \right)} \frac{\partial^2 b^{\prime\prime}_\phi}{\partial \phi^2} = 0, \\
\end{split}
\end{equation}
from which we obtain
\begin{equation}
\begin{split}
&\omega_0 v_{\theta_0} + \mu v_{\phi_0} + \hat{D} h_{\rm v_0} = 0, \\
&\mu v_{\theta_0} + \omega_{\rm B_0} v_{\phi_0} - m h_{\rm v_0} = 0, \\
&\hat{D} v_{\theta_0} + m v_{\phi_0} - \omega_{\rm F} \xi \left( 1 -\mu^2 \right) h_{\rm v_0} = 0,
\end{split}
\label{eq:spherical_mhd_amp_horizontal}
\end{equation}
where the definitions for $\omega_\nu$, $\omega_\eta$ and $\omega_{\rm B_0}$ are identical to the radial-field case, while the definition for $\omega_{\rm B}$ has changed,
\begin{equation}
\begin{split}
\omega_{\rm B} &\equiv \omega_\nu + \frac{i m^2}{\omega_\eta \left( 1 - \mu^2 \right)} \left( \frac{v_{\rm A}}{2 \Omega R} \right)^2 \\
&\approx \omega_\nu + \frac{i m^2}{\omega_\eta} \left( \frac{t_{\rm dyn}}{t_{\rm A}} \right)^2,
\end{split}
\end{equation}
where $v_{\rm A} \equiv \bar{B}_\phi/2\sqrt{\pi \rho}$.  We are again seeking near-equator solutions.

From manipulating the equations for the wave amplitudes using the usual procedure, we obtain
\begin{equation}
\begin{split}
&\left( \omega_{\rm B_0} \hat{D} + \mu m \right) h_{\rm v_0} = \left( \mu^2 - \omega_0 \omega_{\rm B_0} \right) v_{\theta_0}, \\
&\left( \omega_{\rm B_0} \hat{D} - \mu m \right) v_{\theta_0} = \left[ \omega_{\rm B_0} \omega_{\rm F} \xi \left( 1 - \mu^2 \right) - m^2 \right] h_{\rm v_0},
\end{split}
\end{equation}
from which the general governing equation for $v_{\theta_0}$ follows,
\begin{equation}
\begin{split}
&\frac{\partial}{\partial \mu} \left[ \left( 1 - \mu^2 \right) \frac{\partial}{\partial \mu} \right] v_{\theta_0} - \frac{m v_{\theta_0}}{\omega_{\rm B_0}} - \frac{m^2 v_{\theta_0} \omega_0}{\left(1 - \mu^2 \right) \omega_{\rm B_0}} \\
&+ \xi \omega_{\rm F} \left( \omega_0 - \frac{\mu^2}{\omega_{\rm B_0}} \right) v_{\theta_0} \\
&+ \frac{2 \omega_{\rm F} \xi \mu}{\omega_{\rm B_0} \omega_{\rm F} \xi \left( 1- \mu^2 \right) - m^2} \left( \omega_{\rm B_0} \hat{D} - \mu m \right) v_{\theta_0} = 0.
\end{split}
\end{equation}

In the slowly-rotating limit ($\xi \rightarrow 0$), the governing equation for $v_{\theta_0}$ is slightly different from the case with a radial background magnetic field,
\begin{equation}
\left( 1 - \mu^2 \right) \frac{\partial^2 v_{\theta_0}}{\partial \mu^2} - 2\mu \frac{\partial v_{\theta_0}}{\partial \mu} - \frac{m v_{\theta_0}}{\omega_{\rm B_0}} - \frac{\tilde{m}^2 v_{\theta_0}}{\left( 1 - \mu^2 \right)} = 0,
\label{eq:legendre_mhd_hori}
\end{equation}
where
\begin{equation}
\tilde{m} \equiv \left( \frac{\omega_0}{\omega_{\rm B_0}} \right)^{1/2} m.
\end{equation}
The dispersion relations take exactly the same form as in equation (\ref{eq:dispersion_sphere_mhd_slow}), but with different definitions for the separation functions,
\begin{equation}
\begin{split}
\zeta_\pm \equiv& 1 \pm \left( \frac{v_{\rm A}}{2 \Omega R} \right)^2 \frac{m^2}{\omega_{\rm R}^2 + \left( \omega_{\rm I} + m^2/{\cal R}_{\rm B} \right)^2}, \\
\zeta_0 \equiv& \omega_\nu + \left( \frac{v_{\rm A}}{2 \Omega R} \right)^2 \frac{m^4}{{\cal R}_{\rm B} \left[ \omega_{\rm R}^2 + \left( \omega_{\rm I} + m^2/{\cal R}_{\rm B} \right)^2 \right]}.
\end{split}
\end{equation}

Formally, the usual analytical solutions are obtained from equation (\ref{eq:legendre_mhd_hori}) only when $\tilde{m}$ is both an integer and real.  We immediately specialize to steady-state solutions, for which
\begin{equation}
\tilde{m} = \left[ \frac{\omega_\nu}{\omega_\nu + {\cal R}_{\rm B} \left( v_{\rm A} / 2 \Omega R \right)^2 } \right]^{1/2} m.
\end{equation}
In the steady-state limit, $\tilde{m}$ is real but will not generally take on integer values.  However, if $v_{\rm A} / 2 \Omega R \ll 1$, we may assume $\tilde{m} \approx m$.  In this limit, the steady-state solutions take the same form as in equation (\ref{eq:sphere_mhd_steady_slow}).

In the rapidly-rotating limit ($\xi \rightarrow \infty$), the governing equation for $v_{\theta_0}$ is 
\begin{equation}
\frac{\partial^2 v_{\theta_0}}{\partial \tilde{\mu}^2} + \left[ \left( \xi \omega_0 \omega_{\rm F} - \frac{m^2 \omega_0}{\omega_{\rm B_0}} - \frac{m}{\omega_{\rm B_0}} \right) \left( \frac{\omega_{\rm B_0}}{\xi \omega_{\rm F}} \right)^{1/2} - \tilde{\mu}^2 \right] v_{\theta_0} = 0,
\end{equation}
where we have again defined $\tilde{\mu} \equiv \alpha \mu$ and  $\alpha \equiv \left( \xi \omega_{\rm F}/\omega_{\rm B_0} \right)^{1/4}$.  Again using the technique of writing $\omega_{\rm B_0} = \omega_{\rm R} \zeta_- + i ( \zeta_0 + \omega_{\rm I} \zeta_+ )$, we obtain the dispersion relations,
\begin{equation}
\begin{split}
&\xi \omega_{\rm R}^3 \zeta_- - \xi \zeta_- \omega_{\rm R} \left( \omega_{\rm I} - F_0 \right) \left( \omega_{\rm I} + \omega_\nu \right) \\
&- \xi \omega_{\rm R}  \left( \zeta_0 + \omega_{\rm I} \zeta_+ \right) \left( 2 \omega_{\rm I} + \omega_\nu - F_0 \right) - m^2 \omega_{\rm R} \\
&- m - \left( 2 l + 1 \right) \left( \frac{\zeta + \zeta_{\rm R}}{2} \right)^{1/2} = 0, \\
&\xi \omega_{\rm R}^2 \left( \zeta_0 + \omega_{\rm I} \zeta_+ \right) - \xi \left( \zeta_0 + \omega_{\rm I} \zeta_+ \right) \left( \omega_{\rm I} - F_0 \right) \left( \omega_{\rm I} + \omega_\nu \right) \\
&+ \xi \zeta_- \omega_{\rm R}^2 \left( 2 \omega_{\rm I} + \omega_\nu - F_0 \right) - m^2 \left( \omega_{\rm I} + \omega_\nu \right) \\
&- \left( 2 l + 1 \right) \left( \frac{\zeta - \zeta_{\rm R}}{2} \right)^{1/2} = 0,\\
\end{split}
\end{equation}
where we have defined
\begin{equation}
\begin{split}
\zeta_{\rm R} \equiv& \xi \left[ \omega^2_{\rm R} \zeta_- - \left( \omega_{\rm I} - F_0 \right) \left( \zeta_0 + \omega_{\rm I} \zeta_+ \right) \right], \\
\zeta_{\rm I} \equiv& \xi \omega_{\rm R} \left[ \zeta_- \left( \omega_{\rm I} - F_0 \right) + \zeta_0 + \omega_{\rm I} \zeta_+ \right], \\
\zeta =& \left( \zeta_{\rm R}^2 + \zeta_{\rm I}^2 \right)^{1/2}.
\end{split}
\end{equation}
The wave amplitudes and steady-state solutions are identical to the expressions for the radial-field case, as given by equations (\ref{eq:sphere_mhd_amp}) and (\ref{eq:sphere_mhd_steady}), respectively, except that the expression for the generalized friction has changed,
\begin{equation}
\zeta_0 = \omega_\nu + {\cal R}_{\rm B} \left( \frac{v_{\rm A}}{2 \Omega R} \right)^2.
\end{equation}

\section{Applications to Exoplanetary Atmospheres}
\label{sect:apps}

The state of the art of characterizing exoplanetary atmospheres has advanced to the point where 2D infrared maps of the atmosphere may now be obtained, albeit in a non-unique manner \citep{dewit12,majeau12,majeau12b}.  These astronomical observations provide motivation for better understanding the global structure of atmospheres.  Here, we use our shallow water models to elucidate some general theoretical trends.  Since we are mostly interested in the global structure of exoplanetary atmospheres, we examine models with $n=k_{\rm x}=1$ (pseudo-spherical geometry) and $l=m=1$ (spherical geometry).

\subsection{The Effects of Stellar Irradiation and Hydrodynamic Friction}
\label{subsect:results_hydro}

Figure \ref{fig:beta_hydro_forced} shows examples of maps computed using our forced, damped, hydrodynamic shallow water models in pseudo-spherical geometry, which are a consistency check with the work of \cite{matsuno66}, \cite{gill80} and \cite{sp11}.  Due to a difference in the mathematical machinery used to arrive at the same solutions (see \S\ref{subsect:2d_beta_hydro}), our solutions are shifted in longitude ($x$), but this is of no consequence since they are periodic in $x$.  Specifically, \cite{matsuno66}, \cite{gill80} and \cite{sp11} start with a set of stationary (time-independent) equations, derive a key governing equation involving derivatives of the forcing and then perform a series expansion of the forcing to obtain their solutions.  We assume a functional form for the forcing from the beginning, but solve the time-dependent, key governing equation and obtain the stationary state as a final step.

For a free system ($\vert F_0 \vert = \omega_\nu = 0$), we recover the solution of \cite{matsuno66} (see his Figure 7).  For forced solutions with a moderate strength of friction present ($\vert F_0 \vert=1, \omega_\nu=0.1$ in dimensionless units\footnote{Normalized by the reciprocal of the dynamical time scale ($t_{\rm dyn}^{-1}$).}), we recover the familiar chevron-shaped feature published by \cite{matsuno66}, \cite{gill80} and \cite{sp11}.  A noteworthy feature of our approach is that the transition from the free to the forced, damped solutions is smooth with no translation in $x$.  A curious feature of the forced, damped solutions is a ``pinching" effect, which confines the solutions to be closer to the equator for either stronger forcing or weaker friction.  The same effect is seen in the solutions of \cite{sp11} (see their Figure 3).  

The pronounced nature of this pinching effect is an artifact of the $\beta$-plane approximation, partly because the Rossby number (or Lamb's parameter) does not explicitly appear in the solutions.  If one instead examines the solutions in full spherical geometry, one will see that the pinching becomes less pronounced (Figure \ref{fig:sphere_montage}).  The structure of the exoplanetary atmosphere is confined to be near the equator only when the Rossby number is less than unity.  The chevron-shaped feature, witnessed in 3D simulations of atmospheric circulation, is especially prominent when ${\cal R}_0=1$; strong forcing further accentuates it.

\subsection{The Effects of Radial Magnetic Fields and Magnetic Drag}

Next, we examine forced, damped, magnetized atmospheres.  We define our fiducial, hydrodynamic model as having $\vert F_0 \vert=5$ and $\omega_\nu=0.1$ (again in dimensionless units).  These parameter values were arbitrarily chosen to emphasize the chevron-shaped feature.  We then examine the effects of adding magnetic tension and magnetic drag on the atmospheric structure in both pseudo-spherical and spherical geometries.

A fundamental parameter involved is the (square of the) ratio of dynamical to Alfv\'{e}n time scales, which we estimate to be
\begin{equation}
\left( t_{\rm dyn}/t_{\rm A} \right)^2 \sim
\begin{cases}
1\mbox{--}10^2 & \mbox{(vertical/radial)}, \\
10^{-6}\mbox{--}10^{-4}  & \mbox{(horizontal)}. \\
\end{cases}
\label{eq:timescales_estimates}
\end{equation}
where we have adopted parameter values appropriate to hot Jupiters: $L_0 = R \sim 10^{10}$ cm, $H \sim 10^7$ cm, $v_{\rm A} \sim 10^2$--$10^3$ cm s$^{-1}$, $c_0 \sim 10^5$ cm s$^{-1}$.  These values of $v_{\rm A}$ correspond to field strengths of $\sim 1$--10 G and temperatures of $\sim 1500$ K at $\sim 1$ bar.  This diverse range of parameters stems from the difference between the characteristic horizontal and vertical/radial length scales involved and produces a rich variety of atmospheric structures.  These estimates already show that unless unrealistic field strengths are adopted, magnetized atmospheres with purely toroidal magnetic fields resemble their hydrodynamic counterparts.  For this reason, we will only examine models with purely vertical/radial background magnetic fields.  Algebraic intractability prevents us from exploring purely poloidal fields, which may produce markedly different structures.

In Figure \ref{fig:mhd_forced}, we examine examples of magnetized atmospheres with three field strengths: $\sim 1$ G, $\sim 3$ G and $\sim 10$ G.  For the spherical models, we set the Rossby number to be unity.  Magnetic tension and magnetic drag are degenerate effects when specified via $\alpha$, so it is sufficient to hold the magnetic Reynolds number fixed (${\cal R}_{\rm B}=1$) and vary $t_{\rm dyn}/t_{\rm A}$.  Generally, we see that the steady state of the atmosphere looks qualitatively different from its hydrodynamic counterpart.  The pseudo-spherical and spherical solutions are shifted in longitude by some amount, due to the slightly different mathematical approaches used to arrive at the steady-state solutions, but otherwise the computed maps are in qualitative agreement.  The familiar chevron-shaped feature seen in Figure \ref{fig:sphere_montage} is diluted by the enhanced presence of generalized friction ($\zeta_0 \sim 1$--100).  We again see that the pronounced nature of this pinching effect is an artifact of the $\beta$-plane approximation---it is diluted in the spherical models, even though we have set ${\cal R}_0=1$.  As the magnetic field strength increases, the flow transitions from being predominantly zonal (and possessing the chevron-shaped feature) to being predominant meridional.  When the field strength is $\sim 10$ G, the height perturbation field---a proxy for the temperature field---resembles the irradiation profile.  An analogous transition has been witnessed in 3D hydrodynamic simulations, which elucidate the transition from jet- to drag-dominated regimes \citep{showman13}.  The key difference is that, in our shallow water models, the flow converges at the substellar point, opposite from what the 3D models of \cite{showman13} find.

While we have elucidated trends using specific values of parameters, our formalism shows that forcing, hydrodynamic friction, magnetic drag and magnetic tension are degenerate effects that combine to determine the global structure of an exoplanetary atmosphere, at least with the approximations we have taken in deriving our analytical solutions.  This finding informs us that infrared phase curves alone will not suffice to uniquely distinguish between these different effects.  

\subsection{The Effects of Rotation}

In the slowly-rotating limit, forced, damped atmospheres, in the shallow water approximation, behave like purely damped ones in the absence of friction.  The velocity field is unaffected by all forms of friction, including magnetic tension, while the water height (a proxy for the temperature) is shifted in longitude.  Figure \ref{fig:sphere_slow} shows examples of free and forced, damped shallow water models on a sphere, where this phenomenon is clearly seen for both hydrodynamic and magnetized systems.  From the estimates made in equation (\ref{eq:timescales_estimates}), we consider systems with purely toroidal magnetic fields to be uninteresting, since they behave mostly like hydrodynamic systems.  Therefore, we consider only systems with purely radial background magnetic fields (but with horizontal field perturbations present) in Figure \ref{fig:sphere_slow}.  The basic conclusion is that, when rotation is unimportant, all forms of friction simply introduce a phase shift to the shallow water height perturbation.

Next, we ``turn on" rotation by examining models in the rapidly-rotating limit.  We start by focusing on free hydrodynamic models on a sphere.  When rotation becomes rapid, vortices start to appear in the velocity field (Figure \ref{fig:sphere_hydro_fast}).  When the Rossby number is of order unity (${\cal R}_0=0.5$ in our example), the solution resembles that on the $\beta$-plane (Figure \ref{fig:beta_hydro_forced}).  At ${\cal R}_0=0.05$, rotation becomes rapid enough that the atmospheric structure is confined to being near the equator (rotational pinching).

\subsection{Why Hydrodynamic Friction and Magnetic Drag are Fundamentally Different}

\begin{table*}
\label{tab:previous}
\begin{center}
\caption{Comparison to Previous Analytical Work}
\begin{tabular}{lcccc}
\hline
\hline
Reference & spherical geometry? & HD: forcing+friction? & MHD: free? & MHD: forcing+friction? \\
\hline
\hline
\cite{matsuno66} & N & Y & N & N \\ 
\cite{lindzen67} & N & Y & N & N \\
\cite{lh68} & Y & N & N & N \\
\cite{gill80} & N & Y & N & N \\
\cite{spit02} & N & Y & N & N \\
\cite{holton04} & N & N & N & N \\
\cite{kundu} & N & N & N & N \\
\cite{vallis06} & N & N & N & N \\
\cite{zara07} & Y & N & Y & N \\
\cite{hs09} & Y & N & Y & N \\
\cite{sp11} & N & Y & N & N \\
Heng \& Workman (current work) & Y & Y & Y & Y \\
\hline
\hline
\end{tabular}\\
HD: hydrodynamic.  MHD: magnetohydrodynamic.
\end{center}
\end{table*}

Among the sources of friction explored, Rayleigh drag is the easiest to incorporate into any model as it acts equally on all length scales and does not vary across either latitude or longitude.  These properties make it attractive to use Rayleigh drag to mimic magnetic drag, especially when adapting 3D hydrodynamic simulations of atmospheric circulation to include magnetic drag (e.g., \citealt{pmr10a}).  To some extent, this approach is reasonable within the context of our shallow water models, as all sources of friction, including magnetic tension, are included within the generalized friction ($\zeta_0$).  However, even when specifying $\zeta_0$, magnetic drag and magnetic tension act collectively and their overall scale dependence depends on the assumed field geometry.  Furthermore, our 1D models have demonstrated that hydrodynamic friction and magnetic drag have different phase signatures of damping.  Due to the approximations taken, we are unable to explore this property further for our 2D pseudo-spherical and spherical models, but it suggests that hydrodynamic friction and magnetic drag will alter the atmospheric flow in fundamentally different ways in a full-fledged, numerical calculation \citep{batygin13,rogers14}.

\subsection{Qualitatively Altering the Structure of Irradiated Exoplanetary Atmospheres}

Our study has shown that under a wide range of conditions, the chevron-shaped feature believed to be present in the highly-irradiated atmospheres of tidally-locked exoplanets is generic and robust, due to the near-universality of the quantum harmonic oscillator equation governing the meridional velocity.  Conversely, our study has also shown that the key governing equation does not follow that of the quantum harmonic oscillator if:
\begin{itemize}

\item Molecular viscosity or magnetic drag acts non-uniformly across latitude;

\item A poloidal magnetic field is present.

\end{itemize}
(See equations [\ref{eq:parabolic_hydro1}], [\ref{eq:parabolic_mhd_hydrofric1}] and [\ref{eq:parabolic_mhd_fric1}] for the first statement.  See \S\ref{subsect:2d_beta_mhd_free_hori} and \S\ref{subsect:spherical_beta_mhd_toroidal} for the second statement.)  That the meridional velocity ceases to be described by the quantum harmonic oscillator equation, under these conditions, suggests that the chevron-shaped feature will be qualitatively altered in the presence of these effects.

Hints of this behaviour have already been seen in the non-ideal MHD simulations of \cite{batygin13}, who assumed a constant magnetic resistivity and a shallow atmosphere (both assumptions of which we adopt) in the Boussinesq approximation (2D).  The conclusion regarding the importance of the poloidal magnetic field was also reached by \cite{rogers14} via performing 3D, non-ideal MHD simulations in the anelastic approximation.

\section{Discussion \& Summary}
\label{sect:discussion}

\subsection{Summaries}

The mathematical nature of this paper requires that we distinguish between the physical insights gained and the technical advancements made.

\subsubsection{Physical Summary}

\begin{itemize}

\item \textbf{Near-universality:} Atmospheres in the shallow water approximation are fundamentally described by the quantum harmonic oscillator equation, even when they are forced, rotating, magnetized and possess both hydrodynamic and magnetic sources of friction.  This near-universality is broken when either molecular viscosity or magnetic drag acts non-uniformly across latitude; it is also broken in the presence of a poloidal magnetic field.

\item \textbf{Key controlling parameter:} The global structure of an exoplanetary atmosphere is essentially controlled by a single, dimensionless number that we call the ``key controlling parameter" ($\alpha$).  In the hydrodynamic limit, $\alpha$ is directly related to the Prandtl and Rossby numbers.  When magnetic fields are present, $\alpha$ additionally involves magnetic tension and magnetic drag as sources of friction.  In pseudo-spherical geometry, it was previously realized that $1/\alpha^4$ is the Prandtl number in the hydrodynamic limit.  We demonstrate that in full spherical geometry, this description is incomplete as $\alpha$ generally involves the Rossby number as well.

\item \textbf{Global structure of exoplanetary atmospheres:} We are able to solve for the steady state of an atmosphere in the presence of forcing, friction, rotation and magnetic fields.  We use our analytical solutions to elucidate the manifestation of each effect in 2D thermal maps.  Generally, there is degeneracy between the various effects and it will require multi-wavelength data, across multiple epochs, to disentangle them.

\item \textbf{Hydrodynamic friction versus magnetic drag:} Molecular viscosity acts predominantly on small scales, while Rayleigh drag acts equally on all scales.  Magnetic tension and magnetic drag act collectively---whether they favor large scales or are collectively scale-free depends on the field geometry.  Using Rayleigh drag to mimic magnetic drag is akin to asserting that it acts preferentially on a scale that is germane to the problem.  Hydrodynamic friction and magnetic drag possess dissimilar phase signatures and will generally alter the atmospheric flow in qualitatively different ways.

\item \textbf{Rotation:} Rotation, an intrinsically 2D phenomenon, modifies the balance between forcing and friction in a non-trivial manner.  When rotation is unimportant, forced atmospheres, in the shallow water approximation, behave like purely damped ones, as if forcing was absent.  In spherical geometry, rapid rotation acts to confine the global structure of the atmosphere to be near the equator.

\item \textbf{Pinching effect:} In the shallow water approximation, atmospheres experience a pinching effect that is caused by faster rotation, stronger forcing or weaker friction, because all of these effects cause the key controlling parameter ($\alpha$) to take on higher values.  In pseudo-spherical geometry, this pinching effect is more pronounced---it is an artifact of the equatorial $\beta$-plane approximation, partly because it does not explicitly involve the Rossby number.

\item \textbf{Coupling of physical effects:} Forcing, rotation, magnetic fields and sources of friction couple in various ways to modify the frequencies and structures of waves in shallow water systems.

\end{itemize}

\subsubsection{Technical Summary}

\begin{itemize}

\item \textbf{Broad theoretical survey:} Our survey of shallow water models covers a broad range of technical possibilities, exploring: dimensionality (1D and 2D); geometry (Cartesian, pseudo-spherical and spherical); free, forced and damped systems; hydrodynamic and MHD systems; hydrodynamic and magnetic sources of friction.  Generally, shallow water systems may be described by five parameters, albeit with a series of assumptions and caveats (see ``Obstacles to analytical solution"): the forcing ($F_0$), the hydrodynamic friction ($\omega_\nu$), the magnetic Reynolds number (${\cal R}_{\rm B}$), the Rossby number (${\cal R}_0$) and the ratio of dynamical to Alfv\'{e}n timescales ($t_{\rm dyn}/t_{\rm A}$).

\item \textbf{Generalized, complex frequencies:} When generalizing from free, hydrodynamic systems to forced MHD systems with friction, the governing equations, dispersion relations and wave solutions contain complex frequencies that are generalizations of the wave frequency, which is strictly real in free systems.  Four complex frequencies are needed: $\omega_{\rm F}$ (forcing), $\omega_0$ (hydrodynamic friction), $\omega_\eta$ (magnetic drag) and $\omega_{\rm B}$ (magnetic tension and friction).  With each generalization, the mathematical equations retain the same form, except that the wave frequency ($\omega$) in various places is substituted with one of these four complex frequencies.  Generally, $\omega_{\rm B}$ contains a quantity denoted by $\zeta_0$, which we term the ``generalized friction" as it involves all sources of friction, including magnetic tension.

\item \textbf{Slowly- versus rapidly-rotating limits:} In full spherical geometry, the key governing equation has two limiting forms with analytical solutions (Figure \ref{fig:schematic}).  In the slowly-rotating limit, it is the associated Legendre equation, which yields spherical harmonics for the wave solutions.  In the rapidly-rotating limit, it is the quantum harmonic oscillator equation, which yields Hermite polynomials as solutions for the wave amplitudes.  The solutions on the equatorial $\beta$-plane mirror the spherical solutions when the Rossby number is of order unity.  

\item \textbf{Dispersion relations:} For each system, we obtain the dispersion relations: a pair of equations for the real ($\omega_{\rm R}$) and imaginary ($\omega_{\rm I}$) components of the wave frequency, which describe the oscillatory and growing or decaying parts of the wave, respectively.  The $\beta$-plane and spherical derivations mirror each other, except that Lamb's parameter ($\xi$) is generally not unity for the latter.   Deriving the dispersion relations requires the use of De Moivre's formula and a set of separation functions ($\zeta_{\rm R}$, $\zeta_{\rm I}$, $\zeta$, $\zeta_+$, $\zeta_-$ and $\zeta_0$).

\item \textbf{Obstacles to analytical solution:} On the $\beta$-plane, we are unable to proceed analytically unless we ignore the poloidal background magnetic field and its perturbations.  On a sphere, we further restrict ourselves to near-equator solutions.  In all of the models, we require molecular viscosity and magnetic drag to act uniformly across latitude.

\end{itemize}
Table 3 provides an executive summary of the lessons learnt from each shallow water model, in order of increasing sophistication, and lists progressively the approximations needed to render the problem amenable to analytical solution.

\subsection{Comparison to Previous Analytical Work}

To place our present study in context, we provide a comparison to previous analytical work in Table 2.  Generally, the early works on the shallow water system were inspired by studies of the terrestrial atmosphere or ocean and tend to focus on free or forced, damped hydrodynamic systems \citep{matsuno66,lindzen67,lh67,gill80}, including the monographs of \cite{holton04}, \cite{kundu} and \cite{vallis06}.  Most works did not compute the shallow water system in spherical geometry and instead utilized the $\beta$-plane approximation, with the seminal work of \cite{lh68} being a notable exception.

Later works that were inspired by the study of the Sun \citep{gilman00,zara07} or neutron stars \citep{spit02,hs09} tend to focus on free magnetized systems.  Little attention has been paid to studying forced, magnetized systems with friction, since this is an unfamiliar regime for the atmospheric dynamics of these objects.  The first generalization of the shallow water system to exoplanets considered forced, damped hydrodynamic systems on the $\beta$-plane \citep{sp11}.  Furthermore, there has been no previous generalization to consider both vertical/radial and horizontal magnetic fields, as well as both slow and rapid rotation, within a single study.  We have written down the governing equations, dispersion relations and wave solutions for every system, which was previously not done even for forced, damped hydrodynamic systems.

As already mentioned, while it was realized that the meridional velocity is governed by the quantum harmonic oscillator equation in the limit of a free, hydrodynamic, shallow water system (either in pseudo-spherical or spherical geometry), it was previously not demonstrated that this property extended to forced, damped, magnetized, time-dependent systems.

\subsection{Relevance to Atmospheric Circulation Simulations}

Since the most easily characterizable exoplanets are the large, highly-irradiated gas giants, there has been intense interest in understanding the physics of $\sim 1000$--3000 K, partially-ionized atmospheres.  Initial work in the field has focused on adapting 3D general circulation models, traditionally used for studying the relatively quiescent and neutral atmosphere of Earth, towards understanding hot Jupiters (e.g., \citealt{showman09}).  Besides the formidable problem of working in unfamiliar physical and chemical regimes, several concerns have been raised about the numerical issues surrounding such studies, ranging from the ambiguity associated with numerical dissipation \citep{hmp11,tc11} to the possible non-uniqueness of solutions due to differing initial conditions \citep{tc10,ls13}.  Additionally, shocks are expected to exist in these highly-irradiated atmospheres \citep{dd08,lg10,heng12}.  No simulation has succeeded in including non-ideal MHD and shocks in a 3D general circulation model.  The analytical solutions in this study provide a point of reference and a suite of tests for building up to such a simulation, although it should be noted that shallow water systems do not include shocks by definition.

\subsection{The Correspondence Between the Shallow Water and Isothermal Euler Equations}

The shallow water system has a direct correspondence to the mass continuity and isothermal Euler equations in 2D.  Consider $\rho = \rho_0 + \rho^\prime$, where $\rho_0$ is the background mass density and $\rho^\prime$ is the perturbation of the mass density.  When we let $h \rightarrow \rho$ in equation (\ref{eq:height}), it becomes the mass continuity equation in 2D.  Using the ideal gas law ($P = \rho {\cal R}_{\rm gas} T_0$), one can show that $g$ needs to be replaced by ${\cal R}_{\rm gas} T_0/\rho_0$ in the momentum equation, where ${\cal R}_{\rm gas}$ is the specific gas constant and $T_0$ is a constant value of the temperature.  The square of the characteristic velocity is then $c_0^2 = {\cal R}_{\rm gas} T_0$, which is the specific energy.  

In practice, this correspondence means that one may adapt numerical fluid dynamical solvers to mimic shallow water systems and compare them with the analytical solutions presented in this study.

\subsection{Temporal Behavior of Forced, Damped, Magnetized Systems: Power Spectra of Exoplanets and Brown Dwarfs}

In the present study, we have derived a suite of dispersion relations, which describe how the frequencies of Poincar\'{e} and Rossby waves are modified in the presence of forcing, friction and magnetic fields.  Previous mathematical techniques allowed for the dispersion relations to be derived only in the free (unforced) limit \citep{matsuno66,lindzen67,lh68,gill80}.  These dispersion relations need to be solved numerically, which is beyond the scope of the present study.  Their solution allows for time-dependent wave solutions to be constructed, which may aid the interpretation of simulations.  Such solutions are relevant for studying the temporal behavior of exoplanets \citep{showman09,agol10} and brown dwarfs \citep{artigau09,sk13,rm14}.  One may also derive the dispersion relations and wave solutions near the poles \citep{lh68,hs09}; in the present study, we have focused on near-equator solutions.

\acknowledgments
KH acknowledges financial, secretarial and logistical support from the Center for Space and Habitability (CSH) and the Space Research and Planetary Sciences (WP) Division of the University of Bern, as well as grants from the Swiss National Science Foundation (SNSF) and the Swiss-based MERAC Foundation for the Exoclimes Simulation Platform (\texttt{www.exoclime.org}).  We are grateful to S\'{e}bastien Fromang for useful conversations.

\begin{table*}
\label{tab:summary}
\begin{center}
\caption{Executive Summary of Shallow Water Models Explored in this Study}
\begin{tabular}{lll}
\hline
\hline
Name & Section & Salient Properties \\
\hline
\hline
HD: free (1D) & 3.1.1 & Only non-dispersive gravity waves exist. \\
HD: molecular viscosity (1D) & 3.1.2 & Molecular viscosity acts preferentially on small scales. \\
HD: Rayleigh drag (1D) & 3.1.3 & Rayleigh drag acts equally on all scales. \\
HD: hydrodynamic friction (1D) & 3.1.4 & Molecular viscosity and Rayleigh may be represented by a single \\
 & & frequency ($\omega_\nu$). \\
HD: forcing (1D) & 3.1.5 & Forcing decreases the frequency of sinusoidal oscillations. \\
HD: forcing and hydrodynamic friction (1D) & 3.1.6 & Forcing and frictional frequencies need to be equal to create a \\
 & & balanced flow. \\
MHD: free (1D) & 3.2.1 & Magnetic field is out of phase with the velocity and height perturbations. \\
MHD: magnetic drag (1D) & 3.2.2 & Balanced flow does not exist unless $\eta=0$. \\
MHD: forcing (1D) & 3.2.3 & Balanced flow does not exist unless $F_0=0$. \\
MHD: hydrodynamic friction (1D) & 3.2.4 & Hydrodynamic friction retains the same phase signatures even in the presence \\
 & & of a magnetic field. \\
MHD: friction (1D) & 3.2.5 & Magnetic drag and hydrodynamic friction have different phase signatures \\
 &  & and may negate each other in a flow. \\
 MHD: forcing and friction (1D) & 3.2.6 & Forcing, friction and magnetic tension couple via the dispersion relations \\
  & & and wave solutions, even in the absence of rotation. \\
HD: forcing and hydrodynamic friction (2D) & 4.1 & Rotation modifies the coupling between forcing and hydrodynamic friction. \\
MHD: forcing and friction (2D, vertical) & 4.2 & First example where algebra is tedious enough to require the use of \\
 & & separation functions. \\
MHD: forcing and friction (2D, horizontal) & 4.3 & --- \\
Relationship between equations & 5.1 & The Weber and quantum harmonic oscillator equations are related \\
 & & by a transformation of variables. \\
HD: free (2D, $\beta$-plane) & 5.2.1 & Key governing equation is the quantum harmonic oscillator equation. \\
 & & Its solution involves Hermite polynomials. \\
HD: forcing and hydrodynamic friction & 5.2.2 & We recover the same solutions as \cite{matsuno66}, \cite{gill80} \\
(2D, $\beta$-plane) & & and \cite{sp11} if we set $F_0 = - \vert F_0 \vert$.  One has to pick \\
 & & the positive root when using De Moivre's formula. \\
MHD: free (2D, $\beta$-plane, vertical) & 5.3.1 & Key controlling parameter is the ratio of dynamical to Alfv\'{e}n timescales. \\
MHD: forcing and hydrodynamic friction & 5.3.2 & Complex, generalized frequency ($\omega_{\rm B}$) that involves hydrodynamic friction \\
(2D, $\beta$-plane, vertical) & & and magnetic tension is defined to make the algebra tractable.  Problem is \\
 & & intractable unless molecular viscosity is assumed to act uniformly \\
 & & across latitude. \\
MHD: forcing and friction & 5.3.3 & $\omega_{\rm B}$ now involves magnetic drag.  The key controlling parameter ($\alpha$) involves \\
(2D, $\beta$-plane, vertical) & & forcing, friction and magnetic tension.  Magnetic tension and magnetic drag \\
 & & act collectively and preferentially on large scales.  Problem is intractable \\
 & & unless magnetic drag is assumed to act uniformly across latitude. \\
MHD: free (2D, $\beta$-plane, horizontal) & 5.4.1 & Problem is intractable unless the poloidal magnetic field is completely \\
 & & neglected. \\
MHD: forcing and friction (2D $\beta$-plane, horizontal) & 5.4.2 & Magnetic tension and magnetic drag act collectively and equally on all scales. \\
HD: free (2D, spherical) & 6.1.1 & Slowly- and rapidly-rotating limits exist for the analytical solutions, \\
 & & controlled by the Rossby number (or Lamb's parameter).  The former regime \\
 & & yields spherical harmonics as the wave solutions, while the latter regime \\
 & & yields Hermite polynomials for the wave amplitudes. \\
HD: forcing and hydrodynamic friction (2D, spherical) & 6.1.2 & Require molecular viscosity to act uniformly across latitude. \\
 & & Seek near-equator solution. \\
MHD: forcing and friction (2D, spherical, radial) & 6.2 & Solutions mirror those on the $\beta$-plane, except that $\xi \ne 1$ in general. \\
MHD: forcing and friction (2D, spherical, toroidal) & 6.3 & --- \\
\hline
\hline
\end{tabular}\\
Note: all 2D models include rotation.\\
HD: hydrodynamic.  MHD: magnetohydrodynamic.
\end{center}
\end{table*}

\appendix

\section{Useful Expressions and Identities}
\label{append:useful}

Some useful, commonly used expressions include
\begin{equation}
\begin{split}
\omega^2 &= \omega_{\rm R}^2 - \omega_{\rm I}^2 + 2 i \omega_{\rm R} \omega_{\rm I}, \\
\omega^3 &= \omega_{\rm R}^3 - i \omega_{\rm I}^3 - 3 \omega_{\rm R} \omega_{\rm I}^2 + 3 i \omega_{\rm R}^2 \omega_{\rm I}, \\
\hat{D} v_{\theta_0} &= v_0 \alpha ~\exp{\left( - \frac{\tilde{\mu}^2}{2} \right)} ~\left( 2 l \tilde{{\cal H}}_{l-1} - \tilde{\mu} \tilde{{\cal H}}_l \right).
\end{split}
\end{equation}
The expression involving $\hat{D} v_{\theta_0}$ makes the approximation that $(1-\mu^2) \approx 1$.


\label{lastpage}

\end{document}